**Solubility of Rock in Steam Atmospheres of Planets**

Short title: Rock solubility in steam atmospheres

Bruce Fegley, Jr.[1,2], Nathan S. Jacobson[3], K.B. Williams[2],

J.M.C. Plane[4], L. Schaefer[5], and Katharina Lodders[1,2]

[1]Planetary Chemistry Laboratory, McDonnell Center for the Space Sciences

[2]Department of Earth & Planetary Sciences

Washington University, St. Louis, MO 63130 USA

[3]Materials Division, NASA Glenn Research Center, MS106-1, 21000 Brookpark Road, Cleveland, OH 44135 USA

[4]School of Chemistry, National Centre for Atmospheric Science, and School of Earth and Environment, University of Leeds, Leeds LS2 9JT, United Kingdom

[5]Harvard – Smithsonian Center for Astrophysics, 60 Garden Street, Cambridge, MA 02138 USA

Corresponding author: Bruce Fegley Jr., bfegley@wustl.edu

**Abstract.** Extensive experimental studies show all major rock-forming elements (e.g., Si, Mg, Fe, Ca, Al, Na, K) dissolve in steam to a greater or lesser extent. We use these results to compute chemical equilibrium abundances of rocky element – bearing gases in steam atmospheres equilibrated with silicate magma oceans. Rocky elements partition into steam atmospheres as volatile hydroxide gases (e.g., $Si(OH)_4$, $Mg(OH)_2$, $Fe(OH)_2$, $Ni(OH)_2$, $Al(OH)_3$, $Ca(OH)_2$, $NaOH$, $KOH$) and via reaction with HF and HCl as volatile halide gases (e.g., NaCl, KCl, CaFOH, CaClOH, FAl(OH)$_2$) in much larger amounts than expected from their vapor pressures over volatile-free solid or molten rock at high temperatures expected for steam atmospheres on the early Earth and hot rocky exoplanets. We quantitatively compute the extent of fractional vaporization by defining gas/magma distribution coefficients and show Earth's sub-solar Si/Mg ratio may be due to loss of a primordial steam atmosphere. We conclude hot rocky exoplanets that are undergoing or have undergone escape of steam-bearing atmospheres may experience fractional vaporization and loss of Si, Mg, Fe, Ni, Al, Ca, Na, and K. This loss can modify their bulk composition, density, heat balance, and interior structure.

## 1. Introduction.

We investigated the solubility of rocky elements, in particular Mg, Si, and Fe in H$_2$O-rich (henceforth steam) atmospheres and the potential effects of their solubility for composition of hot rocky exoplanets and their atmospheres. Magnesium, silicon, and iron are the three most abundant elements in solar composition material that combine with oxygen to form rock (Lodders 2003). Their atomic abundances on the cosmochemical scale are similar to one another (within 20%) and are $1.03 \times 10^6$ (Mg), $1.00 \times 10^6$ (Si), and $0.848 \times 10^6$ (Fe). Other rock-forming elements that we also consider such as Al ($0.0846 \times 10^6$), Ca ($0.0604 \times 10^6$), Na ($0.0577 \times 10^6$), Ni ($0.049 \times 10^6$), and K ($0.00376 \times 10^6$) are much less abundant and we focus on Mg, Si, and Fe.

Oxygen, Mg, Si, and Fe are also the major elements in the silicate portions of meteorites, the Earth (O + Mg + Si + Fe > 90% by mass), the other three terrestrial planets, and Earth's Moon (see the geochemical analyses for meteorites, the Earth, Moon, Mars, and Venus in Lodders & Fegley 1998, and for Mercury in Nittler et al. 2011). Spectroscopic studies of main sequence F and G stars with near-solar metallicity show constant ratios of Fe, Mg, and Si to one another (see section 3.4.7 in Lodders, Palme & Gail 2009). It is safe to assume that Mg, Si, and Fe are the most abundant rock-forming elements combined with oxygen in rocky exoplanets and the rocky cores of gas-rich and water-rich exoplanets around stars with solar or near-solar metallicity.

The solubility of Mg, Si, and Fe in steam atmospheres is significant. High- pressure steam in equilibrium with quartz + $SiO_2$ – rich melt at 9.5 – 10 kilobars and $\sim$ 1080 C (the upper critical end point in the $SiO_2$ – $H_2O$ system) is $\sim$ 50 mole % silica (Kennedy et al. 1962, Newton & Manning 2008) and molten $SiO_2$ + $H_2O$ are completely miscible at higher temperatures. The significant solubility of Si and other rocky elements in steam (over a wide P – T range) raises interesting possibilities. One is the formation of *potentially* spectroscopically observable gases such as $Si(OH)_4$, $Mg(OH)_2$, $Fe(OH)_2$, $Ni(OH)_2$, $Al(OH)_3$, $Ca(OH)_2$, NaOH, and KOH and their photolysis products. Another is loss of Mg, Si, Fe, Ni, Al, Ca, Na, and K from hot rocky exoplanets that are losing or have lost steam-bearing atmospheres. Significant changes in the relative ratios of Mg, Si, Fe, Ni may alter the bulk composition, density and interior structure of the remnant rocky planet left after loss of an early-formed steam atmosphere. The loss of radioactive [40]K may also affect the heat balance of a remnant rocky planet. The loss of Si, Al, Ca, Na, and K – abundant in Earth's continental crust – may alter the surface composition, mineralogy, and structure of a remnant rocky planet.

Our work is motivated by three disparate developments – (1) observations of over 100 hot rocky exoplanets in recent years, (2) theoretical models of steam atmospheres on the early Earth and rocky exoplanets, and (3) experimental measurements of the solubility of minerals and rocks in steam.

Nearly all of the known hot rocky exoplanets are closer to their host stars than Mercury is to the Sun. All small exoplanets ($R < 2.7$ $R_{Earth}$) with well-constrained masses (as of December 2015) receive at least 10 times more stellar insolation than the Earth (e.g. Fig. 13, Gettel et al. 2015), with correspondingly higher equilibrium temperatures. The hottest of these are planets such as CoRoT-7b and Kepler-10b with equilibrium temperatures greater than 2000 K. However, others, like the newly discovered MEarth planet GJ 1132 b (Berta-Thompson et al. 2015) and the closest and brightest transiting super-Earth HD 219134 b (Motalebi et al. 2015) have lower temperatures of 500 K and 1100 K, respectively. Many of the hot rocky exoplanets lie on a density curve consistent with the composition of the Earth (Dressing et al. 2015). However this population of planets ($R < 2.7$ $R_{Earth}$) also includes objects with densities low enough to require substantial volatile envelopes on top of their solid (or liquid) surface. These include 55 Cancri e, Kepler-454 b, Kepler-11b, Kepler 48-c, HIP 116454b, HD 97658b, and Kepler-10c, which have equilibrium temperatures ranging from ~600 K to greater than 2000 K. New planets in this radius range are being discovered rapidly with K2 (e.g. Vanderburg et al. 2015), and even more plan-ets in short period orbits will probably be discovered following the launches of the Transiting Exoplanet Survey Satellite (TESS) mission and the CHaracterizing Ex-OPlanet Satellite (CHEOPS) mission in 2017. The James Webb Space Telescope (JWST), slated for launch in 2018, should be able to take detailed infrared spectra of these planets' atmospheres.

The planets discussed above are important here because, given their high temperatures and an Earth-like volatile abundance, they could have a steam atmosphere that would generate surface temperatures hot enough to melt silicates. For comparison, (water-poor) Venus has an equilibrium temperature of ~260 K but its atmosphere of ~ 95 bars of $CO_2$ (with much smaller amounts of $SO_2$ and $H_2O$) produces surface temperatures of ~ 740 K. Venus's surface is almost hot enough to melt alkali-rich silicates, e.g., the albite – sodium disilicate eutectic is 767 K (Table 12-1 in Fegley 2013), and all of the planets mentioned above have significantly higher equilibrium temperatures than Venus. Although steam atmosphere conditions on the Earth were likely transient, the lifetime of potential steam atmospheres on the hot rocky exoplanets would be limited only by atmospheric escape. Hydrodynamic escape of hydrogen can also drag along heavier elements – up to Xe – if the outflow is strong enough (e.g., Hunten, Pepin & Walker 1987, Pepin 1997). Therefore, the solubility of rocky elements in steam may lead to elemental fractionation on planets with long-lived steam atmospheres undergoing escape. However we stress our chemical equilibrium calculations are not tied to any particular planet mentioned above, but are meant to map out atmospheric chemistry across a wide P, T range.

Our previous models were about outgassing during planetary accretion and atmospheric chemistry of rocky planets in our solar system and other planetary systems and used chemical equilibrium and chemical kinetic calculations. Schaefer & Fegley (2007, 2010) modeled the composition of the major volatile-bearing gases (H, C, N, O, S) in outgassed atmospheres as functions of temperature and total pressure for the different types of chondritic material (i.e., carbonaceous (CI, CM, CV), ordinary H, L, LL), and enstatite (EH, EL)). Schaefer & Fegley (2009) did chemical equilibrium models of silicate vapor atmospheres on volatile-free hot rocky exoplanets such as CoRoT-7b. Schaefer, Lodders & Fegley (2012) considered vaporization of volatile-bearing hot rocky exoplanets like the Earth using two rocky compositions – Earth's $SiO_2$-rich continental crust and the MgO- and FeO-rich bulk silicate Earth (BSE). The BSE is the composition of Earth's silicate portion before it evolved into the atmosphere, oceans, crust, and mantle. It has a mass of $4.03 \times 10^{24}$ kg, of which the mantle is 99.4%, so the BSE composition is close to that of Earth's mantle.

Outgassing of the two model compositions generated atmospheres rich in steam and $CO_2$ with variable amounts of other gases depending on pressure and temperature (e.g., see Figures 1 – 5, and Table 3 in Schaefer, Lodders & Fegley 2012). The major Mg, Si, and Fe gases in their 100 bar model were $Mg(OH)_2$, SiO, and $Fe(OH)_2$.

At the time the calculations in Schaefer, Lodders & Fegley (2012) were done, a thorough assessment of the thermodynamics of $SiO_2$ solubility in steam and the derived thermodynamic properties of $Si(OH)_4$ gas was unavailable. Fegley (2014) used the recently published $Si(OH)_4$ data of Plyasunov (2011a, 2012) and found $Si(OH)_4$ partial pressures 10,000 times larger than the SiO partial pressure expected from Si vaporization from anhydrous lavas at the same conditions (BSE-like melt at 1873 K in a 100 bar $H_2O$ – $CO_2$ atmosphere). This preliminary result warrants more comprehensive models of rocky element solubility in steam atmospheres.

This paper is organized as follows. Section 2 briefly reviews the history of prior work on steam atmosphere models, describes effects of steam atmospheres on rock melting, and discusses the size of steam atmospheres expected from the current $H_2O$ and $CO_2$ content of Earth's mantle for the early Earth. Section 3 reviews prior experimental and theoretical studies on the solubility of rock-forming elements in steam and focuses on Si, the rocky element that is the most soluble in steam.

Section 4 describes the methods used in our chemical equilibrium calculations. Section 5 compares the solubility of Mg, Si, Fe, Ni, Al, and Ca in steam to the vapor pressure of the pure oxides. Section 6 demonstrates that other gases possibly present in steam atmospheres ($CO_2$, $N_2$, $SO_2$, $O_2$, and $CH_4$) are inert dilutants that do not alter the solubility of Mg, Si, and Fe in steam.

Section 7 describes the results of our chemical equilibrium calculations of metal hydroxide gas abundances in steam atmospheres of hot rocky exoplanets. These calculations take into account chemical interactions with magma oceans on these planets. (We use the terms "rocky elements" and "metals" interchangeably.) The effects of fractional vaporization of rocky elements on the bulk composition of the residual planet are illustrated in several figures and tabulated using gas/magma distribution (i.e., partition) coefficients. We show the Si/Mg ratio in the bulk silicate Earth can be produced by loss of a steam atmosphere with a few % of the BSE mass. Section 7 also describes the effects of stellar UV photolysis on abundances of the major hydrox-ide gases of Mg, Si, and Fe and lists some gases that may be observable spectroscopically. Section 8 discusses some cosmochemical applications of our work and suggests some future avenues. Section 9 summarizes our major conclusions.

## 2. Steam Atmospheres

### 2.1 Historical review

Arrhenius, De & Alfvén (1974) proposed heating during accretion of the Earth degassed water-bearing minerals in the accreted planetesimals and formed a steam atmosphere. The steam atmosphere formed Earth's hydrosphere as the Earth cooled, a process which may have taken $\sim 2.5$ million years (Sleep, Zahnle & Neuhof 2001). Subsequent experiments showed water and $CO_2$ are the two major volatiles formed by impact degassing of CM2 carbonaceous chondritic material during planetary accretion (e.g., Lange & Ahrens, 1982; Tyburczy, Frisch & Ahrens 1986). Chemical equilibrium calculations showed $H_2O$ and $CO_2$ are the two major gases formed by impact degassing of CI, CM2, and CV3 chondritic material (Schaefer & Fegley 2010). Theoretical models of the origin and evolution of an impact generated steam atmosphere on the early Earth were presented by Abe & Matsui (e.g., Abe & Matsui 1985, 1988; Matsui & Abe 1986).

Fegley & Schaefer (2014) modeled a massive ($\sim 1,000$ bar) $H_2O - CO_2 - SO_2$ steam atmosphere on the early Earth and computed gas phase chemical equilibria in it from 2000 – 6000 K. They found thermal dissociation of $H_2O$, $CO_2$, and $SO_2$ produced increasing amounts of OH, $H_2$, CO, $O_2$, H, O, SO with increasing temperature at constant total pressure (see their Figure 5). They also showed a steam atmosphere was significantly more oxidizing with a higher oxygen fugacity ($fO_2$) than the solar nebula and suggested that easily oxidized elements such as Si, Fe, Cr, Mo, W, B, V, would vaporize from the magma ocean as hydroxides (e.g., $Si(OH)_4$, $Fe(OH)_2$, $H_2CrO_4$, $H_2MoO_4$, $H_2WO_4$, $H_3BO_3$) and gaseous oxides of Cr, Mo, V, W. This is potentially important for the early Earth because geochemical signatures may be in the rock record (Fegley, Lodders & Jacobson 2016).

## 2.2 Effects on rock melting

Water vapor and $CO_2$ are greenhouse gases and the development of a massive steam atmosphere and a magma ocean at the planetary surface are closely linked (e.g., Abe & Matsui 1985, 1988, Matsui & Abe 1986; Abe 1993; Abe 2011; Elkins-Tanton 2008; LeBrun et al. 2013; Zahnle, Kasting & Pollack 1988). A sufficiently massive steam atmosphere can heat the surface of a rocky planet to (and above) the melting point of rock (e.g., see the discussion in Zahnle, Kasting & Pollack 1988).

At one bar pressure peridotite, the major rock in Earth's upper mantle, starts to melt at 1120 – 1200 C (1390 – 1473 K, the solidus, $T_{sol}$) and is completely molten by $\sim$ 1970 K (the liquidus, $T_{liq}$) (e.g., see Kushiro, Syono & Akimoto 1968, Takahashi 1986, Takahashi et al. 1993). The bulk composition of peridotite rock from different locales, in particular the Na/Ca ratio, alters the solidus temperature (Green 2015). Peridotite melting has a positive Clapeyron slope $dT_{sol}/dP$ of $\sim$ 12 K kbar$^{-1}$ (120 K GPa$^{-1}$) in the 1 bar – 50 kilobar range (Kushiro, Syono & Akimoto 1968, Green 2015) and the increased pressure caused by the weight of a massive steam atmosphere will increase the melting point. However this is counteracted by the freezing point depression due to the solubility of $H_2O$ (more soluble) and $CO_2$ (less soluble) in sili-cate magmas. The negative $\Delta T$ from the freezing point depression is larger than the positive $\Delta T$ from the increased pressure and the net effect is that the melting point of $H_2O$-saturated peridotite is less than that of dry peridotite, by about 400 degrees at 26 kilobars pressure ($\approx$ 80 km depth, see Figure 1 in Kushiro, Syono & Akimoto 1968). Dissolution of $H_2O$ and $CO_2$ also lowers the freezing points of other molten rocks and minerals and is a general effect that is expected to occur on any rocky exoplanet made of silicates that also contains $CO_2$ and water.

### 2.3 Steam atmosphere on the early Earth

The properties (e.g., mass, composition, lifetime) of a steam atmosphere on a planet depend on several factors such as the total amount of water and other volatiles, fractional amount of the volatiles that are outgassed into the atmosphere, planetary surface temperature, planetary distance from the primary star, and primary star type (e.g., see Hamano, Abe & Genda 2013, Hamano et al 2015). For illustration we briefly discuss possible properties of a steam atmosphere on early Earth.

The mass fraction (in ppm = parts per million) of hydrogen in the bulk silicate Earth is 120 ppm ($\sim$ 1070 ppm as $H_2O$) (Palme & O'Neill 2014). This mass fraction $H_2O$ is equivalent to $\sim 4.3 \times 10^{21}$ kg water versus $\sim 1.7 \times 10^{21}$ kg $H_2O$ in the hydrosphere (oceans + glaciers + freshwater). Thus only about 40% of Earth's total water is outgassed on its surface and additional water $\sim$ 1.6 times that in the hydrosphere remains inside the bulk silicate Earth. Other estimates of water in the BSE are smaller but they still give about one hydrosphere worth of water inside the Earth (Saal et al. 2002; Hirschmann & Dasgupta 2009).

Palme and O'Neill (2014) list 100-ppm carbon ($\sim$ 370 ppm as $CO_2$) in the bulk silicate Earth. Other estimates for the carbon content of the BSE range from 46 – 250 ppm (summarized in Table 6.9 of Lodders & Fegley 1998). Using the Palme & O'Neill (2014) values, mass balance shows outgassing of all hydrogen and carbon in the BSE as $H_2O$ ($4.3 \times 10^{21}$ kg) and $CO_2$ ($1.5 \times 10^{21}$ kg) would give a steam atmosphere with a surface pressure of $\sim$ 1,100 bar composed of $\sim$ 75% steam and 25% $CO_2$ (P = mg, using g = 980.665 cgs). LeBrun et al. (2013) consider a similar range of 100 – 1,000 bars for a steam – $CO_2$ atmosphere on the early Earth.

This calculation is illustrative and assumes the silicate portion of the early Earth had the same composition and mass as the BSE and current surface gravity. Earth's volatile depletion with respect to chondritic material and solar abundances suggests all estimates of its current volatile content are plausibly smaller than its initial endowment (e.g., see pp. 73-77 in Fegley & Schaefer 2014). Although the exact properties of steam atmospheres on the early Earth and hot rocky exoplanets depend on several variables, we explicitly assume steam atmospheres form and we explore their effect on chemistry of rock-forming elements with an emphasis on the major elements Si, Mg, and Fe.

### 3. Past work on the solubility of rocky elements in steam

Extensive experimental work going back to the 1930s shows that most elements found in rocks are soluble in steam (e.g., see Alexander, Ogden & Levy 1963, Maeda, Sasomoto & Sata 1978, Hashimoto 1992 for MgO; Antignano & Manning 2008, Nguyen et al. 2014 for $TiO_2$, Belton & Richardson 1962, Belton & Jordan 1967 for Co, Fe,

Ni; Matsumoto & Sata 1981, Hashimoto 1992 for CaO; Hashimoto 1992, Opila & Myers 2004 for $Al_2O_3$; Meschter, Opila & Jacobson 2013 for a review of all elements; Morey 1957 for $Al_2O_3$, $BaSO_4$, BeO, $CaCO_3$, $CaSO_4$, $Fe_2O_3$, $GeO_2$, NaCl, $Na_2SO_4$, $Nb_2O_5$, NiO, $PbSO_4$, $SiO_2$, $SnO_2$, $Ta_2O_5$, and ZnS; Preston & Turner 1934 and Van Nieuwenberg and Blumendal (1930, 1931a,b) for $SiO_2$; Shen & Keppler 1997, Bureau & Keppler 1999, and Verhoogen 1949 for several minerals). In order of decreasing solar elemental abundances (Lodders 2003) this list of rock-forming elements includes Mg, Si, Fe, Al, Ca, Na, Ni, Cr, Mn, P, K, Ti, Co, Zn, V, Li, Ga, Sr, B, Zr, Rb, Te, Y, Ba, Mo, La (and other rare earth elements REE), Cs, Be, W, and U).

The geological literature contains many experimental studies of the solubility of silica in water, steam, and mixtures of the two and empirical models for total silica solubility because of its importance for processes in Earth's crust and mantle (e.g., Anderson & Burnham 1965, Cruz & Manning 2015, Fournier & Potter 1982, Gunnarsson & Arnórsson 2000, Hunt & Manning 2012, Kennedy 1950, Kennedy et al. 1962, Kitahara 1960, Manning 1994, Morey 1957, Morey & Hesselgesser 1951a,b; Morey, Fournier & Rowe 1962, Newton & Manning 2002, 2003, 2008, Rimstidt 1997, Walther & Helgeson 1977, Weill & Fyfe 1964).

Although significant dissolution of silica in steam was recognized early, the molecular form(s) of the Si-bearing gas(es) in steam remained unknown until Brady (1953) analyzed experimental data of Kennedy (1950), Morey & Hesselgesser (1951a,b), and Straub & Grabowski (1945). Brady inferred orthosilicic acid vapor $Si(OH)_4$ is the major Si-bearing molecule in steam over a wide P – T range. Subsequent work supports his conclusions (e.g., see Mosebach 1957; Wasserburg 1958;

Kitahara 1960; Krikorian 1970; Walther & Helgeson 1977; Hashimoto 1992; Jacobson et al, 2005; Plyasunov 2011, 2012, Zotov & Keppler 2002, and references therein). Silica dissolves in steam primarily via the reaction

$$SiO_2 \text{ (silica)} + 2 H_2O \text{ (steam)} = Si(OH)_4 \text{ (gas)} \qquad (1)$$

In particular we refer the reader to Plyasunov (2011a, 2012). He carefully analyzed ambient pressure transpiration experiments, solubility data for amorphous silica and quartz in water – steam mixtures along the $H_2O$ vapor pressure curve up to the critical point of water (647.096 K), and in steam above the critical point. He computed ideal gas thermodynamic properties and fugacity coefficients for $Si(OH)_4$ gas, partition coefficients for $Si(OH)_4$ between water and steam, and showed reaction (1) accounts for 100% of dissolved silica in steam at densities $\leq 322$ kg m$^{-3}$, the density of $H_2O$ at its critical point (e.g., see Table 3, and Figures 7, 9, 14 in Plyasunov 2012).

## 4. Computational Methods and Data Sources

We did three different sets of calculations – (1) the estimated partial pressures of $Si(OH)_4$ and other Si-O-H gases in steam as a function of P and T from 1573 – 2000 K and $4 \times 10^{-5}$ bar to 1,100 bars, (2) the solubility of pure oxides ($SiO_2$, MgO, "FeO", CaO, $Al_2O_3$, NiO) in steam and (3) the chemistry of a steam atmosphere in equilibrium with a magma ocean. The first set of calculations confirms $Si(OH)_4$ is the major Si-bearing gas in steam at high temperatures up to 1,100 bars pressure, in agreement with the prior experimental and theoretical work cited above. It also shows agreement between calculations done with the IVTAN code at Washington University and with the FactSage code at NASA Glenn. The second set shows the maximum solubility of an oxide in steam and the maximum pressure of the respec-tive hydroxide gas as a function of temperature and steam pressure. The third set of calculations gives the abundances of metal hydroxide gases in the steam atmosphere of an exoplanet. Gas abundances are expressed as mole fractions (X) defined as moles (N) of a gas divided by total moles of all gases in the atmosphere

$$X_i = \frac{N_i}{\sum_{i=1}^{i=N} N_i} \qquad (2)$$

We used the IVTAN code, which is a Gibbs-energy minimization code of the type described by van Zeggern & Storey (1970) to do ideal gas and real gas chemical equilibrium calculations. Thermodynamic data are from the NIST-JANAF Tables (Chase et al. 1999), the IVTAN database (Gurvich et al. 1983, Gurvich et al. 1989-1994), Robie & Hemingway (1995), and other sources cited in the text below. Several hundred compounds of the elements discussed in this paper were included in the chemical equilibrium calculations.

Our first set of calculations (discussed in Section 5.1) uses experimental data for $Si(OH)_4$ gas from Plyasunov (2011a, 2012) and estimated thermodynamic data for other Si – O – H gases from Krikorian (1970) and Allendorf et al. (1995). Krikorian (1970) estimated molecular geometry, bond lengths, and vibrational frequencies for Si – O – H gases by analogy with related compounds and used statistical mechanics (Pitzer & Brewer 1961, chapter 27) to compute free energy functions $[(G^o_T - H^o_0)/T]$. He computed standard enthalpy of reaction values at 0 K from his analysis of data for $SiO_2$ solubility in steam. The combination of the two functions gives the standard Gibbs energy for formation of an ideal gas at one bar pressure from its constituent elements in their reference states as a function of temperature via the relationship

$$\frac{\Delta G_T^o}{T} = \Delta \left[\frac{(G_T^o - H_0^o)}{T}\right] + \frac{\Delta H_0^o}{T} \tag{3}$$

For example, the standard Gibbs energy for formation of $Si_2O(OH)_6$ gas is the Gibbs energy change for the reaction

$$2 \text{ Si (crystal)} + 7/2 \text{ O}_2 \text{ (gas)} + 3 \text{ H}_2 \text{ (gas)} = Si_2O(OH)_6 \text{ (gas)} \tag{4}$$

The change in the Gibbs energy functions for this reaction is

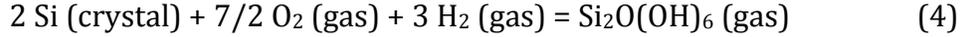

$$\Delta \left[\frac{(G_T^o - H_0^o)}{T}\right] = \left[\frac{(G_T^o - H_0^o)}{T}\right]_{Si_2O(OH)_6} - 2 \left[\frac{(G_T^o - H_0^o)}{T}\right]_{Si(c)} - \frac{7}{2} \left[\frac{(G_T^o - H_0^o)}{T}\right]_{O_2(g)} - 3 \left[\frac{(G_T^o - H_0^o)}{T}\right]_{H_2(g)}$$

$$\tag{5}$$

In contrast, Allendorf et al (1995) used quantum chemistry composite calculations to compute molecular geometry and vibrational frequencies, and then used statistical mechanics to compute Gibbs energy functions for Si – O – H gases. Allendorf et al (1995) computed standard enthalpy of formation values from their quantum chemistry calculations. They then computed the temperature dependent $\Delta G^o_T$ value for a gas using the same equations shown above.

The interactions of $Si(OH)_4$ and the other metal hydroxide gases with $H_2O$ are strongly non-ideal at some P, T conditions and we used fugacity coefficients ($\phi$) for $H_2O$, $Si(OH)_4$, $Mg(OH)_2$, $Fe(OH)_2$, $Ca(OH)_2$, $Ni(OH)_2$, and $Al(OH)_3$ in our real gas calculations. The fugacity coefficients for $H_2O$ were calculated from the equation of state (EOS) for water using the LonerHGK code (Bakker 2009) available from his website (fluids.unileoben.ac.at). Figures 1 and 2 illustrate the extent of non-ideality for $H_2O$ and $Si(OH)_4$ at pressures ≤ 2000 bars where our calculations were done.

Plyasunov (2011b, 2012) used the truncated virial equation of state to derive fugacity coefficients for $B(OH)_3$ and $Si(OH)_4$ in steam. His modeling shows

$$\frac{ln\phi_2^\infty}{ln\phi_1^*} = \frac{2B_{12}}{B_{11}} - 1 = k \qquad (6)$$

The k is an empirical constant, which equals 6.8 ± 0.4 ($2\sigma$) for $Si(OH)_4$ and 5.2 ± 0.30 ($2\sigma$) for $B(OH)_3$, the fugacity coefficient and second virial coefficient for pure steam are $\phi_1^*$ and $B_{11}$, the fugacity coefficient for the second component at infinite dilution in steam is $\phi_2^\infty$, and the second cross virial coefficient for the second component is $B_{12}$. Plyasunov (2011b, 2012) showed the infinite dilution approximation is valid over a wide P, T range for the dilute solutions of $B(OH)_3$ and $Si(OH)_4$ in steam. Based on his modeling, Akinfiev & Plyasunov (2013) propose the empirical constant k for a molecule $MO_n(OH)_p(H_2O)_q$ is given by the formula

$$k = 2(n + p + q) - 1 \qquad (7)$$

This formula gives k = 7 for $Si(OH)_4$ versus the observed value of 6.8 ± 0.4 and k = 5 versus the observed value of 5.2 ± 0.30 for $B(OH)_3$ gas. The dihydroxide gases of Ca, Fe, Mg, and Ni have k = 3. The pressure range in Figure 2 corresponds to the density range in which Plyasunov's fugacity coefficients for $Si(OH)_4$ are valid (see Table 3 and Figures 7, 9, and 14 in Plyasunov 2012).

    We considered the effect of pressure on condensed phases for our calculations of oxide solubility in steam, i.e., the contribution of the VdP term to Gibbs energy in the fundamental equation (dG = VdP – SdT). This is often discussed in terms of thermodynamic activity, "*a*". At one bar pressure the thermodynamic activity of pure condensed phases, such as quartz or molten silica is unity. However pressures greater than one bar increase the thermodynamic activity of condensed phases. Using quartz as an example, its activity (*a*) at higher pressure is given by the thermodynamic relationship (e.g., see pp. 474-476 in Fegley 2013)

$$RTlna = \int_1^P V(T,P)dP \tag{8}$$

This is a perfectly general equation. We evaluated it using the equation

$$V(T,P) = V_{298}^o[1 + \alpha_P(T - 298) - \beta_T(P - 1)] \tag{9}$$

The $V^o_{298}$ is the molar volume of quartz at 298 K and the standard state pressure of one bar, and V (T, P) is the temperature (and pressure) – dependent molar volume of quartz. The units of molar volume are J bar$^{-1}$ mol$^{-1}$, R is the ideal gas constant (R = 8.3145 J bar$^{-1}$ K$^{-1}$ mol$^{-1}$), T is Kelvin temperature, and P is pressure in bars. The isobaric thermal expansion coefficient $\alpha_P$ (K$^{-1}$) (e.g., see pp. 33-34 in Fegley 2013) is

$$\alpha_P = \frac{1}{V}\left(\frac{\partial V}{\partial T}\right)_P = \left(\frac{\partial lnV}{\partial T}\right)_P \tag{10}$$

The isothermal compressibility coefficient ($\beta_T$ bar$^{-1}$) (e.g., see pp. 34-35 in Fegley 2013) is defined as

$$\beta_T = -\frac{1}{V}\left(\frac{\partial V}{\partial P}\right)_T = -\left(\frac{\partial lnV}{\partial P}\right)_T = \frac{1}{K} \tag{11}$$

The K in this equation is the isothermal bulk modulus. Hemingway et al. (1998) give the molar volume, isobaric thermal expansion coefficient ($\alpha_P$) for quartz, and isothermal compressibility as $V^o_{298}$ = 2.269 J bar$^{-1}$ mol$^{-1}$,

$$\alpha_P = 4.48{\times}10^{-5} + 6.3{\times}10^{-9}(T - 298) \tag{12}$$

 and $\beta_T$ = 2.7 × 10$^{-6}$ bar$^{-1}$. We used analogous equations to compute activity as a function of pressure for the stable silica polymorph at ambient temperature (quartz, cristobalite, molten SiO$_2$), and the other solid and liquid oxides we considered. The input data are from Holland & Powell (2011), Fei (1995), and Linard et al (2008).

Within its calibration range, the MELTS code (described next) incorporates the effect of pressure on activity and no further calculations were necessary for oxide activities in the silicate magmas for the continental crust and bulk silicate Earth.

We used the pMELTS (version 5.6.1) and rhyolite-MELTS (version 1.02) codes (Ghiorso & Sack 1995, Ghiorso et al. 2002, Gualda et al. 2012) to calculate the activities of rock-forming oxides for both the BSE and continental crust compositions. The activity of an oxide is proportional to its mole fraction (X) in the melt (a = $\gamma \cdot$ X) and the proportionality constant is the activity coefficient ($\gamma$). The calculated activities were input into the IVTAN code along with the compositions of the BSE (or continental crust), and fugacity coefficients for $H_2O$ and the metal hydroxide gases to model chemical equilibria between the steam atmosphere and magma ocean. The MELTS programs are Gibbs energy minimization codes using regular solution models for silicate liquids and coexisting mineral phases as a function of temperature, pressure, and oxygen fugacity. In some runs we set the oxygen fugacity ($fO_2$) equal to that of the steam atmosphere by varying the $Fe^{2+}/Fe^{3+}$ ratio of the starting composition at each temperature step. The MELTS program gives activities of selected mineral components in the melt (e.g., $Si_4O_8$, $Mg_2SiO_4$, $Fe_2SiO_4$, $Al_4O_6$, $Ca_2Si_2O_6$, $NiSi_{0.5}O_2$, $NaSi_{0.5}O_{1.5}$, $KAlSiO_4$). Carmichael et al. (1977) and Ghiorso & Carmichael (1980) discuss the reasons for using mineral instead of oxide components. Using thermodynamic data from Berman (1988) and Robie & Hemingway (1995), we converted activities of the molten mineral components used in the MELTS program to activities of molten oxides of interest ($SiO_2$, $Al_2O_3$, $MgO$, $FeO$, $CaO$, $Na_2O$, $K_2O$, $NiO$).

We compared results of the MELTS programs with the FactSage code, which is a Gibbs energy minimization code that uses the quasichemical model to describe thermodynamic properties of multicomponent oxide melts. FactSage has been extensively tested and validated against experimental data but it is generally optimized for molten oxide systems important in metallurgy and materials science (Bale et al 2002). There is generally good agreement with results from the MELTS and FactSage codes as subsequently mentioned throughout the paper.

In response to a question by the referee about whether we are assuming only oxides in the magma – which we think other readers may share – we comment briefly on thermodynamic modeling of molten oxides and silicates (i.e., the magma ocean). Molten silicates conduct electricity and are ionic in nature (e.g., see the classic work of Bockris et al 1948, 1952a,b). However thermodynamic data are unavailable for the actual ionic species in the melts. Fortunately, thermodynamic modeling of solutions (molten silicates in this case) does not have to use the actual species present in the solutions and any choice of components can be made. In their discussion of components and solutions, Pitzer & Brewer (1961) noted "The components are the substances of fixed composition which can be mixed in varying amounts to form the solution. For thermodynamic purposes, the choice of components of a system is often arbitrary and depends upon the range of conditions for the problem being considered." The two codes we used chose molten minerals (MELTS) or molten oxides (FactSage) as components for their melt models. A series of papers by developers of the two codes (e.g., Ghiorso & Sack 1995, Ghiorso et al. 2002, Gualda et al. 2012, Pelton & Blander 1984, Blander & Pelton 1984) show how closely the two codes repro-duce experimental measurements for molten silicates. We discuss our results in terms of the activities of oxides in the magma ocean but this does not necessarily mean that $SiO_2$, MgO, FeO, and so on are the actual species present. In some cases the use of negative amounts of components is advantageous for thermodynamic models of solid solution in mica and amphibole minerals (Korzhinskii 1959, Thompson 1981). A simple example illustrating use of negative components is given by Fegley (2013, pp. 236-237).

## 5. Results for Pure Oxides

### 5.1 Partial pressures of Si(OH)₄ and other Si-O-H gases in steam

We now describe our first set of calculations. As discussed earlier in Section 3, silica dissolves in steam primarily via

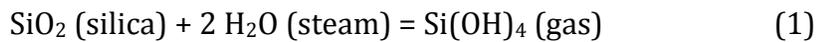

$$SiO_2 \text{ (silica)} + 2 H_2O \text{ (steam)} = Si(OH)_4 \text{ (gas)} \qquad (1)$$

However, Hildenbrand & Lau (1994, 1998) reported SiO, $SiO_2$, SiO(OH), and $SiO(OH)_2$ but not $Si(OH)_4$ in a gas-leak Knudsen cell study of liquid silica reacting with water vapor near 2000 K at $P_{H2O} \sim 4 \times 10^{-5}$ bars. They proposed silica dissolved in steam via the reactions

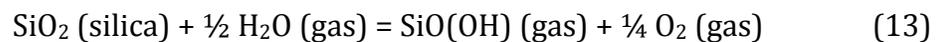

$$SiO_2 \text{ (silica)} + \tfrac{1}{2} H_2O \text{ (gas)} = SiO(OH) \text{ (gas)} + \tfrac{1}{4} O_2 \text{ (gas)} \qquad (13)$$

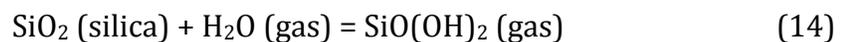

$$SiO_2 \text{ (silica)} + H_2O \text{ (gas)} = SiO(OH)_2 \text{ (gas)} \qquad (14)$$

Earlier, Krikorian (1970) proposed reaction (13) was important at 1760 K and 0.5 – 1 bar steam pressure. This proposal was based on his estimated thermodynamic properties for SiO(OH), $SiO(OH)_2$, $Si(OH)_4$, and the work of Elliot (1952) on silica vaporization in steam gas mixtures. He also concluded reaction (1) was im-portant at 600 – 900 K and 1 – 100 bars steam pressure, at much higher pressures and lower temperatures than studied by Hildenbrand & Lau (1994, 1998).

Hashimoto (1992) used the transpiration method to study the reaction of silica with $H_2O$ – $O_2$ gas mixtures at 1373 – 1773 K and~ 1 bar pressure and found evidence for only reaction (1) and $Si(OH)_4$ gas. Opila, Fox & Jacobson (1997) used a high pressure sampling mass spectrometer to study reaction of silica with $H_2O$ – $O_2$ gas mixtures at 1473 – 1673 K and one bar total pressure. They found $Si(OH)_4$ was the major Si-bearing gas and concluded $SiO(OH)_2$ was much less abundant under these conditions. Jacobson et al (2005) did a transpiration study of silica reacting with $H_2O$ – Ar gas mixtures at 1073 – 1728 K and one bar pressure. They found $Si(OH)_4$ was the major Si-bearing gas, and that $SiO(OH)_2$ was much less abundant under their experimental conditions. Jacobson et al (2005) derived thermodynamic data for both gases.

As the pressure of steam increases, silica may also dissolve via reactions such as

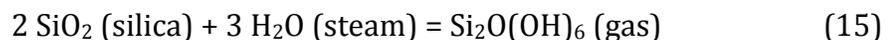

$$2\ SiO_2\ (silica) + 3\ H_2O\ (steam) = Si_2O(OH)_6\ (gas) \tag{15}$$

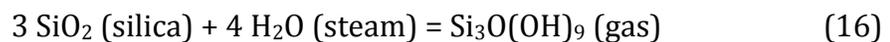

$$3\ SiO_2\ (silica) + 4\ H_2O\ (steam) = Si_3O(OH)_9\ (gas) \tag{16}$$

The dimer $Si_2O(OH)_6$, trimer $Si_3O(OH)_9$, and higher polymers may become increasingly important at water-like densities (e.g., Gerya et al. 2005, Krikorian 1970, Newton & Manning 2002, Tossell 2005, Zotov & Keppler 2002). However the exact P, T conditions at which the different polymers become important are not clear.

For example, Krikorian (1970) also proposed $Si_2O(OH)_6$ is the major Si-bearing gas in steam at 600 – 900 K and 100 – 1000 bars pressure and that $Si_2O(OH)_6$ and $Si(OH)_4$ are about equally important at 1350 K in the 2 – 7 kilobar range. This pro-posal was based on his estimated thermodynamic data for $SiO(OH)$, $SiO(OH)_2$, $Si(OH)_4$, and $Si_2O(OH)_6$.

However, Zotov & Keppler (2002) concluded $Si_2O(OH)_6$ only became important at higher pressures than proposed by Krikorian (1970). They measured Raman spectra of dissolved silica species in saturated aqueous solutions of quartz and observed $Si_2O(OH)_6$ at pressures above 5 – 6 kilobars. Their calculated concentrations of $Si(OH)_4$ and $Si_2O(OH)_6$ show significant amounts of $Si_2O(OH)_6$ at the high pressures they studied. For example at 973 K and 5.6 ± 0.9 kilobar pressure, ~37 mole % of total dissolved silica is present as $Si_2O(OH)_6$, increasing to ~ 55 mole % at 10.6 ± 2.3 kilobars. These high concentrations of $Si_2O(OH)_6$ are in high pressure steam with water-like densities of 780 – 940 kg m$^{-3}$. Water-rich fluids with densities in this range may be important at the boundary between the atmospheres and rocky interiors of planets such as Uranus and Neptune in our solar system (e.g., see the models in Fegley & Prinn 1986) and $H_2O$-rich exoplanets. Water-rich fluids such as $H_2O$ – $H_2$ and/or $H_2O$ – $CO_2$ may be relevant to Uranus- and Neptune-like exoplanets in other planetary systems. We note that the solubility of high pressure polymorphs (e.g., coesite and stishovite) in water-rich fluids may be applicable to the atmosphere – "surface" interface inside water-rich exoplanets analogous to Neptune but we do not consider this topic further here.

We used the experimental values for thermodynamic properties of $Si(OH)_4$ gas (Plyasunov 2011a, 2012), the partly experimental and partly estimated properties for $SiO(OH)_2$ gas (Allendorf et al. 1995, Jacobson et al. 2005) and the estimated thermodynamic properties for $SiO(OH)$ gas (Allendorf et al. 1995) and $Si_2O(OH)_6$ gas (Krikorian 1970) to calculate the partial pressures of all four species for four sets of P, T conditions: (A) 2000 K and $4 \times 10^{-5}$ bar, (B) 1673 K and 1 bar, (C) 1500 K and 270 bar, and (D) 2000 K and 1,100 bar. These conditions correspond to the experiments of Hildenbrand & Lau (1994, 1998), Opila, Fox & Jacobson (1997), a steam atmosphere produced by vaporization of all water in Earth's oceans (e.g., Zahnle, Kasting & Pollack 1988), and a steam atmosphere produced by complete outgassing of all $H_2O$ and $CO_2$ in the BSE.

The results of our chemical equilibrium calculations are summarized in Table 1. They show Hildenbrand & Lau (1994, 1998) are correct that $Si(OH)_4$ is unimportant and $SiO(OH)$ and $SiO(OH)_2$ are more abundant at 2000 K and $4 \times 10^{-5}$ bar. However, we compute SiO (92%) and $SiO_2$ (8%) are the major gases under their experimental conditions. Second we find $Si(OH)_4$ is the major species at the other three sets of P, T conditions. For example, at 2000 K the crossover point where the abundances of SiO and $Si(OH)_4$ become equal is 0.23 bars with $Si(OH)_4$ being the major gas at higher pressures. It remains the major gas until much higher pressures. Table 1 shows the $Si_2O(OH)_6/Si(OH)_4$ ratio is $< 9 \times 10^{-4}$ in the 1,100 bar steam atmosphere. Other calculations in Section 5.2.1, show $Si(OH)_4$ is the major species in steam at 2 kilobars at temperatures $\geq 1300$ K, where the $H_2O$ density is $\leq 322$ kg m$^{-3}$.

### 5.2 Vapor pressure and solubility in steam of $SiO_2$, MgO, and Fe oxides

We now describe the results of our second set of calculations. Figures 3 – 11 compare the total vapor pressures of the pure oxides (black curves) with solubility of the oxide in steam (red curves). The error bars on the red curves correspond to the uncertainties in the standard Gibbs energies of $Si(OH)_4$, $Mg(OH)_2$, $Fe(OH)_2$, $Ca(OH)_2$, $Al(OH)_3$, and $Ni(OH)_2$ and are described in the figure captions. All of these figures cover the same temperature range of 288.15 K – 3500 K. The lower temperature of 288.15 K (15 C) is the global average surface temperature on the Earth. The upper temperature of 3500 K is above the estimated surface temperatures of all known hot rocky exoplanets and above the one bar melting points of essentially all minerals and rocks (except $ThO_2$, which melts at 3640 ± 40 K Ackermann et al. 1963). As discussed in Section 2.2, at one bar dry peridotite starts to melt at 1390 – 1473 K and is completely molten by $\sim$ 1970 K.

We show oxide solubility in steam along the $H_2O$ vapor pressure curve up to the critical point of pure water at 647.096 K (Wagner and Pruss 2002) and then at a constant steam pressure of 220.64 bars, which is the pressure at the critical point (called the critical isobar in our discussion below). The solubility of each oxide in steam is the sum of the partial pressures of all gases of the respective element (e.g., all Si-bearing gases for $SiO_2$, all Mg-bearing gases for MgO, and all Fe-bearing gases for Fe oxides). Likewise, the total vapor pressure of each pure oxide is the sum of the partial pressures of all gases in the saturated vapor in equilibrium with the solid or molten oxide, e.g., $Mg + O_2 + O + MgO + Mg_2$ for MgO.

The vapor pressure curves were calculated using the IVTAN code and database (Gurvich et al. 1983; Gurvich et al. 1989-1996). We emphasize the vapor pressure curves are calculated from the temperature – dependent standard Gibbs free energies of the solid and gases. With one exception discussed later ($Fe_3O_4$), the curves are not extrapolations of high temperature vapor pressure data. We compare the IVTAN code calculations for vapor pressures of the pure oxides to representative values from other calculations and measurements where data are available. Vapor pressures were measured by Knudsen effusion mass spectrometry (KEMS, Chervonnyi et al. 1977, Drowart et al. 1960, Grimley, Burns & Ingrham 1961, Kazenas et al. 1983, 1985, Kazenas & Tagirov 1995, Samoilova & Kazenas 1995) and manometry (Salmon 1961). Oxygen fugacities (partial pressures) were measured using solid-state zirconia sensors (Blumenthal & Whitmore 1961, Jacobsson 1985, O'Neill 1988, O'Neill & Pownceby 1993). We refer the reader to the experimental and/or theoretical papers cited for each oxide for details of the experimental measurements and /or calculations.

In our discussion below we use 2000 K – just above the liquidus temperature of peridotite – as a reference temperature for comparing oxide solubility in steam and the vapor pressure of the pure oxide. Our 2000 K reference temperature is well within the range of sub-stellar equilibrium temperatures for several hot rocky exoplanets (e.g., $\sim$ 1475 K for Kepler-36b, $\sim$ 1570 K for Kepler-93b, 2425 K for CoRoT-7b, $\sim$ 2670 K for 55 Cnc e, and $\sim$3010 K for Kepler-10b) (Kite et al 2016).

### 5.2.1 Silica

Silica is the most abundant oxide in Earth's continental crust ($\sim$ 69 mole %) and the second most abundant oxide in the bulk silicate Earth ($\sim$ 40 mole %, see Tables 2 and 3). It also has the highest solubility in steam of rocky oxides. Figure 3 compares the vapor pressure of solid and liquid (T ≥ 1996 K) $SiO_2$ (the black curve) with the solubility of silica in steam (the red curve). The silica vapor pressure curve is simpler to explain and we discuss it first.

Silica vaporization produces a mixture of gases with an O/Si ratio of 2, as in $SiO_2$. The vapor pressure ($P_{vap}$) is the sum of partial pressures of all gases in the mixture

$$P_{vap} = P_{SiO} + P_{O_2} + P_O + P_{SiO_2} + P_{Si} + P_{O_3} + P_{Si_2} + P_{Si_3} \qquad (17)$$

At 2000 K the total vapor pressure over liquid $SiO_2$ is $1.54 \times 10^{-5}$ bar and the vapor is dominantly composed of SiO (61%), $O_2$ (26%), O (8.5%) and $SiO_2$ (4.5%). Liquid silica "boils" at 3130 K where the total vapor pressure is one bar and the vapor is dominantly composed of SiO (57%), $O_2$ (24%), $SiO_2$ (10%), and O (9%). All other gases (including ions) are less abundant than these four major gases. Measured (Kazenas et al. 1985, blue points) and calculated (Krieger 1965, green points) vapor pressures of $SiO_2$ (s, liquid) agree with the calculated vapor pressure (black curve) from the IVTAN code.

In contrast, the amount of silica dissolved in steam corresponds to a significantly higher pressure (at the same temperature) than the vapor pressure curve until very high temperatures ($\sim$ 3000 K). The total pressure ($P_{\Sigma Si}$) of silica dissolved in steam is dominated by $Si(OH)_4$ until very high temperatures where either SiO or $SiO_2$ reaches the same abundance. The exact temperature depends on the total steam pressure, and is 2000 K ($P_{steam}$ = 0.23 bars), 2200 K ($P_{steam}$ = 1 bar), 2500 K ($P_{steam}$ = 10 bar), and > 3000 K ($P_{steam}$ = 100 bar). At 2000 K, the total pressure of silica dissolved in steam along the critical isobar is $\sim$ 0.59 bar, all of which is $Si(OH)_4$ gas. This is $\sim$ 38,000 times higher the vapor pressure of silica at the same temperature.

As discussed in Sections 3 and 5.1, dissolution of silica ($SiO_2$) in steam primarily proceeds via formation of orthosilicic acid vapor $Si(OH)_4$

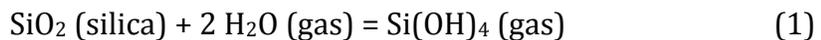

$$SiO_2 \text{ (silica)} + 2 H_2O \text{ (gas)} = Si(OH)_4 \text{ (gas)} \qquad (1)$$

The equilibrium constant for reaction (1) is

$$K_1 = \frac{f_{Si(OH)_4}}{a_{SiO_2} f_{H_2O}^2} \tag{18}$$

The fugacity ($f_i$) of each gas is the product of its partial pressure ($P_i$) and fugacity coefficient ($\phi_i$). The fugacity coefficient equals unity for an ideal gas and is either > 1 or < 1 for a real gas. The thermodynamic activity ($a_i$) of silica is unity at one bar pressure for pure silica and is proportional to its mole fraction in silicate magma. The proportionality constant is the activity coefficient ($\gamma_.$), which is unity for an ideal solution and is either > 1 or < 1 for a non-ideal solution.

We can rewrite the equilibrium constant expression for reaction (1) as

$$K_1 = \frac{\phi_{Si(OH)_4}}{\phi_{H_2O}^2} \cdot \frac{P_{Si(OH)_4}}{P_{H_2O}^2} \cdot \frac{1}{a_{SiO_2}} \tag{19}$$

The partial pressure of silicic acid vapor is thus

$$P_{Si(OH)_4} = K_1 \cdot a_{SiO_2} \cdot P_{H_2O}^2 \cdot \frac{\phi_{H_2O}^2}{\phi_{Si(OH)_4}} \cdot \tag{20}$$

The equilibrium constant $K_1$ varies with temperature and is calculated from the standard Gibbs free energy of reaction via

$$K_1 = exp\left(-\frac{\Delta_r G^o}{RT}\right) \tag{21}$$

The standard Gibbs free energy of reaction $\Delta_r G^o$ is for reaction (1) with ideal gases at one bar pressure. It was calculated from thermodynamic data for $Si(OH)_4$ (g) given by Plyasunov (2011a, 2012) and thermodynamic data for $H_2O$ (g) and $SiO_2$ (s,liq) from thermodynamic data compilations (Chase et al. 1998; Gurvich et al 1983).

The equilibrium constant expression for reaction (1) shows the amount of $Si(OH)_4$, given by its mole fraction $X_{Si(OH)_4}$, is proportional to the total pressure ($P_T$):

$$X_{Si(OH)_4} = P_T \cdot K_1 \cdot a_{SiO_2} \cdot X_{H_2O}^2 \cdot \frac{\phi_{H_2O}^2}{\phi_{Si(OH)_4}} \qquad (22)$$

Thus under otherwise constant conditions, more silica will dissolve in steam at a higher total pressure and more $Si(OH)_4$ will be produced.

Figure 4 shows the $Si(OH)_4$ mole fractions and mass % silica solubility along isobars from 1 – 2,000 bar total (steam) pressure. The proportionality deduced from equation (22) holds very well in the 1 – 2,000 bar range, e.g., at 2000 K, in going from 1 – 3 – 10 – 30 – 100 – 300 – 1,000 – 2,000 bars the $Si(OH)_4$ mole fraction increases by factors of 3.0, 10.1, 31.7, 101, 292, 803, and 1,656 times, respectively. Deviations from the exact linear proportionality are due to small changes with temperature and pressure of the product

$$a_{SiO_2} \cdot \frac{\phi_{H_2O}^2}{\phi_{Si(OH)_4}} \qquad (23)$$

For example, at 2000 K and 2000 bars, this product equals 0.828 (thus giving 2000 = 1656/0.828 for the increase in the $Si(OH)_4$ mole fraction from 1 – 2000 bars pressure). The expected linear proportionality is also affected by thermal dissociation of steam to $H_2$ and $O_2$ at high temperature and low pressure, which slightly decreases the steam mole fraction.

With the exception of temperatures ≤ 1300 K on the 2-kilobar isobar, all calculations on the graph are at mass density ≤ 322 kg m$^{-3}$, the density at the critical point of water. This is the density range in which Plyasunov's fugacity coefficients for $Si(OH)_4$ are valid (e.g., Table 3 and Figures 7, 9, and 14 in Plyasunov 2012). The three green points show the measured $SiO_2$ solubility in steam at 2 kilobars pressure (Anderson & Burnham 1965) at 1000, 1100, and 1200 K where the mass densi-ty is larger than 322 kg m$^{-3}$. These points smoothly blend into the 2-kilobar curve at 1300 K where the steam mass density decreases to the critical value.

Figures 3 and 4 also give the maximum amount of Si(OH)$_4$ in a steam atmosphere at a given pressure and temperature. Figure 4 also shows the mass percentage of SiO$_2$ in the gas as a function of pressure and temperature. The activity of pure silica is greater than that of SiO$_2$ dissolved in a silicate melt at the same temperature and total pressure, otherwise pure silica would precipitate out of the melt. For example, at 2000 K the SiO$_2$ activity in a melt with the composition of the bulk silicate Earth (Table 3, henceforth BSE magma) is

$$a_{SiO_2}(\text{BSE}) = X_{SiO_2}\gamma_{SiO_2} \sim (0.40)(0.7) \sim 0.3 \qquad (24)$$

and the SiO$_2$ activity in a melt with the composition of the continental crust (Table 2, henceforth CC magma) is

$$a_{SiO_2}(\text{CC}) = X_{SiO_2}\gamma_{SiO_2} \sim (0.69)(0.85) \sim 0.6 \qquad (25)$$

versus an activity of unity for pure silica. The activity coefficients in Equations (24) and (25) are computed with the MELTS codes discussed in Section 4. The FactSage code gives similar values for silica activities of $\sim 0.2$ for the BSE and $\sim 0.55$ for the CC magma at 2000 K. Thus, the Si(OH)$_4$ partial pressure over the BSE magma is $\sim$ 0.3 times that over pure silica and the Si(OH)$_4$ partial pressure over the CC magma is $\sim$ 0.6 times that over pure silica at the same total pressure of steam.

### 5.2.2 Periclase (MgO)

Magnesium oxide is $\sim$ 48 mole % of the bulk silicate Earth but only $\sim$ 6 mole % of the continental crust. Periclase is the mineralogical name for pure MgO that occurs naturally, and we use that name for pure MgO. However most of the MgO in the BSE and CC is a constituent of other minerals. Figure 5 compares the vapor pressure of solid and liquid (T ≥ 3100 K) MgO (the black curve) and its solubility in steam (the red curve).

Vaporization of MgO produces a mixture of gases with a Mg/O ratio of unity. At 2000 K the vapor pressure is ∼ $5.9 \times 10^{-6}$ bars and the vapor is dominantly composed of Mg (61%), $O_2$ (24%), O (13%), and MgO (2%). The measurements (blue points) of Kazenas et al. (1983) and the calculations (green points) of Krieger (1966a) agree with the IVTAN calculations (black curve) for the vapor pressure.

Laboratory studies show MgO dissolution in steam proceeds primarily via formation of $Mg(OH)_2$ gas (Alexander, Ogden & Levy 1963; Maeda, Sasomoto & Sata 1978; Hashimoto 1992)

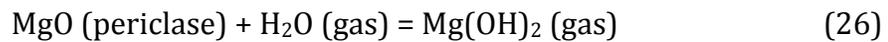

$$MgO \text{ (periclase)} + H_2O \text{ (gas)} = Mg(OH)_2 \text{ (gas)} \qquad (26)$$

This is the reaction along the red curve at T ≥ 780 K in Figure 5. However at T ≤ 780 K (the slight kink in the red curve) the solubility of MgO in steam is limited by precipitation of $Mg(OH)_2$ (brucite). This is the P, T point where the MgO (periclase) – $Mg(OH)_2$ (brucite) univariant curve intersects the solubility curve for MgO in steam. Our calculated P, T point for this intersection agrees with the measured (Kennedy 1956) position of the periclase – brucite univariant curve. Below this point the partial pressure of $Mg(OH)_2$ in steam equals the vapor pressure of brucite:

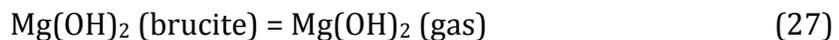

$$Mg(OH)_2 \text{ (brucite)} = Mg(OH)_2 \text{ (gas)} \qquad (27)$$

At 2000 K, the $Mg(OH)_2$ gas partial pressure in steam is ∼ 0.01 bars. This is ∼ 1,750 times larger than the vapor pressure of MgO at the same temperature.

The equilibrium constant for MgO dissolution in steam via reaction (26) is

$$K_{26} = \frac{P_{Mg(OH)_2}}{P_{H_2O}} \cdot \frac{\phi_{Mg(OH)_2}}{\phi_{H_2O}} \cdot \frac{1}{a_{MgO}} \qquad (28)$$

Rearranging this equation (28) shows the abundance (mole fraction) of $Mg(OH)_2$ gas is independent of total pressure:

$$X_{Mg(OH)_2} = K_{26} \cdot a_{MgO} \cdot X_{H_2O} \cdot \frac{\phi_{H_2O}}{\phi_{Mg(OH)_2}} \cdot \qquad (29)$$

Calculations from 1 – 1,000 bars total pressure confirm the near constancy of the abundance of $Mg(OH)_2$ gas along an isotherm. At 2000 K, the $Mg(OH)_2$ mole fraction varies from $4.64 \times 10^{-5}$ ($P_T \sim P_{steam} = 1$ bar) to $4.68 \times 10^{-5}$ ($P_T \sim P_{steam} = 1,000$ bars).

Figure 5 also gives the maximum amount of $Mg(OH)_2$ in a steam atmosphere at a given pressure and temperature. The activity of pure MgO is greater than that of MgO dissolved in a silicate melt at the same temperature and total pressure, otherwise pure periclase would precipitate out of the melt. For example, at 2000 K the MgO activity in BSE magma is $\sim 0.2$ (MELTS) to $\sim 0.3$ (FactSage) and the MgO activity in CC magma is $\sim 0.01$ (FactSage) to $\sim 0.04$ (MELTS) versus an activity of unity for pure MgO.

### 5.2.3 Iron oxides.

Iron oxides are minor constituents of the bulk silicate Earth and continental crust (5.90 mole % in the BSE and 2.60 mole % in the CC). Figure 6 compares the solubility of "FeO" (denoting wüstite, which is actually $Fe_{1-x}O$ with a temperature – dependent Fe/O ratio close to 0.95) in steam (red curve) and the $\Sigma$ Fe and $O_2$ partial vapor pressures (black curves). We first discuss the vapor pressure curves. Wüstite and the other two iron oxides vaporize incongruently (e.g., Brewer & Mastick 1951; Chizhikov et al. 1971; Shchedrin et al. 1978; Kazenas & Tagirov 1995). This means that the Fe/O atomic ratio in the vapor is different than that in the solid (or liquid). The black curves are the partial vapor pressures of Fe gases ($P_{\Sigma Fe} \sim P_{Fe} \sim P_{vap}$) and $O_2$ over pure metal-saturated "FeO" (wüstite) at T = 843 – 1650 K and liquid "FeO" at T ≥ 1650 K. The lower temperature bound is the wüstite eutectoid temperature. Wüstite is unstable at T ≤ 843 K with respect to a mixture of iron metal and $Fe_3O_4$ (magnetite) and it decomposes to this mixture at 843 K. Below 843 K the black curves are the partial vapor pressures of Fe and $O_2$ over metal saturated magnetite. Several comparisons to experimental data are shown on the graph. The blue and green squares are solid-state zirconia sensor $fO_2$ measurements by O'Neill (1988) for iron – wüstite and iron – magnetite, respectively. The yellow squares are solid-state zirconia sensor (i.e., emf) $fO_2$ measurements by O'Neill & Pownceby (1993) for iron – wüstite.  The black triangle is a set of $fO_2$ measurements for liquid "FeO" by Knudsen effusion mass spectrometry by Kazenas & Tagirov (1995). As Figure 6 shows, the Fe partial vapor pressure is significantly larger than the $O_2$ partial vapor pressure (i.e., the oxygen fugacity, $fO_2$). At 2000 K the vapor pressure of liquid "FeO" is $\sim$ 0.0004 bars ($P_{vap} \sim P_{Fe}$).

The red curve is the total amount of Fe in all forms ($P_{\Sigma Fe} = P_{Fe(OH)2} + P_{FeOH} + P_{FeO(OH)} + P_{Fe} + P_{Fe2} + P_{FeO} + P_{FeO2} + P_{FeH}$) dissolved in steam. $Fe(OH)_2$ is the dominant gas at all temperatures shown. Representative error bars corresponding to ± 30 kJ/mol uncertainty in the $Fe(OH)_2$ gas data (Gurvich et al. 1983) are shown on the red curve. Thermodynamic calculations predict "FeO" dissolution in steam occurs as

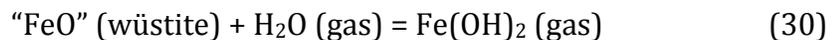

$$\text{"FeO" (wüstite)} + H_2O \text{ (gas)} = Fe(OH)_2 \text{ (gas)} \qquad (30)$$

The analogous reaction involving FeO (gas) is well known (Farber, Harris & Sri-

vastava 1974, Rollason & Plane 2000), and Belton & Richardson (1962) showed Fe

metal dissolved in steam via an analogous reaction to equation (30). At 2000 K the

Fe(OH)$_2$ gas partial pressure in steam is $\sim$ 0.09 bars, which is $\sim$ 220 times larger than the vapor pressure of liquid "FeO".

The equilibrium constant expression for reaction (30) is

$$K_{30} = \frac{\phi_{Fe(OH)_2}}{\phi_{H_2O}} \cdot \frac{P_{Fe(OH)_2}}{P_{H_2O}} \cdot \frac{1}{a_{FeO}} \tag{31}$$

The partial pressure and mole fraction of Fe(OH)$_2$ vapor are thus given by

$$P_{Fe(OH)_2} = K_{30} \cdot a_{FeO} \cdot P_{H_2O} \cdot \frac{\phi_{H_2O}}{\phi_{Fe(OH)_2}} \tag{32}$$

$$X_{Fe(OH)_2} = K_{30} \cdot a_{FeO} \cdot X_{H_2O} \cdot \frac{\phi_{H_2O}}{\phi_{Fe(OH)_2}} \cdot \tag{33}$$

Equation (33) shows the mole fraction of Fe(OH)$_2$ gas is independent of total pres-

sure. Calculations from 1 – 1,000 bars total pressure confirm the near constancy of

the abundance of Fe(OH)$_2$ gas. At 2000 K, the Fe(OH)$_2$ mole fraction only varies from

$4.25 \times 10^{-4}$ (P$_T$ $\sim$ P$_{steam}$ = 1 bar) to $4.27 \times 10^{-4}$ (P$_T$ $\sim$ P$_{steam}$ = 1,000 bars).

Figure 7 compares the solubility of Fe$_3$O$_4$ (magnetite) in steam (red curve) and

the Fe (g) and O$_2$ partial vapor pressures (black curves). Magnetite vaporization

produces significantly more Fe gas than oxygen until high temperatures. The Fe and

O$_2$ partial pressures are equal at $\sim$ 1540 K and O$_2$ is dominant at higher tempera-

tures. The partial vapor pressure curves for Fe and O$_2$ are for metal-rich Fe$_3$O$_4$ and

liquid Fe$_3$O$_4$ (at T$\geq$ 1870 K) and are computed from the partial molal Gibbs energies

of oxygen and Fe metal-rich Fe$_3$O$_4$ in equilibrium with wüstite from 843 – 1573 K

tabulated by Spencer & Kubaschewski (1978), i.e.,

$$RTlnf_{O_2} = 2\Delta G_O \tag{34}$$

$$RTlnf_{Fe} = \Delta G_{Fe} \tag{35}$$

The reason for doing this is as follows. The $O_2$ partial vapor pressure of $Fe_3O_4$ coexisting with wüstite is for the reaction:

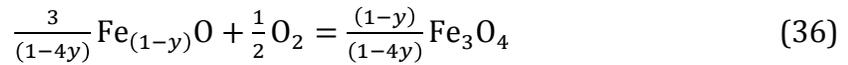

$$\frac{3}{(1-4y)}Fe_{(1-y)}O + \frac{1}{2}O_2 = \frac{(1-y)}{(1-4y)}Fe_3O_4 \tag{36}$$

The wüstite composition along the phase boundary (843 – 1697 K) is different than that of metal-rich wüstite, and varies significantly with temperature. Neither JANAF nor IVTAN (nor any other compilation we know of) tabulate the necessary thermodynamic data to do calculations. We extrapolated the partial vapor pressure curves from 1573 K to higher temperatures. The pink circles (Jacobsson 1985) and green squares (O'Neill 1988) are solid-state zirconia sensor $fO_2$ measurements. These data sets are on our calculated $O_2$ partial vapor pressure curve. The two blue triangles are $O_2$ partial pressures read off the Fe – O phase diagram of Muan & Osborn (1965). They are slightly higher than our extrapolated $O_2$ curve. Below 843 K the curves are the same as in Figure 6 because $Fe_3O_4$ coexists with Fe metal in this range.

The red curve is analogous to the one in Figure 6. It shows the partial pressure of $Fe(OH)_2$ in steam due to dissolution of $Fe_3O_4$ via the reaction

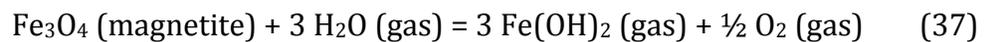

$$Fe_3O_4 \text{ (magnetite)} + 3 H_2O \text{ (gas)} = 3 Fe(OH)_2 \text{ (gas)} + ½ O_2 \text{ (gas)} \tag{37}$$

As discussed below, Belton & Richardson (1962) showed $Fe_2O_3$ dissolves in steam via an analogous reaction. At 2000 K the $Fe(OH)_2$ partial pressure in steam due to dissolution of magnetite is ∼ 0.02 bars, while the partial vapor pressure of Fe over liquid $Fe_3O_4$ is ∼ 2,000 times smaller and is about $10^{-5}$ bars.

Figure 8 compares the solubility of $Fe_2O_3$ (hematite) in steam with the Fe and $O_2$ partial vapor pressures of hematite and liquid $Fe_2O_3$ (T ≥ 1895 K). Hematite vaporizes to almost pure $O_2$ with very little Fe. Figure 8 shows two vapor pressure curves for $Fe_2O_3$ – one is the $O_2$ partial pressure and the other is the sum of the pressures of all Fe-bearing gases (Fe + FeO + $FeO_2$ + $Fe_2$). The blue points (manometry – Salmon 1961), green points (emf – Jacobsson 1985), and pink points (emf – Blumenthal & Whitmore 1961) on the $O_2$ curve are measurements of the $O_2$ partial pressure by two different methods.

The red curve is analogous to the one in Figure 6. It shows the partial pressure of $Fe(OH)_2$ in steam due to dissolution of $Fe_2O_3$ via the reaction

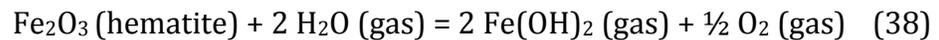

$$Fe_2O_3 \text{ (hematite)} + 2\ H_2O \text{ (gas)} = 2\ Fe(OH)_2 \text{ (gas)} + \tfrac{1}{2}\ O_2 \text{ (gas)} \quad (38)$$

Belton & Richardson (1962) studied reaction (38) and the analogous reaction with iron metal:

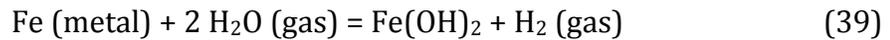

$$Fe \text{ (metal)} + 2\ H_2O \text{ (gas)} = Fe(OH)_2 + H_2 \text{ (gas)} \qquad (39)$$

At 2000 K the $Fe(OH)_2$ partial pressure in steam due to dissolution of hematite is ~ 0.016 bars. The partial vapor pressure of Fe-bearing gases over liquid $Fe_2O_3$ is dominated by $FeO_2$ and is ~ 800 times smaller (~ $2 \times 10^{-5}$ bars). The partial vapor pressure of Fe (g) is only $1.5 \times 10^{-9}$ bars.

We focus on "FeO" dissolution in steam, reaction (30), because our MELTS and FactSage calculations show FeO is the major Fe species in the BSE and CC magmas at the oxygen fugacity ($fO_2$) of the steam atmospheres, e.g., at 2000 K the $FeO/Fe_2O_3$ activity ratio in the BSE magma is ~ 155 and ~ 20 in the CC magma.

Figure 6 also gives the maximum amount of $Fe(OH)_2$ in a steam atmosphere at a given pressure and temperature. The activity of pure "FeO" is greater than that of FeO dissolved in a silicate melt at the same temperature and total pressure, otherwise pure wüstite would precipitate out of the melt. For example, at 2000 K the FeO activity in BSE magma is $\sim 0.11$ (FactSage) to $\sim 0.14$ (MELTS) and the FeO activity in CC magma is $\sim 0.06$ (FactSage) to $\sim 0.15$ (MELTS) versus an activity of unity for pure wüstite.

### 5.3 Vapor pressure and solubility in steam of less abundant oxides

*5.3.1 Calcium oxide*

Calcium oxide (CaO, calcia, lime) is a minor constituent of Earth's continental crust ($\sim 6.5\%$) and BSE ($\sim 3.4\%$). Figure 9 compares the vapor pressure of solid and liquid (T ≥ 3172 K) CaO (black curve) and its solubility in steam (red curve), which is limited by precipitation of solid and liquid $Ca(OH)_2$ at temperatures up to 1550 K.

Lime vaporizes congruently to a mixture of gases with a Ca/O ratio of unity. Our calculated vapor pressure curve agrees with measurements (blue circles, Samoilova & Kazenas 1995) and calculations (green triangles, Krieger 1967). At 2000 K the vapor pressure is $\sim 3.6 \times 10^{-7}$ bars and the vapor is dominantly composed of Ca (55%), O (35%), and $O_2$ (10%). In contrast the total pressure of all Ca-bearing gases dissolved in steam is $\sim 3.1 \times 10^{-2}$ bars, about 84,000 times larger.

Calcium dihydroxide [$Ca(OH)_2$] is the major Ca species in steam. It forms via the reaction (Matsumoto & Sata 1981, Hashimoto 1992)

$$CaO \text{ (lime, liq)} + H_2O \text{ (g)} = Ca(OH)_2 \text{ (g)} \qquad (40)$$

The equilibrium constant expression for this reaction is

$$K_{40} = \frac{P_{Ca(OH)_2}}{P_{H_2O}} \cdot \frac{\phi_{Ca(OH)_2}}{\phi_{H_2O}} \cdot \frac{1}{a_{CaO}} \tag{41}$$

Rearranging equation (41) shows the mole fraction of $Ca(OH)_2$ gas is independent of the total pressure,

$$X_{Ca(OH)_2} = K_{40} \cdot a_{CaO} \cdot X_{H_2O} \cdot \frac{\phi_{H_2O}}{\phi_{Ca(OH)_2}} \tag{42}$$

However at T ≤ 1550 K, the solubility of CaO in steam and thus the partial pressure of $Ca(OH)_2$ gas is controlled by precipitation of $Ca(OH)_2$ (portlandite). This occurs at the P, T point where the CaO (lime) – $Ca(OH)_2$ (portlandite) univariant curve intersects the solubility curve for CaO in steam. Below this point the partial pressure of $Ca(OH)_2$ gas equals the vapor pressure of portlandite:

$$Ca(OH)_2 \text{ (portlandite, liquid)} = Ca(OH)_2 \text{ (gas)} \tag{43}$$

Calculations at 2000 K from 1 – 338 bars total pressure confirm the near constancy of the abundance of $Ca(OH)_2$ gas. At this temperature, the $Ca(OH)_2$ mole fraction varies from $1.39 \times 10^{-4}$ ($P_T \sim P_{steam}$ = 1 bar) to $1.43 \times 10^{-4}$ ($P_T \sim P_{steam}$ = 338 bars). Liquid $Ca(OH)_2$ forms at $P_{steam} \geq 338$ bars and the partial pressure of $Ca(OH)_2$ is controlled by the vapor pressure of liquid $Ca(OH)_2$ at $P_{steam} \geq 338$ bars at 2000 K.

Figure 9 also gives the maximum amount of $Ca(OH)_2$ in a steam atmosphere at a given pressure and temperature. The activity of pure CaO is greater than that of CaO dissolved in a silicate melt at the same temperature and total pressure, otherwise pure lime would precipitate out of the melt. For example, at 2000 K the CaO activity in BSE magma is $\sim 5.5 \times 10^{-4}$ (MELTS) to $\sim 8.3 \times 10^{-4}$ (FactSage) and the CaO activity in CC magma is $\sim 1.9 \times 10^{-4}$ (FactSage) to $\sim 6.4 \times 10^{-4}$ (MELTS) versus an activity of unity for pure lime.

*5.3.2 Aluminum sesquioxide*

Aluminum oxide ($Al_2O_3$, alumina, corundum) comprises ~10% of Earth's continental crust and ~ 2.3% of the BSE. Figure 10 compares the vapor pressure of solid $Al_2O_3$ (corundum) and liquid (T ≥ 2327 K) $Al_2O_3$ (the black curve) and its solubility in steam (the red curve). Corundum vaporizes to a mixture of gases with an Al/O ratio of 2/3. We compare the calculated vapor pressure curve to experimental data and other calculations. The blue circles are laser vaporization measurements of the vapor pressure of liquid $Al_2O_3$ (Hastie, Bonnell & Schenk 2000), the pink triangles (Drowart et al. 1960) and green squares (Chervonnyi et al 1977) are KEMS measurements of the vapor pressure of $Al_2O_3$ (corundum), and the cyan triangles are calculations by Krieger (1966b).

At 2000 K the vapor pressure of corundum is ~ 1.4 × 10^{-8} bars and the vapor is dominantly composed of O (56.4%), Al (35.2%), AlO (6.3%), $Al_2O$ (1.1%), and $O_2$ (1.0%). In contrast, the partial pressure of $Al(OH)_3$ in steam at 2000 K is ~ 0.02 bars, about 1,400,000 times higher.

Hashimoto (1992) and Opila & Myers (2004) showed the dissolution of $Al_2O_3$ in steam proceeds via

$$Al_2O_3 \text{ (alumina)} + 3\, H_2O \text{ (gas)} = 2\, Al(OH)_3 \text{ (gas)} \quad (44)$$

The equilibrium constant expression for reaction (44) is

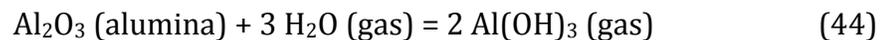

$$K_{44} = \frac{P^2_{Al(OH)_3}}{P^3_{H_2O}} \cdot \frac{\phi^2_{Al(OH)_3}}{\phi^3_{H_2O}} \cdot \frac{1}{a_{Al_2O_3}} \quad (45)$$

Rearranging equation (45) shows the mole fraction of $Al(OH)_3$ gas depends upon the square root of the total pressure:

$$X_{Al(OH)_3} = \left(K_{44} \cdot P_T \cdot a_{Al_2O_3}\right)^{1/2} \cdot X_{H_2O}^{3/2} \cdot \left(\frac{\phi_{Al(OH)_3}^3}{\phi_{H_2O}^2}\right)^{1/2} \qquad (46)$$

However, at T ≤ 642 K, the solubility of $Al_2O_3$ in steam and thus the $Al(OH)_3$ partial pressure is limited by precipitation of AlO(OH) (diaspore)

AlO(OH) (diaspore) + $H_2O$ (gas) = $Al(OH)_3$ (gas)         (47)

The kink in the red curve in Figure 10 is at 642 K, which is the P, T point where the diaspore – corundum univariant curve intersects the solubility curve for corundum in steam. Our calculated P, T point for this intersection is 9 degrees higher than the measured value of 633 ± 7 K (Fyfe & Hollander 1964, Haas 1972, Kennedy 1959). This small difference is within the uncertainty of the thermodynamic data. Our calculations used $\Delta_f H_{298}^o = -1001.3 \pm 2.2$ kJ mol[-1] and $S_{298}^o = 35.3 \pm 0.2$ J mol[-1] K[-1] for AlO(OH) from Robie & Hemingway (1995), heat capacity measurements of Perkins et al (1979), and V=V(T) data from Pawley, Redfern & Holland (1996).

Figure 10 also gives the maximum amount of $Al(OH)_3$ in a steam atmosphere at a given pressure and temperature. The activity of pure $Al_2O_3$ is greater than that of $Al_2O_3$ dissolved in a silicate melt at the same temperature and total pressure, otherwise pure corundum would precipitate out of the melt. For example, at 2000 K the $Al_2O_3$ activity in BSE magma is ∼ $6.3 \times 10^{-5}$ (FactSage) to ∼ $7.4 \times 10^{-3}$ (MELTS) and the $Al_2O_3$ activity in CC magma is ∼ 0.013 (MELTS) to ∼ 0.037 (FactSage) versus an activity of unity for pure corundum. The calculated $Al_2O_3$ activity values in silicate melt from the two codes disagree because the MELTS code does not consider solid or liquid $MgAl_2O_4$ (spinel), which is an important Al-bearing component in the FactSage calculations. We used the MELTS results in our calculations.

*5.3.3 Nickel oxide*

Nickel oxide (NiO) is a trace constituent of the continental crust ($\sim$ 0.006%) and the

BSE ($\sim$ 0.17%) and occurs as the mineral bunsenite or as a minor component of

other minerals. Figure 11 compares NiO solubility in steam (red curve) with the Ni

and $O_2$ partial vapor pressures of bunsenite and liquid NiO (T ≥ 2228 K). The points

on the vapor pressure curves are measured partial vapor pressures ($P_{Ni}$ + $P_{NiO}$)

(Grimley, Burns & Ingrham 1961, cyan circles; Kazenas & Tagirov 1995, dark red

diamonds), measured $O_2$ (O'Neill & Pownceby 1993, green squares), and calculated

$O_2$ partial vapor pressures (Hemingway 1990, blue triangles).

At 2000 K the partial vapor pressures of Ni and NiO sum up to $\sim$ 3.0 × 10$^{-4}$ bars.

In contrast the total pressure of all Ni-bearing gases dissolved in steam is $\sim$ 0.19

bars, about 630 times larger, and is 98% Ni(OH)$_2$ gas and 2% NiOH gas.

Belton & Jordan (1967) measured Ni(OH)$_2$ gas formation from Ni metal reacting

with water vapor. Based on their work, Ni(OH)$_2$ gas forms via the reaction

$$\text{NiO (bunsenite, liq)} + H_2O \text{ (g)} = \text{Ni(OH)}_2 \text{ (g)} \qquad (48)$$

The equilibrium constant expression for reaction (48) is

$$K_{48} = \frac{P_{Ni(OH)_2}}{P_{H_2O}} \cdot \frac{\phi_{Ni(OH)_2}}{\phi_{H_2O}} \cdot \frac{1}{a_{NiO}} \qquad (49)$$

Rearranging equation (49) shows the mole fraction of Ni(OH)$_2$ gas is independent of the total pressure,

$$X_{Ni(OH)_2} = K_{48} \cdot a_{NiO} \cdot X_{H_2O} \cdot \frac{\phi_{H_2O}}{\phi_{Ni(OH)_2}} \qquad (50)$$

In contrast to other oxides (e.g., CaO, MgO), precipitation of $Ni(OH)_2$ does not occur at low temperatures – at least according to thermodynamic data tabulated by NIST – and the $Ni(OH)_2$ partial pressure is always limited by solubility of NiO in steam.

The red curve in Figure 11 also gives the maximum amount of $Ni(OH)_2$ in a steam atmosphere at a given pressure and temperature. The activity of pure NiO is greater than that of NiO dissolved in a silicate melt at the same temperature and total pressure, otherwise pure NiO would precipitate out of the melt. Assuming ideality for NiO dissolved in silicate melts, lower limits to the NiO activity are given by its mole fraction in the BSE ($\sim 0.002$) and CC ($\sim 6 \times 10^{-5}$) magmas versus an activity of unity for pure NiO. Holzheid, Palme & Chakraborty (1997) found NiO activity coefficients $\gamma = 2.7 \pm 0.5$ in silicate melts. This would increase the NiO activity by that factor ($a_{NiO} = \gamma \cdot X_{NiO}$), but our conclusion remains unchanged – Figure 11 gives the maximum $Ni(OH)_2$ gas pressure.

Finally, Belton & Jordan (1967) showed that $Co(OH)_2$ also exists. However cobalt has about one tenth the abundance of nickel in the BSE and our calculations found $Co(OH)_2$ is a very minor gas that we do not discuss further.

### 5.3.4 Sodium and potassium oxides

Sodium oxide ($Na_2O$) and potassium oxide ($K_2O$) are too reactive to occur in nature. Thus we did compare their vapor pressures to their solubility in steam. We discuss the chemistry of sodium and potassium chemistry in steam atmospheres equilibrated with magma oceans in Sections 7.3 – 7.4.

### 5.4 Summary of oxide solubility in steam

Table 4 summarizes our results in Figures 3 – 11 for the partial pressures of metal hydroxide gases in steam at 220.64 bars pressure for three selected temperatures (1000, 1500, 2000 K). The relative solubility (or volatility) of the major rock-forming oxides in steam varies somewhat as a function of temperature but $SiO_2$ is always the most soluble (volatile) oxide, "FeO" is the 2[nd] or 3[rd] most soluble (volatile), and MgO is always the least soluble (volatile).

### 6. Solubility of $SiO_2$, MgO, and Fe oxides in Steam-bearing Atmospheres

Steam atmospheres are not pure water vapor and they contain other gases due to thermal dissociation of steam (e.g., $H_2$, OH, H, $O_2$, O) and the outgassing of other volatiles from rocky material. Schaefer, Lodders & Fegley (2012) and Fegley & Schaefer (2014) computed the major H-, C-, N-, and S-bearing gases as a function of pressure and temperature for hot rocky exoplanets with compositions like the BSE or continental crust, see Figures 7-8 of Schaefer, Lodders & Fegley (2012) and Figure 5 in Fegley & Schaefer (2014). They found the major gases in steam atmospheres with pressures ≥ 1 bar and surface temperatures ≤ 2000 K, are $H_2O$, $CO_2$, $N_2$, $SO_2$, and $H_2$, $O_2$, and that CO may also be present.

Kuts (1967) studied the effects of $N_2$, $CO_2$, and $O_2$ on solubility of amorphous silica in steam at 708 – 913 K and 1 – 15 atmospheres. He found silica solubility in the gas mixtures was the same as in pure steam at the same temperature and total steam pressure. Thus $N_2$, $CO_2$, and $O_2$ were inert in the P, T range he studied.

We calculated the effects of a second gas on solubility of $SiO_2$, MgO, and FeO in steam as a function of composition at 300 bars total pressure and 1500 K. These conditions apply to a steam atmosphere formed by vaporization of an ocean of water on the early Earth (Zahnle, Kasting & Pollack 1988). Figures 12 – 14 show our results for binary mixtures of steam with other abundant gases (e.g., $N_2$, $CO_2$, $H_2$, $SO_2$, $O_2$, and $CH_4$) in steam atmospheres according to published calculations (Schaefer, Lodders & Fegley 2012, Fegley & Schaefer 2014). The different points on the graphs indicate the different binary mixtures. In the case of $SiO_2$ the points form a straight line given by

$$\sqrt{X_{Si(OH)_4}} \propto X_{H_2O} \tag{51}$$

as predicted from the equilibrium constant expression for equation (1). Figures 13 and 14 show straight lines given by

$$X_{Mg(OH)_2} \propto X_{H_2O} \tag{52}$$

$$X_{Fe(OH)_2} \propto X_{H_2O} \tag{53}$$

as expected from the equilibrium constant expressions for equations (26) and (30). In all three cases the second gases are inert dilutants, as expected from Kuts (1967).

## 7. Chemical equilibria between steam atmospheres and magma oceans

Now we discuss our third set of calculations for the partial pressures of metal hydroxide gases formed by reactions of steam atmospheres and magma oceans having either the composition of the continental crust (CC, Table 2) or the bulk silicate Earth (BSE, Table 3). We first describe the temperature range over which the magma oceans exist. Next we discuss the major gases in the steam atmospheres, then all the metal hydroxide gases together, and then we consider the relative importance of hydroxide and halide gases for Si, Mg, Fe, Al, Ca, Na, and K. A series of plots are needed to display the complex chemistry in the steam atmospheres.

### 7.1 Solidus and liquidus temperatures for the BSE and CC magmas

The solidus temperature where the first melt forms is the minimum temperature where magma can exist. A magma ocean with fluid-like behavior exists at T ≥ the lock-up temperature ($T_{lock}$) where the melt fraction is ≥ (10 – 40)% (Abe 1993). At $T_{sol}$ ≤ T < $T_{lock}$ the magma ocean has much higher viscosity, has solid-like behavior, and contains less water (per unit mass) than a fully molten magma ocean. The liquidus temperature is the maximum temperature where solid rocks exist. The maximum temperature for existence of a magma ocean is the critical curve along which the liquid – vapor distinction vanishes (e.g., see the discussion in chapter 6 of Rowlinson & Swinton 1982). Estimates for the critical temperature of pure silica range from ∼ 4,700 to ∼ 13,500 K (Table 1 in Melosh 2007) and it is plausible that the critical curves for the continental crust and bulk silicate Earth are within the same temperature range. These temperatures are much higher than the estimated surface temperatures of hot rocky exoplanets and we do not consider the critical temperature of magma oceans further in this paper.

#### 7.1.1 Continental crust magma ocean

We consider a magma ocean with the composition of the average continental crust (CC, Table 2). To first approximation, the continental crust is granitic and Goranson (1932) reported a solidus < 1173 K and a liquidus of 1323 ± 50 K for Stone Mountain granite at ambient pressure (∼ one bar). MELTS predicts the continental crust solidus is 1197 K and the liquidus is 1415 K where orthopyroxene solid solu-tion [$(Mg,Fe)_2Si_2O_6$] is the last phase to melt. The FactSage program (with the SLAGA database) predicts a solidus of 1169 K and a liquidus of 1578 K where hematite ($Fe_2O_3$) is the liquidus phase. All these values are at one bar pressure.

As discussed in Section 2.2 the net effect of a steam atmosphere is to lower the solidus temperature of a magma ocean because $H_2O$ dissolution in the molten silicate depresses the freezing point more than the atmospheric mass increases it (via the positive Clapeyron slope). The calculated solidus temperatures (from the MELTS code) are 873 K for a 270 bar steam atmosphere and 809 K for the 1100 bar steam atmosphere (825 bars $H_2O$, 275 bars $CO_2$).

### 7.1.2 Bulk silicate Earth magma ocean

The calculated solidus and liquidus temperatures for the bulk silicate Earth (BSE) composition in Table 3 are 1267 – 1973 K (MELTS) and 1310 – 1938 K (FactSage with SLAGA database). Jennings & Holland (2015) used the THERMOCALC code (Powell, Holland & Worley 1998) and the database of Holland & Powell (2011) and computed values of 1393 – 2053 K for the KLB-1 peridotite. Forsterite – rich olivine solid solution [$(Mg,Fe)_2SiO_4$] is the liquidus phase in all computations. For comparison measured values for the KLB-1 peridotite are 1393 – 1973 K (Takahashi et al 1993). The freezing point depressions due to $H_2O$ dissolution in magma give solidus temperatures of 1206 K and 1173 K respectively for the 270 and 1100 bar steam atmospheres.

### 7.1.3 Comparison of MELTS and FactSage results for melting temperatures

The agreement of the calculated melting temperatures is good for the BSE composition but only satisfactory for the CC composition. However, it is about as good as the agreement of calculated values with measurements. The bulk silicate Earth (less so) and continental crust (more so) compositions are far removed from the optimized compositions in the FactSage databases and the calculated melting temperatures are probably accurate to only ±(50 – 100) K. Thus the solidus and liquidus temperatures from the MELTS codes are probably more realistic.

## 7.2 Major gases

Figure 15 shows the abundances of major gases in steam atmospheres (with pressures of 270 and 1100 bars) in chemical equilibrium with magmas formed by the bulk silicate Earth (BSE) and continental crust (CC). In order of decreasing abundance (mole fractions $X \sim 0.8 – 0.01$), the major gases in steam atmospheres equilibrated with CC magmas are $H_2O > CO_2 > O_2 > HF \sim SO_2 > (HCl, OH, CO)$. The sequence in steam atmospheres equilibrated with BSE magmas is $H_2O > CO_2 > SO_2 \sim H_2 > CO > (HF, H_2S, HCl, SO)$. There are a number of gases with mole fractions $X \sim 0.01 – 0.001$ including NaCl, NO, $N_2$, $SO_3$ (on the 1100 bar CC magma plot in Figure 15), $S_2$, $Si(OH)_4$, and $Mg(OH)_2$ (for the BSE plots). In general the metal hydroxide gases have lower abundances with mole fractions $X \sim 10^{-3}$ to $10^{-7}$ (see below).

### 7.2.1 Molecular oxygen

Molecular oxygen is the third most abundant gas in steam atmospheres equilibrated with CC magmas, but it is not nearly as abundant ($X_{O2} << 10^{-3}$) in steam atmospheres equilibrated with BSE magmas. Earth's mantle (99.4% of the BSE by mass) is dominated by ferrous iron with a $Fe^{3+}/\Sigma$ Fe ratio of $\sim 0.04$ to $\sim 0.11$ depending on the samples analyzed and the technique used (e.g., see Canil et al. 1994, Claire, Catling & Zahnle 2006). The lower value of $\sim 0.04$ is from Mössbauer spec-troscopic analyses of relatively unaltered samples of upper mantle rocks by Canil et al. (1994). The upper value of $\sim 0.11$ is from wet chemical analyses of glasses in mid-ocean ridge basalts (MORB) by Bézos & Humler (2005). In contrast the Earth's continental crust is richer in ferric iron with a $Fe^{3+}/\Sigma$ Fe ratio of $\sim 0.25$ (Claire, Catling & Zahnle 2006).

The dichotomy between the oxidation states of iron in Earth's crust and mantle is due to life on Earth. Most of the oxygen produced by photoautotrophs on Earth has been consumed by geochemical reactions that produced hematite, other $Fe^{3+}$-bearing minerals, and sulfate minerals in Earth's continental crust with only $\sim 4\%$ of all oxygen produced residing in the atmosphere today (Warneck 1989). For example, the sedimentary rocks in Earth's crust are significantly richer in ferric iron ($Fe^{3+}/\Sigma$ Fe $\sim 0.44$) than the entire crust because of oxidation during weathering (e.g., Yaroshevsky 2006). The continental crust would be much less oxidized and hence richer in FeO-bearing minerals in the absence of life on Earth. However abiotic production of oxygen, e.g. via solar UV photolysis of $CO_2$ – as on Mars today – would still provide an oxidant for production of ferric iron in the crust of an extrasolar rocky planet.

### 7.2.2 Comparison to our prior work

The results in Figure 15 agree with our prior work (Schaefer, Lodders & Fegley 2012) for the major gases in steam atmospheres equilibrated with CC and BSE magmas with a few differences caused by the total pressures, volatile element abundances, and silicate magma solution models used in the calculations. The higher total pressures used in this work (270 or 1100 bars) than before (100 bars) give smaller abundances of gases produced by thermal dissociation of $H_2O$, $CO_2$, and $SO_2$ as illustrated in Figure 8 of Schaefer, Lodders & Fegley (2012).

This work and our prior study use two different compilations for the composition of the bulk silicate Earth. The major element composition is very similar but the volatile element abundances can be different. For example, Schaefer, Lodders & Fegley (2012) used Kargel & Lewis (1993) who recommended hydrogen, carbon and sulfur abundances of 54.7, 65, and 274 ppm, respectively, corresponding to H/C and S/C molar ratios of ~ 10 and 1.6 in the BSE. In this work we use Palme & O'Neill (2014) who recommend 120 ppm H, 100 ppm C, and 200 ppm S corresponding to H/C and S/C molar ratios of ~ 14 and 0.75, respectively. This leads to $H_2O/CO_2$ ratios ~ 40% larger and $SO_2/CO_2$ ratios ~ 50% smaller than before in steam atmospheres equilibrated with BSE magmas (compare to Figure 2 of Schaefer, Lodders & Fegley 2012). The abundances of N, F, and Cl are similar in the two compilations – 0.88, 20.7, and 36.4 ppm, respectively, in Kargel & Lewis (1993) versus 2, 25, and 30 ppm, respectively, in Palme & O'Neill (2014). The slightly different gas abundances in this work and our prior study are well within the range of variations shown in Figures 3 – 5 of Schaefer, Lodders & Fegley (2012), which explore sensitivity of steam atmosphere chemistry to the elemental abundances used for calculations.

### 7.3 Metal hydroxide and halide gas chemistry

Figure 16a-d shows chemical equilibrium abundances of $Si(OH)_4$ and the other metal hydroxide gases in steam atmospheres of 270 and 1100 bars pressure equilibrated with BSE or CC magma. The abundance trends for the metal hydroxide gases as a function of pressure and composition are given by the equations in Sections 5.2

and 5.3 for $Si(OH)_4$ and the other hydroxide gases. For example, the $Si(OH)_4$ mole fraction at 1100 bars is $\sim 5 \times$ higher than at 270 bars mainly because of the higher pressure (a factor of $\sim 4.1 \times$), but also because of the slightly larger $H_2O$ mole fraction, and the slightly larger $H_2O$ fugacity coefficient in the 1100 bar steam atmosphere. Likewise the $Al(OH)_3$ mole fraction at 1100 bars is $\sim 2 \times$ higher than at 270 bars because it is proportional to the square root of the total pressure (see equation (46) in Section 5.3.2). However the $Mg(OH)_2$, $Ca(OH)_2$, $Fe(OH)_2$, and $Ni(OH)_2$ abundances are nearly the same because their mole fractions are independent of the total pressure, e.g., see equations (29), (42), and (50), and are affected only by the smaller changes in the mole fraction and fugacity coefficient of $H_2O$.

Figures 17-20 show the hydroxides are the major gases of Si, Mg, Fe, and Ni in the 270 bar steam atmosphere equilibrated with BSE magma and this is also true for the other three cases studied (270 bars steam atmosphere equilibrated with CC magma and the 1100 bar steam atmosphere equilibrated with BSE and CC magma). However, the hydroxides are not the major Na and K gases. Figures 21 and 22 show NaCl and KCl are the major Na and K gases and NaF and KF are also important in the 270 bar steam atmosphere equilibrated with BSE magma. Sodium and potassium chlorides and fluorides are also important in the other three cases studied. The alkali halides are also important in the 100 bar steam atmosphere modeled by Schaefer, Lodders & Fegley (2012) – see their Figures 1 and 2.

Figures 23 and 24 show mixed halide – hydroxide gases ($CaClOH$, $CaFOH$ and $FAl(OH)_2$, $F_2AlOH$, $ClAl(OH)_2$) are important for Ca and Al in the 270 bar steam atmosphere equilibrated with BSE magma. The mixed halide – hydroxide gases are the major species for Ca (see Figure 24) while they are about as important as Al(OH)$_3$ for Al in Figure 23. Again, similar results were obtained for the other cases not shown (270 bar steam atmosphere with CC magma and 1100 bar steam atmos-pheres with BSE or CC magma). Finally under some conditions (e.g., T < 2200 K in the 1100 bar steam atmosphere equilibrated with CC magma) FeCl$_2$ may be more abundant than Fe(OH)$_2$ gas.

Hydrogen chloride and HF are the major halogen – bearing gases in the steam atmospheres considered here and by Schaefer, Lodders & Fegley (2012). This is consistent with HCl and HF being the major Cl – and F – bearing gases in terrestrial volcanic gases, which are generally dominated by steam (Symonds et al 1994).

The relative abundance of hydroxide and halide gases in H$_2$O – rich systems with

HCl, and HF are controlled by exchange equilibria (pp. 68-73 in Hastie 1975), e.g.,

$$HCl\ (g) + NaOH\ (g) = NaCl\ (g) + H_2O\ (g) \tag{54}$$

The equilibrium constant expression for reaction (54) and analogous exchange equi-libria involving other halides (e.g., NaF, KCl, KF, CaClOH, CaFOH, FAl(OH)$_2$) are inde-pendent of total pressure, e.g.,

$$K_{54} = \frac{P_{H_2O}P_{NaCl}}{P_{HCl}P_{NaOH}} = \frac{X_{H_2O}X_{NaCl}}{X_{HCl}X_{NaOH}} \tag{55}$$

Rearranging equation (55) shows the molar ratio of NaOH to NaCl is proportional to the molar ratio of H$_2$O to HCl in the gas,

$$\frac{X_{NaOH}}{X_{NaCl}} = \frac{X_{H_2O}}{X_{HCl}} \cdot \frac{1}{K_{54}} \tag{56}$$

At 2000 K, the equilibrium constant K$_{54}$ = 3,365, the H$_2$O/HCl ratio is ~ 72 (Figure

15), and the NaOH (g)/NaCl (g) molar ratio is ~ 0.02 in the 270 bar steam atmos-phere equilibrated with BSE magma. A much larger $H_2O/HCl$ ratio $\geq 3,365$, the value of the equilibrium constant for reaction (54), is needed for $NaOH/NaCl \sim 1$. Similarly Hastie (1975) found that $H_2O/HCl$ ratios $\sim 10^3 - 10^4$ are needed for equal abundances of metal hydroxide and halide gases and showed that this is due primarily to differences in the M – OH and M – Cl or M – F bond dissociation energies whereas the entropy changes for the exchange reactions between hydroxides and chlorides or between hydroxide and fluorides are fairly constant for many systems.

### 7.4 Rocky Element Distribution between Magmas and Steam Atmospheres

In section 5 we demonstrated the high solubility of rocky elements in pure steam and our calculations in section 7.3 showed the equilibrium abundances of these elements in steam atmospheres as a function of temperature, pressure, and magma ocean composition. Figures 16 and 25 illustrate the important point that fractional vaporization of rocky elements from magma oceans equilibrated with steam atmospheres changes the elemental compositions of gas and magma from the original bulk composition. Our chemical equilibrium calculations show that essentially *all* rocky element ratios in the steam atmospheres are fractionated from those in the parental magmas (except at the intersection points with the original BSE or CC composition in Figures 26-28, and their analogs for other rocky element pairs). For example consider Si, Mg, and Fe in the bulk silicate Earth. Table 3 shows MgO, $SiO_2$ and FeO comprise > 90% of the BSE. Figures 26 and 27 show the molar Si/Mg and Si/Fe ratios in steam atmospheres equilibrated with magma initially having BSE composition (shown as the horizontal line). The Si/Mg ratio in the gas varies by almost a factor of 10,000 and the Si/Fe ratio by nearly a factor of 30 from 2000 – 3000 K.

At surface temperatures $\sim$ 2000 K, the Si/Mg and Si/Fe ratios in the gas are higher than the original source composition, whereas at 3000 K the ratios are lower than the original ones for both total pressures (see Figures 26 and 27). This happens because more Si partitions into the steam atmospheres at temperatures around 2000 K than either Fe or Mg. Partitioning of Fe and Mg becomes much more favorable at higher temperatures. Figures 25a and 25b illustrate this point and also show that partitioning of Si and Fe is fairly similar at lower temperatures so the Si/Fe ratio in steam atmospheres equilibrated with BSE magmas is less fractionated than the Si/Mg ratios in the gas.

Figure 28 is another example; it shows fractional vaporization of Si and Ca from the continental crust ($SiO_2$ + CaO > 78% of the crust, Table 2). Depending on the P, T conditions, Figure 28 shows the Si/Ca ratio varies by nearly a factor of 100 from the initial composition (shown by the horizontal line). The Si/Ca ratio in the steam atmospheres goes from much larger (at $\sim$ 2000 K) to much smaller (at $\sim$ 3000 K) than the initial ratio because significantly more Ca partitions into the gas as temperature increases. Figures 25-28 demonstrate that in principle significant changes in planetary (or crustal) bulk composition, density, and internal structure are possible if atmospheric loss occurs from hot rocky planets with steam atmospheres.

We now examine the distribution (or partitioning) of rocky elements between steam atmospheres and magma oceans on hot rocky exoplanets in more detail and explore the relative importance of halide and hydroxide gases for this partitioning. The molar distribution coefficient D in Figure 25 is defined as

$$D = \frac{N_g^E}{N_m^E} \tag{57}$$

The moles of an element "E" in the gas ($N_g^E$) is the total mole fraction of an element

in the gas times the total moles of gas, e.g., for sodium

$$N_g^{Na} = \left( X_{Na} + X_{NaF} + X_{NaCl} + X_{NaOH} + X_{NaH} + 2X_{(NaCl)_2} + \cdots \right) N_{gas} \tag{58}$$

The moles of element "E" remaining in the magma ($N_m^E$) is the total number of moles in the system ($N_T^E$, i.e., the total number of moles of element "E" input into the calcu-lation) minus the amount in the gas, e.g.,

$$N_m^{Na} = N_T^{Na} - N_g^{Na} \tag{59}$$

The total number of moles of an element "E" is given in Table 2 for the continental crust and in Table 3 for the bulk silicate Earth. Note that the values for $Na_2O$ (and also for $K_2O$, $Al_2O_3$, and $Fe_2O_3$) in Tables 2 and 3 have to be multiplied by two be-cause Equations (57) – (59) are counting moles of atoms for each element.

Figures 25b and 25d show the molar distribution coefficients (D values) for the

1100 bar steam atmospheres equilibrated with the BSE and CC magmas, and Figure

25c shows the D values for the 270 bar steam atmosphere equilibrated with CC

magma. All of these D values are computed with F and Cl included in the calculations

(our nominal models). Figure 25a is slightly different because it shows two sets of D

values for the 270 bar steam atmosphere equilibrated with BSE magma: either with

(solid curves) or without (dashed curves) any chlorine or fluorine.

The solubility of Cl- and F-bearing gases (e.g., HCl, HF, NaCl, NaF, KCl, KF, etc.) in high temperature silicate magmas is poorly known and the calculations without chlorine or fluorine in the system (i.e., zero moles of Cl and F input into the calcula-tions) simulate complete solubility of the halogens in the magma ocean. (We do not say complete solubility of F and Cl in the magma ocean is reasonable, but it is one endmember case with the other being all Cl and F in the steam atmosphere as in our nominal models.)

The largest effects in Figure 25a are observed for K, Na, Ca, and Al in roughly this order because of the importance of KCl, NaCl, CaClOH and AlF(OH)$_2$ gases. The actual distribution coefficients for these elements will lie between the extremes indicated by the solid line (all F and Cl in the gas) and dashed line (all F and Cl in the magma). The solid and dashed lines are much closer for the other rocky elements (Ni, Mg, and Fe ~ Si) because the hydroxides are their major gases. Table 5 lists the D values for the four plots in Figure 25.

We now show how the molar distribution coefficients (D values) in Table 5 are related to partition coefficients (k$_D$) written in terms of concentrations. The k$_D$ values are very useful for geochemical modeling and allow us to easily compute how much rocky element concentrations and ratios in the residual planet(s) vary from those in the original planet(s) as a function of the amount of atmosphere lost.

Rearranging Equation (59) gives the total number of moles ($N_T^E$) of any element "E" in the steam atmosphere – magma ocean system:

$$N_T^E = N_m^E + N_g^E \tag{60}$$

We rewrite Equation (60) in terms of concentrations (C) and mass fractions (F)

$$N_T^E = C_m^E F_m + C_g^E F_g = C_m^E F_m + C_g^E (1 - F_M) \tag{61}$$

Mass balance requires the mass fractions sum to unity, which allows the substitution of $(1 - F_M)$ for $F_g$ made above. We now combine Equations (57) and (61) to obtain two very useful relationships

$$D \equiv \frac{N_g^E}{N_m^E} = \frac{C_g^E}{C_m^E}\left(\frac{1-F_m}{F_m}\right) = k_D \left(\frac{1-F_m}{F_m}\right) \tag{62}$$

$$k_D \equiv \frac{C_g^E}{C_m^E} = D \left(\frac{1-F_m}{F_m}\right)^{-1} \tag{63}$$

The IVTAN code results show the mass fractions of magma are fairly constant from 2000 – 3000 K and have values of 0.998 for the BSE and 0.982 for the continental crust. The $k_D$ in Equations (62) and (63) is the partition coefficient in terms of molar concentrations. We use the Si/Mg ratio in the bulk silicate Earth to illustrate geochemical modeling with the partition coefficients defined in Equation (63).

### 7.5 Origin of the Si/Mg Ratio in the Bulk Silicate Earth

The Si/Mg molar ratio in the bulk silicate Earth is fairly well constrained – the mean value ($\pm 1\sigma$) from eight well known geochemical tabulations is $\sim 0.82 \pm 0.03$ and a recent recommended value is $\sim 0.83$ (Palme & O'Neill 2014). The Si/Mg ratio in the BSE has aroused considerable interest over time because it is about 15% smaller than the solar Si/Mg ratio of $\sim 0.97$ (Lodders 2003) and because Si and Mg are the two most abundant rocky elements combined with oxygen in the Earth. The difference between the BSE and solar composition ratios may reflect different Si/Mg ratios in Earth's upper and lower mantle (unlikely), incorporation of Si into Earth's core (plausible), and/or fractional vaporization and loss of Si and Mg during accretion of the Earth (e.g., discussion in Palme & O'Neill 2014). Based on the results in Figures 26-28 and as an illustrative example, we explicitly assume the sub-solar

Si/Mg ratio of the BSE is due solely to fractional vaporization into and subsequent loss of a steam atmosphere on the early Earth.

We need to use solar – normalized element ratios in our calculations. We refer to these normalized ratios as CI – normalized ratios because the solar elemental abundances of rocky elements are best determined by chemical analyses of CI chondrites (e.g., see Lodders 2003). Using Equation (63) the solar (CI chondrite) – normalized Si/Mg concentration ratio of 0.853 (= 0.828/0.971) in the BSE is given by

$$\left(\frac{C^{Si}}{C^{Mg}}\right)_{BSE/CI} \equiv \left(\frac{Si}{Mg}\right)_{BSE/CI} = \frac{k_D^{Si}+F_{BSE}\left(1-k_D^{Si}\right)}{k_D^{Mg}+F_{BSE}\left(1-k_D^{Mg}\right)} \qquad (64)$$

The $F_{BSE}$ in Equation (64) is the mass fraction of the bulk silicate Earth and $F_g$ is the mass fraction of the steam atmosphere. These two mass fractions sum to unity. The reader may ask – what about the core? The bulk silicate Earth and the steam atmosphere are all we need to consider for any lithophile (rock – loving) element ratio where neither element goes into the metallic core to any appreciable extent. (For a contrary view see O'Rourke & Stevenson (2016) who speculate that 1 – 2 mass % Mg may dissolve into Earth's core.)  Here we focus on Si/Mg, but we also consider Si/Al, Si/Ca, Si/Na, and Si/K. Figure 27 shows the Si/Fe ratio of the BSE could be changed by vaporization and atmospheric loss, but the complete mass balance for iron also requires consideration of Earth's core.

We now find which values of $F_g = (1 – F_{BSE})$ satisfy Equation (64) as a function of temperature for the 270 and 1100 bar steam atmospheres. The intersection of each curve with the horizontal line in Figure 26 gives the maximum temperature at which atmospheric loss from the 270 or 1100 bar steam atmosphere will lower Earth's

Si/Mg ratio. These upper temperature limits are ~ 2590 K for the 270 bar steam atmosphere and ~ 2920 K for the 1100 bar steam atmosphere. Table 5 and Equation (63) give the appropriate $k_D$ values for each case.

At 270 bars and 2000 K, loss of a steam atmosphere having ~ 1.3 % of the mass of the BSE reproduces the CI – normalized Si/Mg ratio (0.854 vs. 0.853). These conditions correspond to a fully molten magma ocean 30 degrees above its liquidus temperature of ~ 1970 K and a steam atmosphere that can be produced by vaporizing Earth's oceans. The atmospheric mass fraction that needs to be lost to match the Si/Mg ratio decreases slightly with decreasing temperature. It increases proportional to pressure (because $Si(OH)_4$ gas comprises ~ 100% of gaseous silicon and its mole fraction is proportional to pressure as shown in Equation (22) and discussed in Sections 5.2.1 and 7.3).

For example at 270 bars $F_g$ decreases to ~ 1.2 % at 1970 K. Figure 15 shows the steam atmospheres equilibrated with the BSE are ~ 75 % $H_2O$ by *mass*, so $F_g$ of 1 – 1.2 % corresponds to losing ~ 7 – 8 times Earth's present water inventory (0.107 % of the BSE by mass). But at 1100 bars and 2000 K loss of a steam atmosphere having ~ 5.5% of the BSE mass – or nearly 40 times present Earth's water inventory – is required to match the Si/Mg ratio.

As noted by the referee, correlated depletions due to the enhanced volatility of otherwise refractory elements in a steam atmosphere may lead to characteristic signatures not produced by vaporization from volatile – free magmas (i.e., different than found by Fegley & Cameron 1987, Leger et al 2011, Ito et al 2015). We computed CI – normalized ratios for other important lithophile elements under the same conditions as those discussed above for the Si/Mg ratio (i.e., 2000 K, 270 bars, $F_g$ = 0.013). The calculated ratios of Si/Ca (0.852) and Si/Al (0.880) are ~ 20% larger than the observed ratios of Si/Ca (0.730) and Si/Al (0.720). Calcium and aluminum have about the same depletion factor in the BSE, and it is important that Ca and Al are more volatile than Mg in the steam atmosphere. This is the opposite of their behavior in a solar composition gas where Al is about as refractory as Ca, and both are more refractory than Mg. It is also the opposite of their behavior for vaporization of anhydrous magmas where Al and Ca are also more refractory than Mg and remain in the residual magma after evaporative loss (e.g., Figure 6 of Fegley & Cameron 1987, Figure 5 of Leger et al 2011).

Likewise, the calculated CI – normalized Si/Na (1.4) and Si/K (3.8) ratios produced by vaporization into and subsequent loss of a steam atmosphere are also of the right size as the observed ratios of Si/Na (3.9) and Si/K (4.2) if the $k_D$ values for the halogen-free system are used (Figure 25a). However, the loss of Na and K can be 10-20 times higher if we use $k_D$ values for the system with evaporation of alkali hydroxides and halides. We did not estimate the potential loss for Fe with a steam atmosphere because the Fe in the source composition will also distribute between the BSE and the core, which makes the modeling very complex. These preliminary calculations show it is possible to match the Si/Mg ratio in the BSE, but more detailed modeling is required to determine the optimal conditions (P, T, mass fraction lost) that give the best match to the observed ratios of lithophile elements in the bulk silicate Earth. This is beyond the scope of this paper and will be done elsewhere.

A larger water inventory on the early Earth is plausible because the chondritic building blocks of the Earth contain more $H_2O$ than the present-day Earth. Lodders (2003) gives $\sim 2.1\%$ H by mass in CI chondrites, which corresponds to $\sim 18.8\%$ $H_2O$. This is $\sim 175$ times more water than in the BSE today. On average other types of chondritic material (i.e., carbonaceous, enstatite, and ordinary) also contain several times more water than Earth (Figure 2 of Fegley & Schaefer 2014). Geochemical models postulating extensive volatile loss from early Earth date back at least to Ringwood (1966). Hydrodynamic escape is a possible mechanism that has been explored with emphasis on noble gases (e.g., Section 6.2 of Porcelli & Pepin 2000 and references therein).

Our calculations for the 270 and 1100 bar steam atmospheres show that fractional vaporization and subsequent loss of rocky elements in a steam atmosphere can explain Earth's Si/Mg ratio and give upper limits for atmospheric mass loss because some of the "missing" Si may have dissolved in Earth's core.

Recent work (e.g., Kurosaki et al 2014, Lopez et al 2012, Lopez & Fortney 2013, 2014) indicates stellar UV driven mass loss is important for the evolution of hot rocky exoplanets. Detailed modeling of photo-evaporative atmospheric loss is beyond the scope of our paper but we do briefly consider stellar UV photolysis of Fe, Mg, and Si hydroxide gases.

### 7.6 Photodissociation of $Fe(OH)_2$, $Mg(OH)_2$ and $Si(OH)_4$

The geometries of the three molecules were first optimized using the range-separated CAM-B3LYP functional (Yanai et al 2004) and the 6-311+g(2d,p) basis set within the Gaussian 09 suite of programs (Frisch et al 2009). Figure 29 illustrates the ground-state geometries of the molecules. The energies and transition oscillator strengths for vertical transitions to the first 30 excited states were then computed using the time-dependent (TD) density functional method (Scalmani et al 2006). The long-range correction in the CAM-B3LYP functional makes it suitable for modeling electron excitations to high-lying orbitals. We have found that for small metal-containing molecules such as AlO, MgCl, NaOH and SiO where experimental data are available for comparison, the TD/CAM-B3LYP theory is much better than Hartree-Fock (HF) theory followed by configuration interaction singles with perturbative doubles correction CIS(D) for excited states, which is a widely-used alternative approach (Frisch et al 2009).

Figure 30 illustrates the calculated photolysis cross section as a function of wavelength ($\sigma(\lambda)$) for the three species. Their photodissociation coefficients (J) were then computed from the relation

$$J_1 = \int_{\lambda_1}^{\lambda_2} \sigma(\lambda)\Phi(\lambda)d\lambda \tag{65}$$

where $\Phi(\lambda)$ is the stellar actinic flux at 1AU (taken as the solar value), and the integration is from the thermodynamic threshold (indicated by arrows in Figure 30) to 120 nm. The resulting $J_1$ values, for the top of the atmosphere for a solar flux at 1 AU, are listed in Table 6. These values are plausibly upper limits because they neglect the UV opacity of the steam atmosphere. A comparison of the $J_1$ values in Table 6 and in Table 4 of Schaefer, Lodders & Fegley (2012) shows $Fe(OH)_2$ and $Mg(OH)_2$ are about as photochemically labile as $H_2S$ ($J_1 \sim 3.3 \times 10^{-3}$) and that $Si(OH)_4$ is about as labile as $O_2$ ($J_1 \sim 4.86 \times 10^{-6}$). The "true" photochemical lifetimes for $Si(OH)_4$, $Mg(OH)_2$, and $Fe(OH)_2$ are expected to be much longer because of the absorption and scattering by other gases (e.g., $H_2O$, $CO_2$, $SO_2$, $O_2$) in the steam atmospheres.

The photolysis products may be easier to observe than the parental metal hydroxide (M-OH) gases for several reasons. The vibrational frequencies for M – OH bonds lie in the far infrared, e.g., Spinar & Margrave (1958) observed the strongest absorption at 22.7 – 23.1 μm in the saturated vapor over NaOH (liquid). Belton & Jordan (1967) estimated the M – OH bending in $Fe(OH)_2$ at ~ 8.55 μm and the estimated wavelengths for $Ca(OH)_2$, $Mg(OH)_2$, $Ni(OH)_2$, $Al(OH)_3$, and $Si(OH)_4$ are similar (e.g., see Allendorf et al 1995, Gurvich et al 1989 – 1994, and Chase 1999). The O – H stretching frequencies in metal hydroxide gases are estimated to be in the same region as those in water vapor (~ 2.7 – 2.8 μm) and are probably masked by water bands in steam atmospheres. On the other hand, Mg, Na, and $Si^{2+}$ are observed escaping from the hot Jupiter HD209458b (Vidjal-Madjar et al 2013 and references therein), and these gases may also be observable on hot rocky exoplanets with steam atmospheres. Another reason is that the photochemical equilibrium abundances of the metal hydroxide gases may be small compared to the abundances of their photoproducts. We suggest that Al, Ca, Fe, Mg, Ni, and Si (and/or their ions) may be easier to see than the parental metal hydroxide gases. More detailed modeling is beyond the scope of this paper and is not discussed here.

## 8. Possible Cosmochemical Applications of our Work

Our work on rocky element solubility in steam is potentially relevant to several other problems including the chemistry of Uranus – and Neptune – like exoplanets (mentioned earlier in Section 5.1), chemistry during formation of the Earth and Moon (e.g., Fegley & Schaefer 2014, Fegley, Lodders & Jacobson 2016), transport and fractionation of elements on initially hydrous primitive bodies such as asteroids during heating (e.g., by [26]Al), and alkali loss from oxidized, alkali poor basaltic asteroids such as the parent bodies for the angrite and eucrite meteorites.

The DAWN mission confirmed the asteroid 4 Vesta is very likely the parent body for eucrites (e.g., see Consolmagno et al 2015 and references therein) while the angrite parent body is not yet identified. However in both cases the basaltic meteorites are severely depleted in sodium with average Na/Al mass ratios of ∼ 0.05 (eucrites) and ∼ 0.003 (angrites), which is only ∼ 8 % (eucrites) and ∼ 0.5% (angrites) of the CI – chondritic Na/Al ratio of ∼ 0.6 (CI – Lodders 2003, eucrites – Kitts & Lodders 1998, angrites – Keil 2012). The other alkalis (K, Rb, Cs) are also severely depleted relative to CI chondritic abundances (Mittlefehldt 1987). The ideas to explain the alkali depletions on the eucrite parent body (EPB = 4 Vesta) include formation of the EPB from volatile – depleted material or loss of volatiles later in the history of the EPB. Lodders (1994) briefly considered thermal escape of alkali hydroxides from the EPB and gave the sequence CsOH (most volatile) > RbOH > KOH > NaOH > LiOH (least volatile). Vaporization and loss of alkalis from a steam atmosphere on the angrite and eucrite parent bodies may occur under conditions similar to those that oxidize iron, e.g., via the schematic reaction

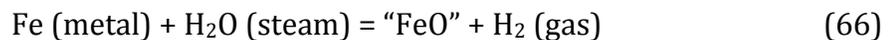

$$Fe \text{ (metal)} + H_2O \text{ (steam)} = \text{"FeO"} + H_2 \text{ (gas)} \tag{66}$$

The angrites and eucrites are FeO-rich, with ∼ 18 % (eucrites) and ∼ 20 % FeO (angrites). Our modeling could be used to study chemistry on the angrite and eucrite parent bodies as noted by the referee.

## 9. Summary

    The major conclusion of our work is that the bulk composition, density, heat balance, and interior structure of rocky planets that are undergoing or have undergone escape of steam-bearing atmospheres may be significantly altered by fractional vaporization and subsequent loss of rocky elements such as Si, Mg, Fe, Ni, Al, Ca, Na, and K that are soluble in steam. In other words, atmospheric loss may alter the composition of the rocky planet left behind if rock dissolves in the atmosphere as is true for rock in steam. This conclusion is based on chemical equilibrium calculations that show rocky element solubility in pure steam and "steam" atmospheres equilibrated with silicate magmas as a function of P, T, and composition.

The chemical equilibrium calculations use tabulated thermodynamic data for hydroxide gases of rocky elements ($Si(OH)_4$, $Mg(OH)_2$, $Fe(OH)_2$, $Ca(OH)_2$, $Al(OH)_3$, and $Ni(OH)_2$) from JANAF and elsewhere (e.g., Plyasunov 2011a, 2012). In turn, the hydroxide gas thermodynamic data are based on extensive experimental studies of the solubility of the major rock-forming elements (e.g., Si, Mg, Fe, Ca, Al, Na, K) in steam. We also show that halide and mixed halide – hydroxide gases of Na, K, Ca, and Al contribute significantly to the solubility of these elements in steam atmospheres. Our conclusions are potentially testable by measurements of planetary mass and radius and *possibly* by spectroscopic observations of the metal hydroxide gases and/or their photolysis products.

## Acknowledgments

BF conceived the idea, integrated the models, and wrote much of the paper with help from KL, LS, and the other authors. KL developed the partition coefficient mod-eling and did the calculations for changes in Si/Mg ratio due to atmospheric loss. BF, NSJ, and KBW did chemical equilibrium calculations with the IVTAN & MAGMA, FactSage, and MELTS codes, respectively. JMCP did quantum chemical calculations for thermochemical and photochemical reactions described in the text. BF and KL were supported by grant AST-1412175 from the NSF Astronomy Program and by the NASA EPSCOR Program Grant NNX13AE52A (BF). The NASA EPSCOR Program and NASA Glenn Research Center supported NSJ. KBW was supported by the McDonnell Center Roger B. Chaffee Fellowship, JCMP was supported by European Research Council (project number 291332 – CODITA), and LS was supported by the Simons Foundation. We thank Andrey Plyasunov for helpful discussions and his tabular data for orthosilicic acid vapor, Bob Pepin for helpful discussions on his model, Beth Opilia for comments on Figure 29, and the anonymous referee for their helpful comments which led us to clarify and expand our discussion of chemical interactions between steam atmospheres and magma oceans and its possible effects for planetary compositions.

Treatise on Geochemistry 2nd ed. Chapter 3.1

## Figure Captions

Figure 1. Fugacity coefficient ($\phi$ = f/P) isobars for steam from 1000 – 3000 K. The $\phi$ values are unity within 0.5% at 10 bars and within 0.02% at P < 10 bars and are calculated from the equation of state for water using the Loner HGK code of Bakker (2009). Compare the ordinate in this graph will that on Figure 2.

Figure 2. Fugacity coefficient ($\phi$ = f/P) isobars for $Si(OH)_4$ from 1000 – 3000 K. The $\phi$ values are calculated as described by Akinfiev & Plyasunov (2013). Compare the ordinate in this graph with that on Figure 1.

Figure 3. The enhanced volatility of silica due to its solubility in steam at $P_{steam}$ = saturated vapor pressure of water up to 647 K, then $P_{steam}$ = 220.64 bars – the critical pressure of water. The total amount of gaseous silicon in all forms ($P_{\Sigma Si}$) is plotted for silica in steam (red curve) and for the vapor pressure of pure silica (solid, liquid) (black curve). Silica vaporizes to a mixture of gases ($SiO + O_2 + O + SiO_2 + Si$) that has the same Si/O ratio as silica. Krikorian (1970) notes the solubility of silica in steam may be limited by precipitation of hydrated silica (e.g., $SiO_2 \cdot \frac{1}{2}H_2O$) below 475 K, but thermodynamic data for silica hydrates are very uncertain and their precipitation is not shown in this graph. The total pressure ($P_{\Sigma Si}$) of Si-bearing gases in steam is dominated by $Si(OH)_4$ until very high temperatures where SiO and $SiO_2$ also become important. The exact temperature depends on the total steam pressure. Representative error bars corresponding to ± 3.0 kJ/mol uncertainty in $\Delta_r G^o$ for reaction (1) (Plyasunov 2011a, 2012) are shown on the red curve. The white points

are the results from Table 6 of Plyasunov (2012), the red curve are our calculations for the same reaction using his data. Measured (Kazenas et al. 1985, blue points) and calculated (Krieger 1965, green points) vapor pressures of $SiO_2$ (s, liquid) agree with the calculated vapor pressure from the IVTAN code.

Figure 4. (H4SiO4pp.spw) Silica solubility isobars in steam from 1 – 2,000 bars pressure. Solubility is expressed as mass % silica in steam and as the mole fraction of orthosilicic acid vapor $Si(OH)_4$. The 1000 – 1200 K points on the 2000 bar isobar are above the maximum density of 322 kg m$^{-3}$ at which the $Si(OH)_4$ fugacity coefficients are reliable (see text and Plyasunov 2012). The three green points from Anderson & Burnham (1965) show the "true" solubility of silica at these points and they blend smoothly into the 2000 bar isobar at 1300 K where $\rho_{steam} \sim 322$ kg m$^{-3.}$

Figure 5. The enhanced volatility of MgO (periclase) in steam is compared to the vapor pressure of pure MgO (solid, liquid). The total amount of gaseous magnesium in all forms ($P_{\Sigma Mg}$) is plotted for MgO dissolved in steam and for the vapor pressure of pure MgO (solid, liquid). The red curve is the total amount of Mg in all forms ($P_{\Sigma Mg} = P_{Mg(OH)2} + P_{MgOH} + P_{Mg} + P_{MgO} + P_{MgH}$) dissolved in steam. The solubility of MgO in steam is limited by precipitation of $Mg(OH)_2$ at temperatures below 780 K. Representative error bars corresponding to $\pm$ 20 kJ/mol uncertainty in $\Delta_r G^o$ for reaction (26) are shown on the red curve. The black curve is the vapor pressure ($P_{\Sigma Mg}$) of pure MgO (solid, liquid). Periclase vaporizes to a mixture of gases (Mg + $O_2$ + O + $O_3$ + MgO + $Mg_2$) that has the same Mg/O ratio as MgO. Measured (Kazenas et al. 1983, blue points) and calculated (Krieger 1966, green points) vapor pressures of MgO (s, liquid) agree with the calculated vapor pressure from the IVTAN code.

Figure 6. The enhanced volatility of "FeO" (wüstite) in steam is compared to the vapor pressure of pure "FeO" (solid, liquid). The red curve is the total amount of Fe in all forms ($P_{\Sigma Fe} = P_{Fe(OH)2} + P_{FeOH} + P_{FeO(OH)} + P_{Fe} + P_{Fe2} + P_{FeO} + P_{FeO2} + P_{FeH}$) dissolved in steam. The black curves are the partial vapor pressures of Fe gases ($P_{\Sigma Fe} \sim P_{Fe} \sim P_{vap}$) and $O_2$ of pure metal-saturated "FeO" (wüstite, liquid) at T ≥ 843 K, the wüstite eutectoid temperature. Below 843 K the black curves are the partial vapor pressures of Fe and $O_2$ over metal saturated magnetite. The blue and green squares are solid-state zirconia sensor $fO_2$ measurements by O'Neill (1988) for wüstite and magnetite, respectively. The yellow squares are solid-state zirconia sensor $fO_2$ measurements by O'Neill & Pownceby (1993) for wüstite.  The black triangle is a set of $fO_2$ measurements for liquid FeO by Knudsen effusion mass spectrometry by Kazenas & Tagirov (1995). Representative error bars corresponding to ± 30 kJ/mol uncertainty in the data for Fe(OH)$_2$ gas (Gurvich et al. 1983) are shown on the red curve.

Figure 7. The enhanced volatility of $Fe_3O_4$ (magnetite) in steam is compared to the vapor pressure of pure $Fe_3O_4$ (solid, liquid). The red curve is the total amount of Fe in all forms ($P_{\Sigma Fe} = P_{Fe(OH)2} + P_{FeOH} + P_{FeO(OH)} + P_{Fe} + P_{Fe2} + P_{FeO} + P_{FeO2} + P_{FeH}$) dissolved in steam. The black curves are the partial vapor pressures of Fe and $O_2$ of metal-rich $Fe_3O_4$ (magnetite) and liquid $Fe_3O_4$ (T ≥ 1870 K). The pink circles (Jacobsson 1985) and green squares (O'Neill 1988) are solid-state zirconia sensor $fO_2$ measurements.  The two blue triangles are $O_2$ partial pressures read off the Fe – O phase diagram of Muan & Osborn (1965). Representative error bars corresponding to ± 30 kJ/mol uncertainty in the data for Fe(OH)$_2$ gas (Gurvich et al. 1983) are shown on the red curve.

Figure 8. The enhanced volatility of $Fe_2O_3$ (hematite) in steam is compared to the vapor pressure of pure $Fe_2O_3$ (hematite) and liquid $Fe_2O_3$ (T ≥ 1895 K). The red curve is the total amount of Fe in all forms ($P_{\Sigma Fe} = P_{Fe(OH)2} + P_{FeOH} + P_{FeO(OH)} + P_{Fe} + P_{Fe2} + P_{FeO} + P_{FeO2} + P_{FeH}$) dissolved in steam. The two black curves are the vapor pressure of $O_2$ and of all Fe gases ($P_{\Sigma Fe}$) of pure $Fe_2O_3$ (solid, liquid). The yellow circles are calculated using the $\Delta G°$ equation of Hemingway (1990) for the $fO_2$ of coexisting magnetite + hematite. The blue, green, and pink points are measurements by Salmon (1961), Jacobsson (1985), and Blumenthal & Whitmore (1961) of the $O_2$ partial pressure of hematite saturated with magnetite (i.e., along the magnetite – hematite phase boundary in the Fe – O phase diagram). Representative error bars corresponding to ± 30 kJ/mol uncertainty in in the data for $Fe(OH)_2$ gas (Gurvich et al. 1983) are shown on the red curve.

Figure 9. The enhanced volatility of CaO (lime) in steam is compared to the vapor pressure of pure CaO (lime) and liquid CaO (T ≥ 3172 K). The red curve is the total amount of Ca in all forms ($P_{\Sigma Ca} = P_{Ca(OH)2} + P_{CaOH} + P_{Ca} + P_{CaO} + P_{CaH} + P_{Ca2}$) dissolved in steam. The solubility of CaO in steam is limited by precipitation of $Ca(OH)_2$ at temperatures below 1550 K. Representative error bars corresponding to ± 15 kJ/mol uncertainty in the data for $Ca(OH)_2$ gas are shown on the red curve. The black curve is the vapor pressure ($P_{\Sigma Ca}$) of pure CaO (solid, liquid). Lime vaporizes to a mixture of gases (Ca + $O_2$ + O + $O_3$ + CaO + $Ca_2$) that has the same Ca/O ratio as lime. Our calculated vapor pressure curve agrees with experimental data (blue circles, Samoilova & Kazenas 1995) and calculations (green triangles, Krieger 1967).

Figure 10. The enhanced volatility of $Al_2O_3$ (corundum) in steam is compared to the vapor pressure of pure $Al_2O_3$ (corundum) and liquid $Al_2O_3$ (T ≥ 2327 K). The red curve is the total amount of Al in all forms ($P_{\Sigma Al} = P_{Al(OH)3} + P_{Al(OH)2} + P_{AlOH} + P_{HAlO2} + P_{HAlO} + P_{AlH} + P_{AlH2} + P_{AlH3} + P_{Al} + P_{AlO} + P_{Al2O} + P_{Al2O2} + P_{AlO2} + P_{Al2O3} + P_{Al2}$) dissolved in steam. At T ≤ 642 K the solubility of $Al_2O_3$ in steam and thus the $Al(OH)_3$ partial pressure is limited by precipitation of $AlO(OH)$ (diaspore). Representative error bars corresponding to ± 15 kJ/mol uncertainty in the data for $Al(OH)_3$ gas are shown on the red curve. The black curve is the vapor pressure ($P_{\Sigma Al}$) of pure $Al_2O_3$ (solid, liquid). Corundum vaporizes to a mixture of gases ($Al + AlO + Al_2O + Al_2O_2 + AlO_2 + Al_2O_3 + Al_2 + O + O_2 + O_3$) that has the same Al/O ratio as $Al_2O_3$. The blue circles are laser vaporization measurements of the vapor pressure of liquid $Al_2O_3$ (Hastie et al 2000), the pink triangles (Drowart et al. 1960) and green squares (Chervonnyi et al 1977) are KEMS measurements of the vapor pressure of $Al_2O_3$ (corundum), and the cyan triangles are calculations by Krieger (1966b).

Figure 11. The enhanced volatility of NiO in steam is compared to the vapor pressure of NiO (bunsenite) and liquid NiO (T ≥ 2228 K). The red curve is the total amount of Ni in all forms ($P_{\Sigma Ni} = P_{Ni(OH)2} + P_{NiOH} + P_{Ni} + P_{NiO} + P_{NiH} + P_{Ni2}$) dissolved in steam. Representative error bars corresponding to ± 20 kJ/mol uncertainty in the data for $Ni(OH)_2$ gas are shown on the red curve. The black curves are the partial vapor pressures of Ni and $O_2$ of pure metal-saturated NiO. The green squares are the solid-state $fO_2$ measurements of O'Neill & Pownceby (1993) and the blue triangles are calculated $fO_2$ values using the $\Delta G°$ equation of Hemingway (1990) for NiO formation. The cyan circles are Ni partial vapor pressures measured by KEMS by Grim-

ley et al (1961). The dark red diamonds are KEMS Ni partial vapor pressures of Ka-zenas & Tagirov (1995).

Figure 12. The calculated solubility at 300 bar and 1500 K of $SiO_2$ in steam and in binary gas mixtures formed by steam plus a second gas. The different points show the different gases. The square root of the $Si(OH)_4$ mole fraction is proportional to the steam mole fraction as predicted by the equilibrium constant expression (22).

Figure 13. The calculated solubility at 300 bar and 1500 K of MgO in steam and in binary gas mixtures formed by steam plus a second gas. The different points show the different gases. The $Mg(OH)_2$ mole fraction is proportional to the steam mole fraction as predicted by the equilibrium constant expression (29).

Figure 14. The calculated solubility at 300 bar and 1500 K of FeO in steam and in binary gas mixtures formed by steam plus a second gas. The different points show the different gases. The $Fe(OH)_2$ mole fraction is proportional to the steam mole fraction as predicted by the equilibrium constant expression (33).

Figure 15. Abundances of the major gases in steam atmospheres (270, 1100 bars) in

chemical equilibrium with bulk silicate Earth (BSE) and continental crust (CC)

magmas. The same color-coding is used in all graphs.

Figure 16. Chemical equilibrium abundances of the major metal hydroxide gases in

270 and 1100 bar steam atmospheres equilibrated with the BSE and CC magmas.

Figure 17. Abundances of the major Si-bearing gases in a 270 bar steam atmosphere

in chemical equilibrium with BSE magma.

Figure 18. Abundances of the major Mg-bearing gases in a 270 bar steam atmos-

phere in chemical equilibrium with BSE magma.

Figure 19. Abundances of the major Fe-bearing gases in a 270 bar steam atmosphere in chemical equilibrium with BSE magma. The curves for FeCl and FeF plot on top of each other and only that for FeCl is shown on the graph.

Figure 20. Abundances of the major Ni-bearing gases in a 270 bar steam atmosphere in chemical equilibrium with BSE magma.

Figure 21. Chemical equilibrium abundances of the major Na-bearing gases in a 270 bar steam atmosphere in equilibrium with the BSE magma.

Figure 22. Chemical equilibrium abundances of the major K-bearing gases in a 270 bar steam atmosphere in equilibrium with the BSE magma.

Figure 23. Chemical equilibrium abundances of the major Al-bearing gases in a 270 bar steam atmosphere in equilibrium with the BSE magma.

Figure 24. Chemical equilibrium abundances of the major Ca-bearing gases in a 270 bar steam atmosphere in equilibrium with the BSE magma.

Figure 25. Gas/magma molar distribution coefficients (D) for rocky elements in the 270 or 1100 bar steam atmospheres equilibrated with BSE magma (a,b) and CC magma (c,d). The dashed curves in panel (a) are calculations without any halide gases, i.e. all F and Cl dissolved in the BSE magma ocean. The D values are defined in Equation (57) in the text.

Figure 26. Fractional vaporization of Si and Mg from BSE magma into steam atmospheres with pressures of 270 bars (red) or 1100 bars (blue). The solid and dashed lines are calculations with or without halide gases. The molar Si/Mg ratios in the molten BSE magma and in the steam atmosphere are shown. The atmospheric Si/Mg ratio is greater than that in the BSE above the horizontal line and less than that in the BSE below the horizontal line. Thus atmospheric loss will either deplete (cooler surface temperature) or enrich (hotter surface temperature) the residual rocky planet in Si relative to Mg. The crossover temperature depends on pressure of the steam atmosphere.

Figure 27. Fractional vaporization of Si and Fe in 270 and 1100 bar steam atmospheres in equilibrium with the BSE composition of Palme & O'Neill (2014). The atomic Si/Fe ratios in the molten BSE magma and in the steam atmosphere are shown. The atmospheric Si/Fe ratio is greater than that in the BSE above the horizontal line and less than that in the BSE below the horizontal line. Thus atmospheric loss will either deplete (cooler surface temperature) or enrich (hotter surface temperature) the residual rocky planet in Si relative to Fe. The crossover temperature depends on pressure of the steam atmosphere.

Figure 28. The fractional vaporization of Si and Ca from CC magma (Table 2) into 270 bar (red) and 1100 bar (blue) steam atmospheres as a function of temperature. The Si/Ca ratio in the CC magma is the horizontal black line. The atmospheric Si/Ca ratio is greater than that in the CC above the horizontal line and less than that in the CC below the horizontal line. Thus atmospheric loss will either deplete (cooler surface temperature) or enrich (hotter surface temperature) the residual rocky planet in Si relative to Ca. The crossover temperature depends on pressure of the steam atmosphere.

Figure 29. Optimized geometries at the CAM-B3LYP/6-311+g(2d,p) level of theory. Scale: the Fe-O bond length in $Fe(OH)_2$ is 1.77 Å.

Figure 30. Calculated absorption cross sections as a function of wavelength for Fe(OH)$_2$ (red line), Mg(OH)$_2$ (green line) and Si(OH)$_4$ (blue line). The arrows in corresponding colors indicate the thermodynamic threshold for photolysis.

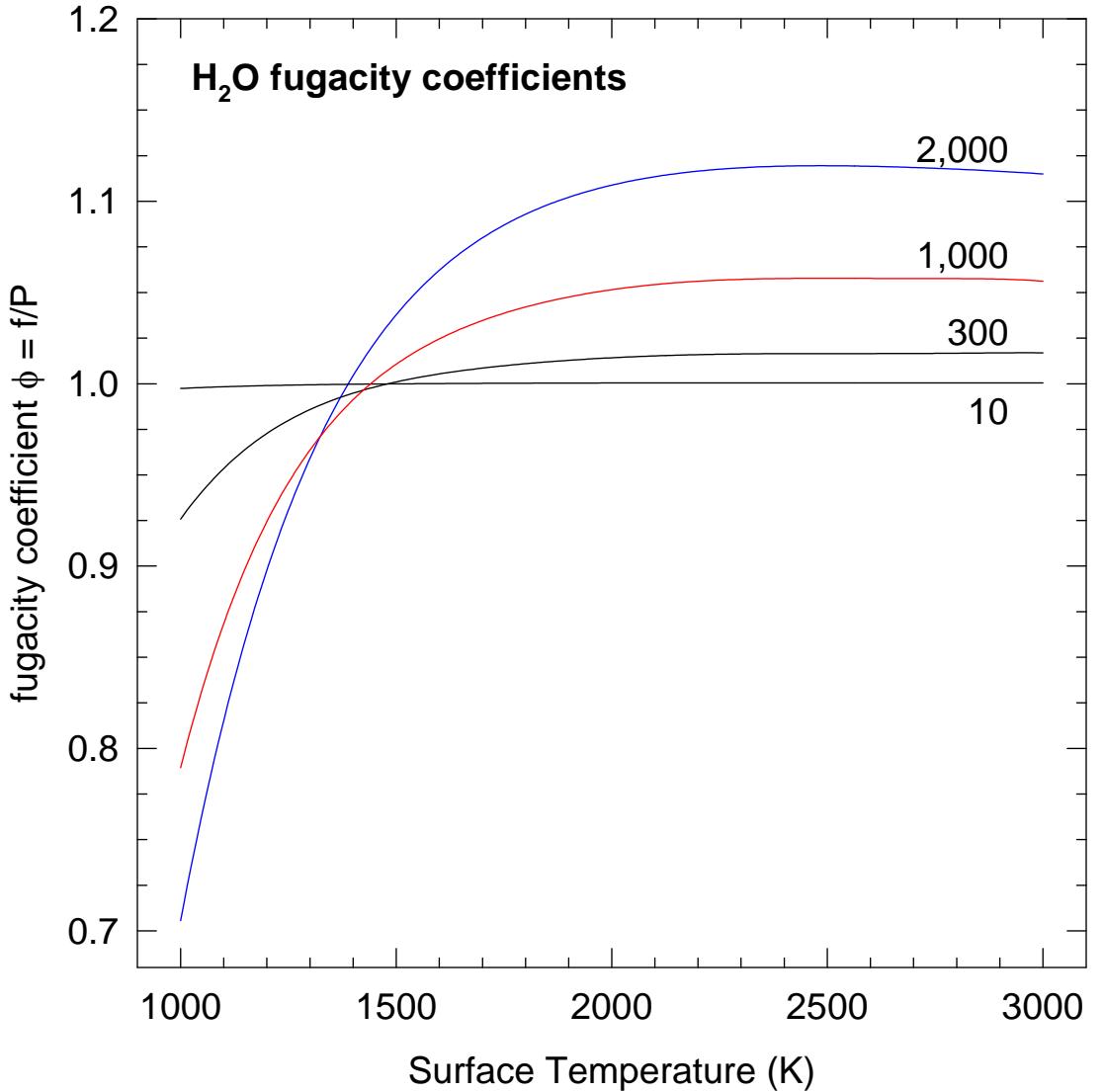

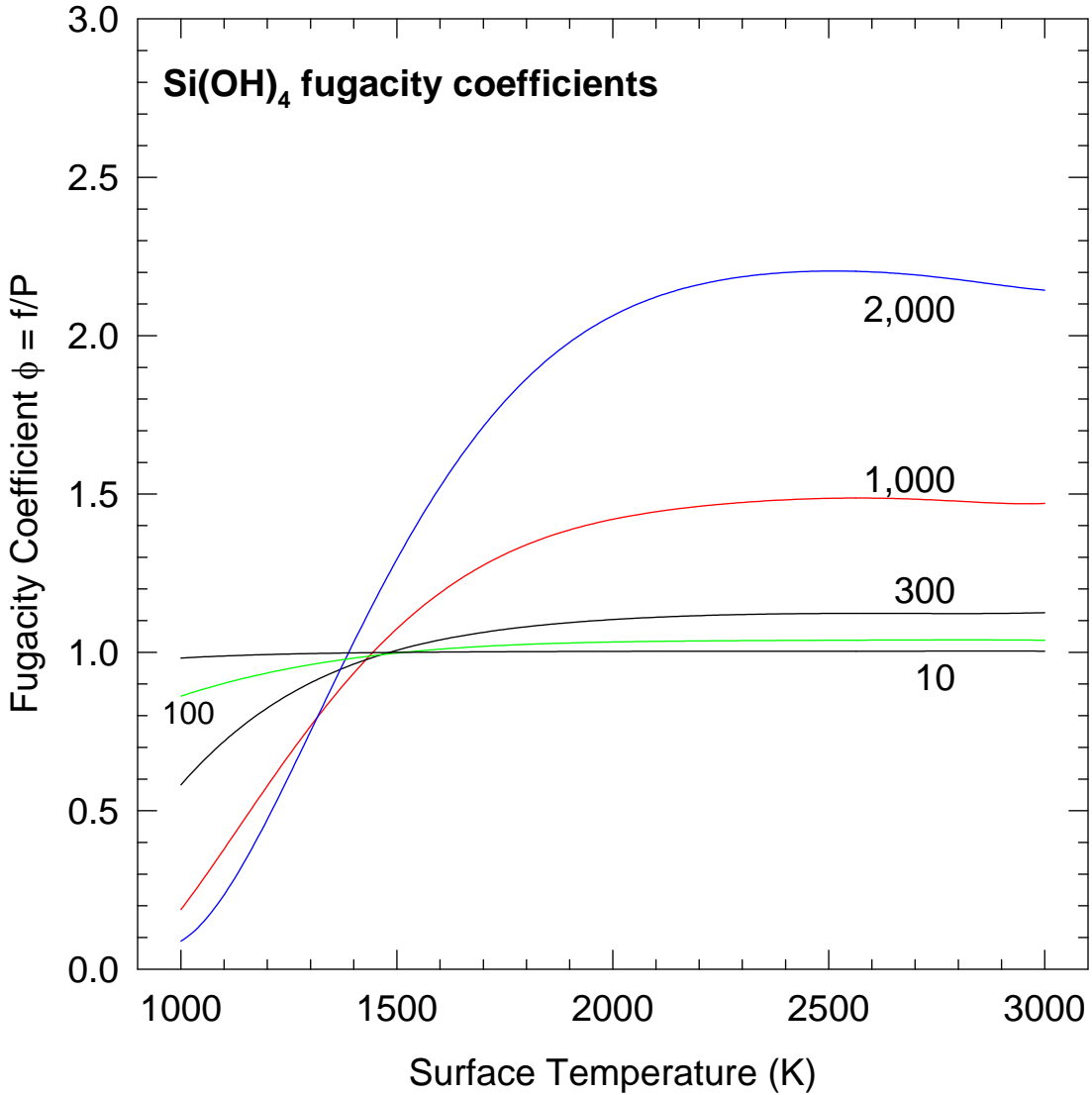

Surface Temperature (Celsius)

silica solubility in steam
red - this work
white - Plyasunov (2012)

vapor pressure
of pure $SiO_2$ (s,liq)

$\log_{10} P_{\Sigma Si}$ (bar)

10,000/T (K)

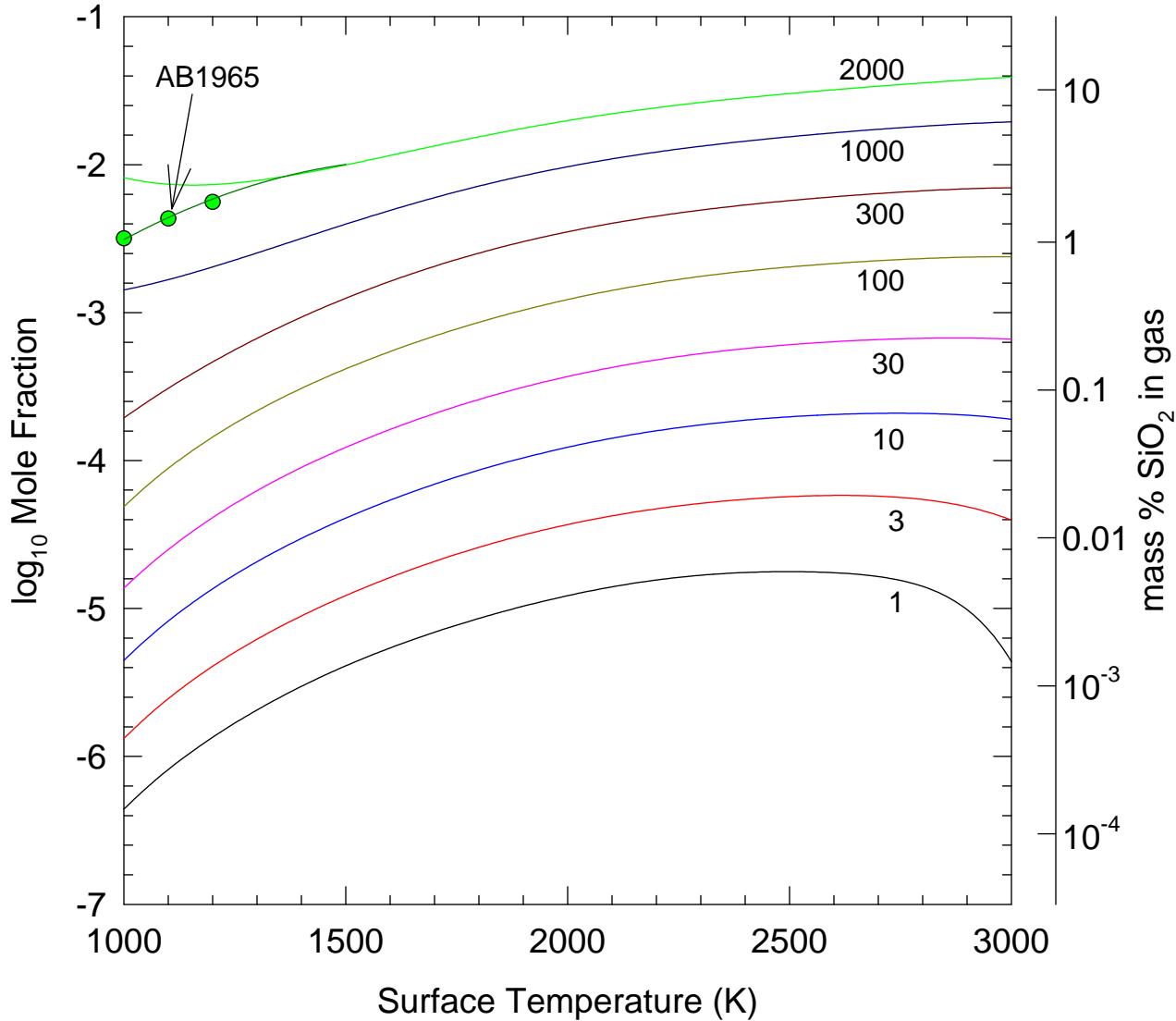

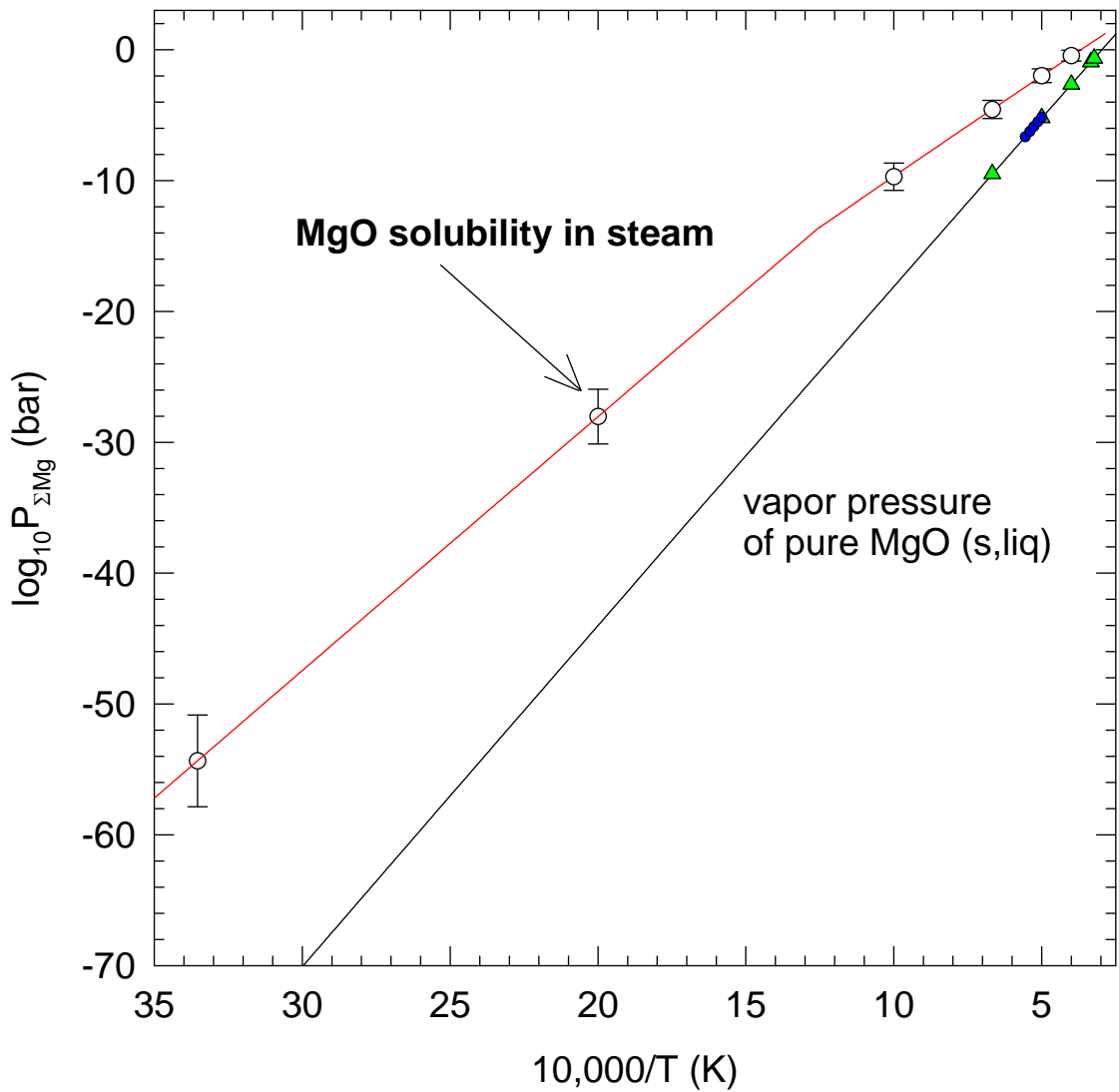

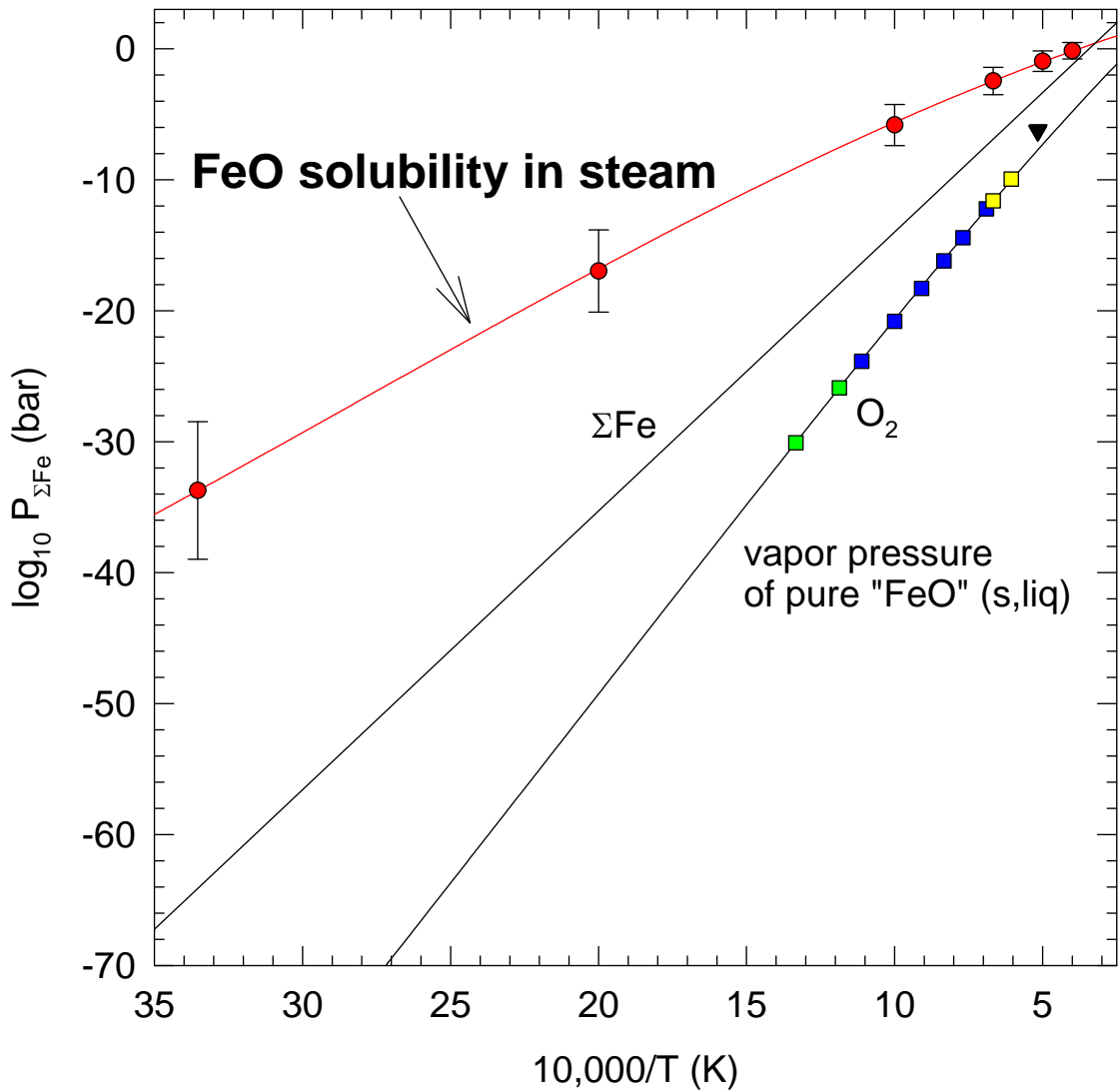

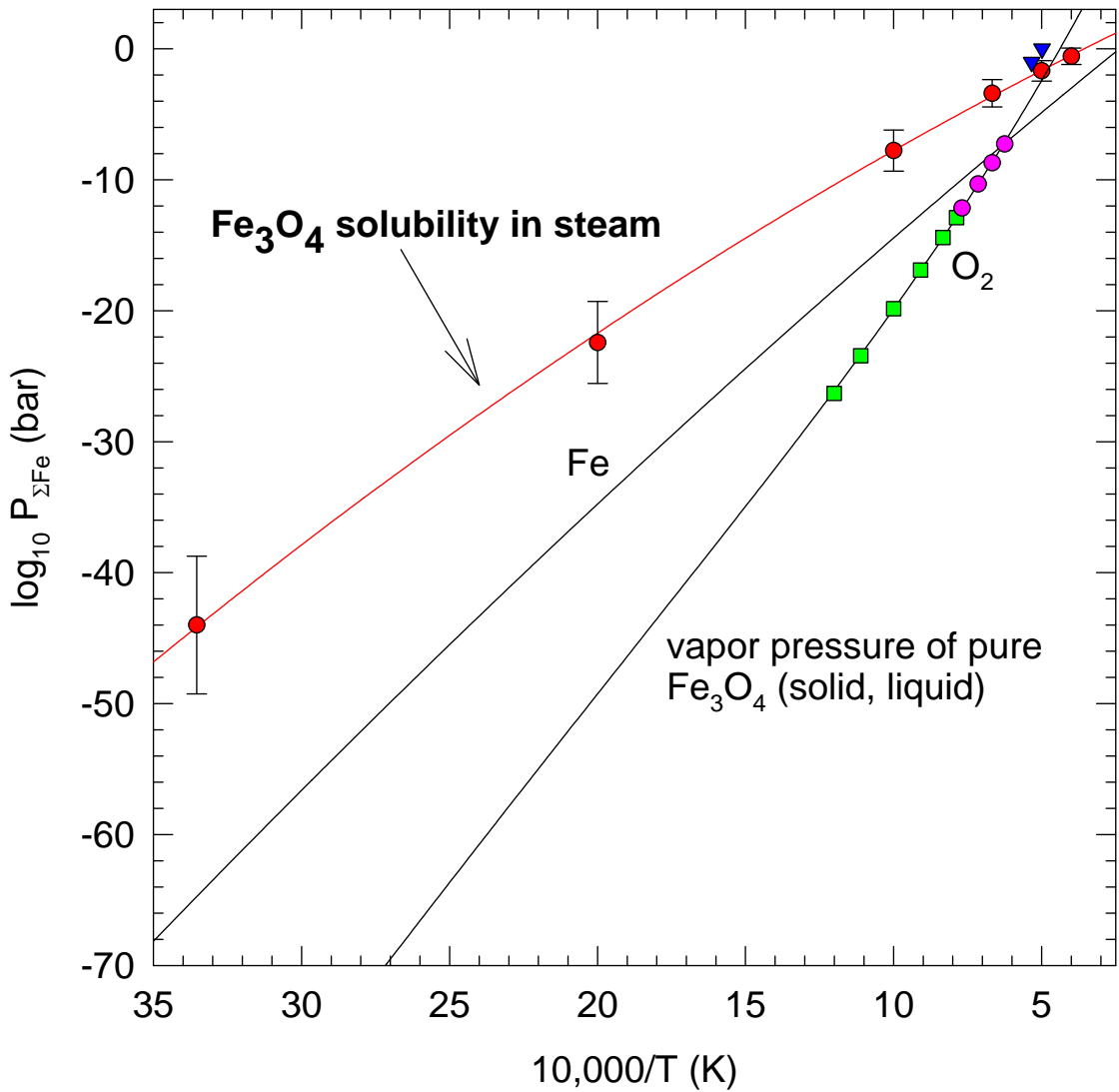

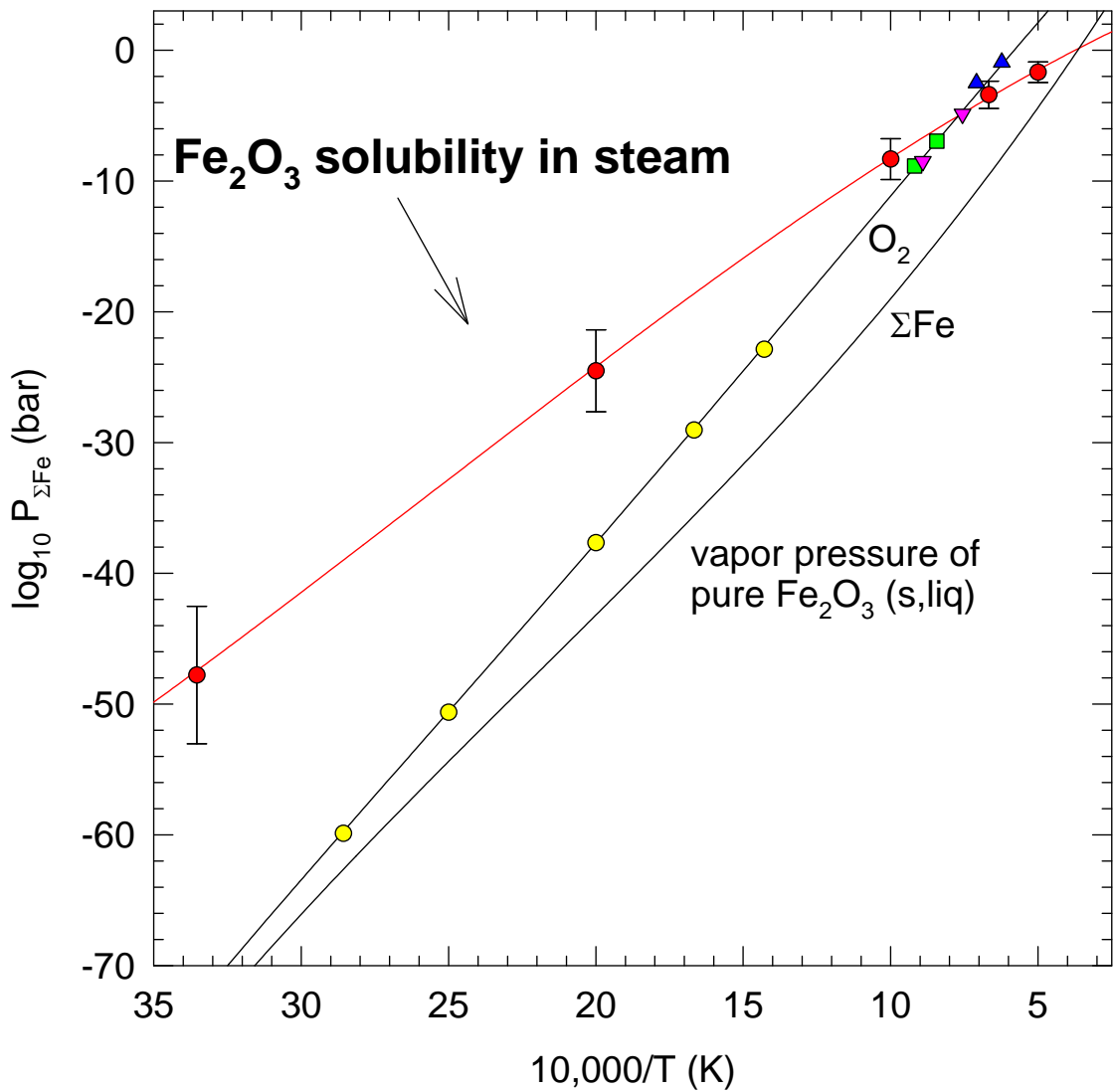

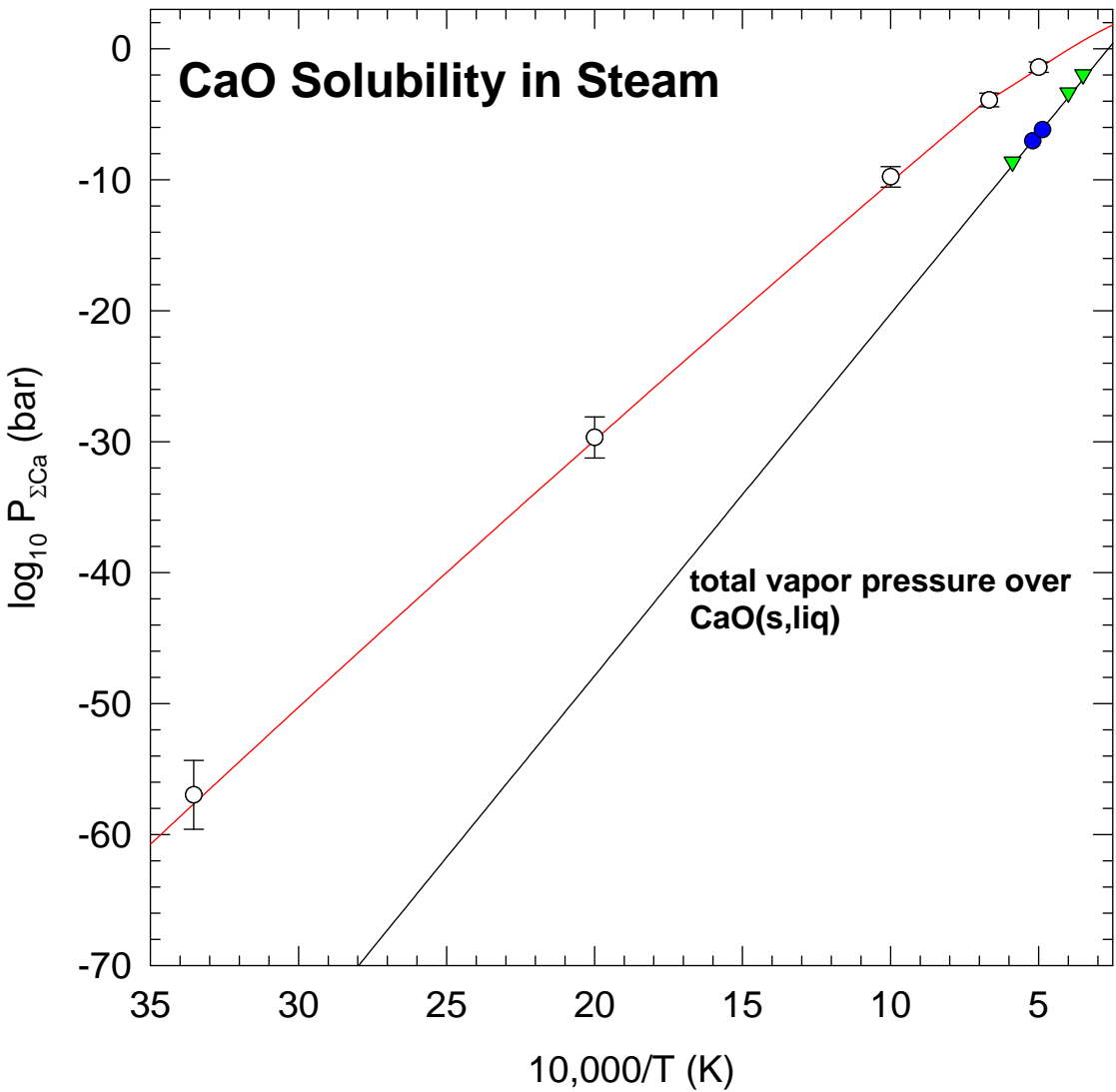

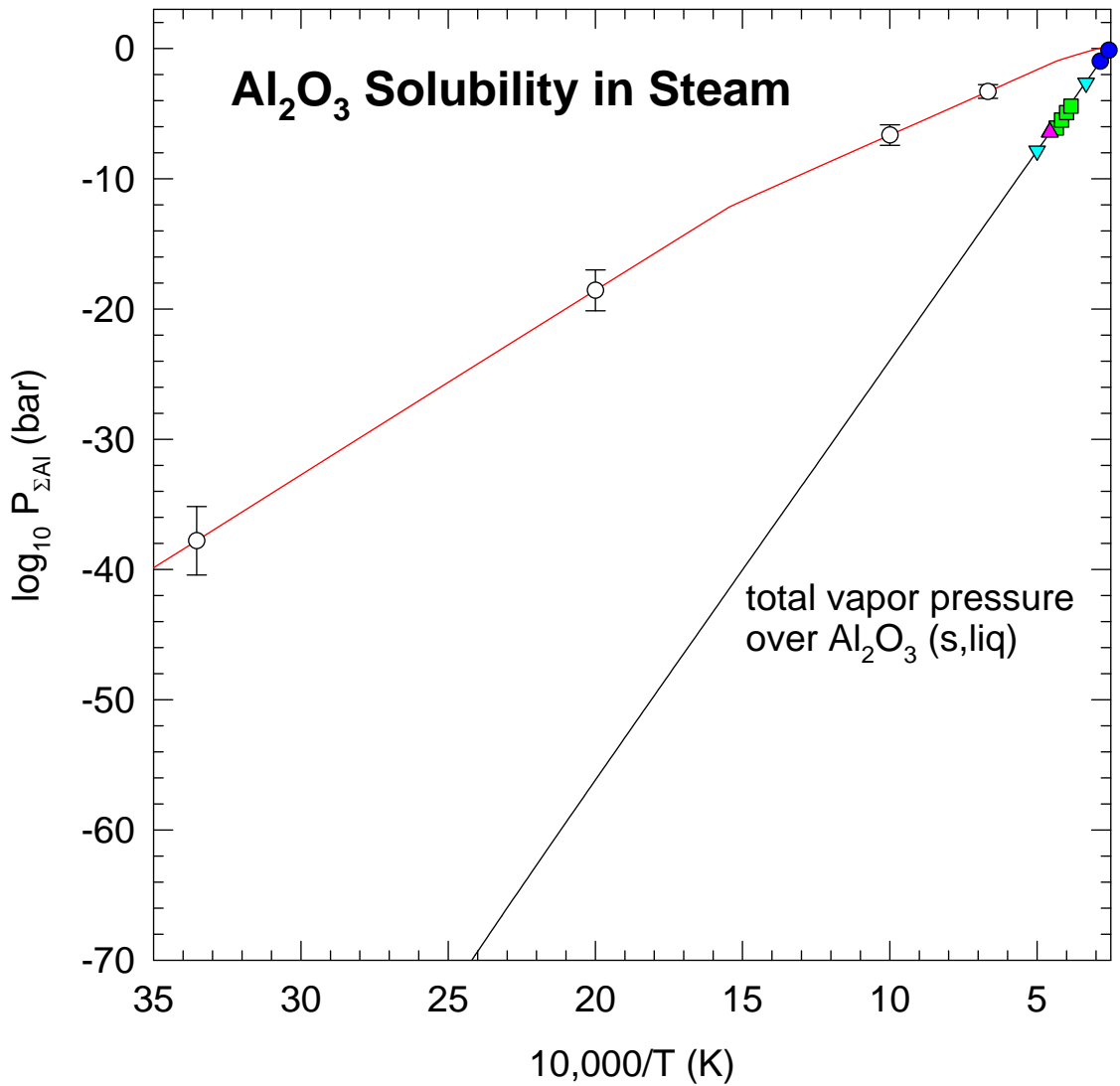

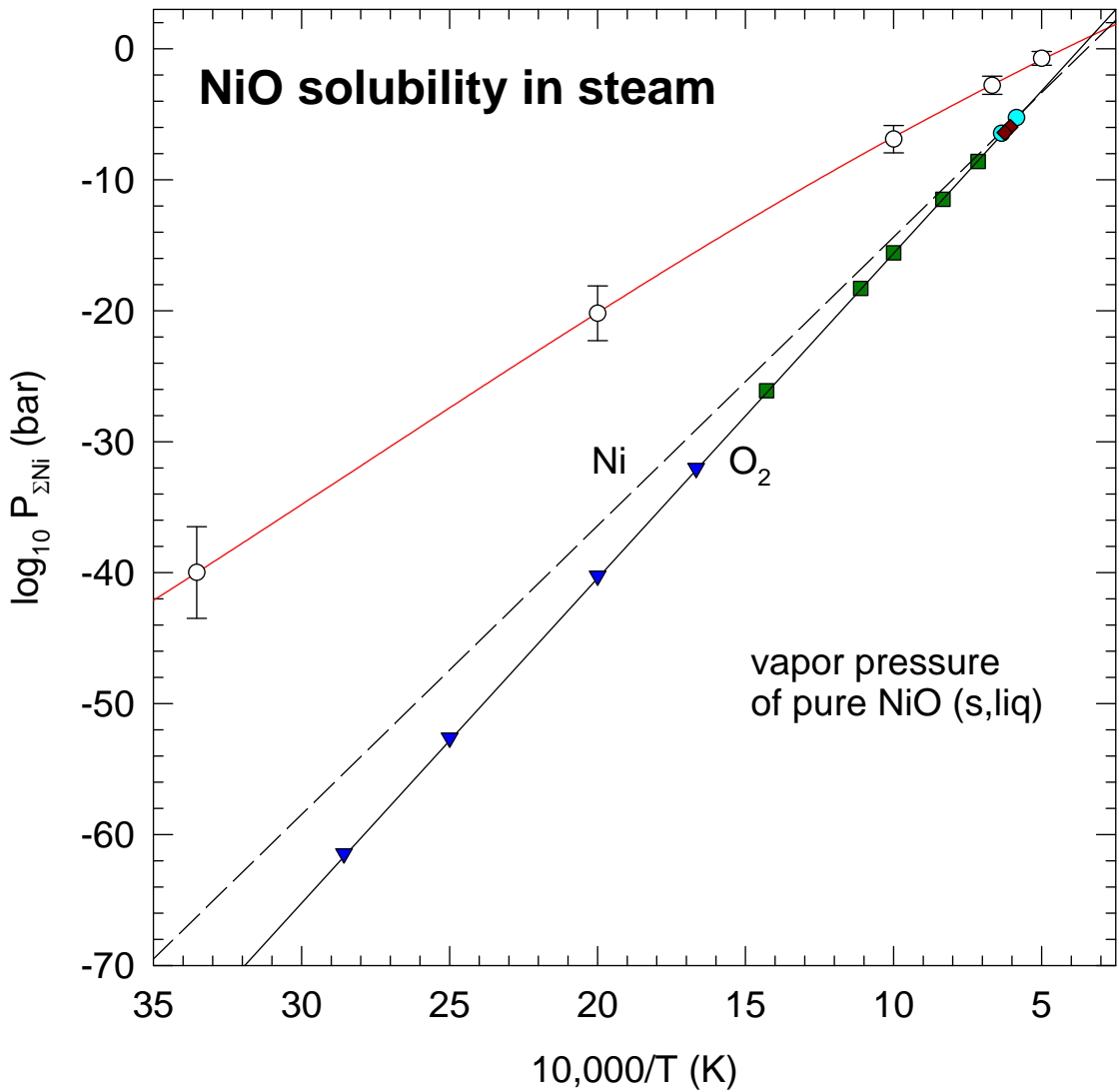

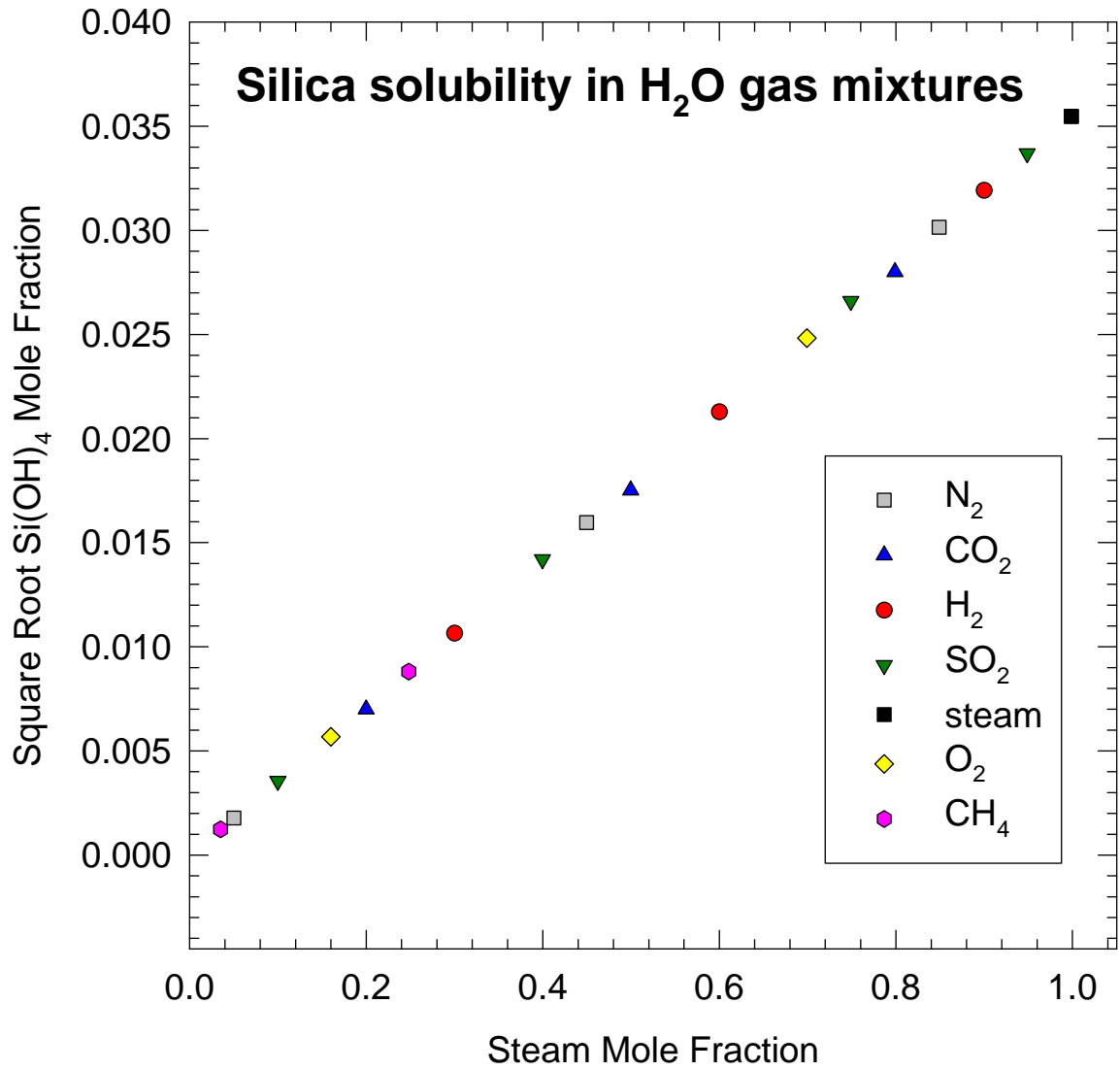

Silica solubility in H$_2$O gas mixtures

Legend:
- N$_2$
- CO$_2$
- H$_2$
- SO$_2$
- steam
- O$_2$
- CH$_4$

300 bars 1500 K

SiGasMix.spw

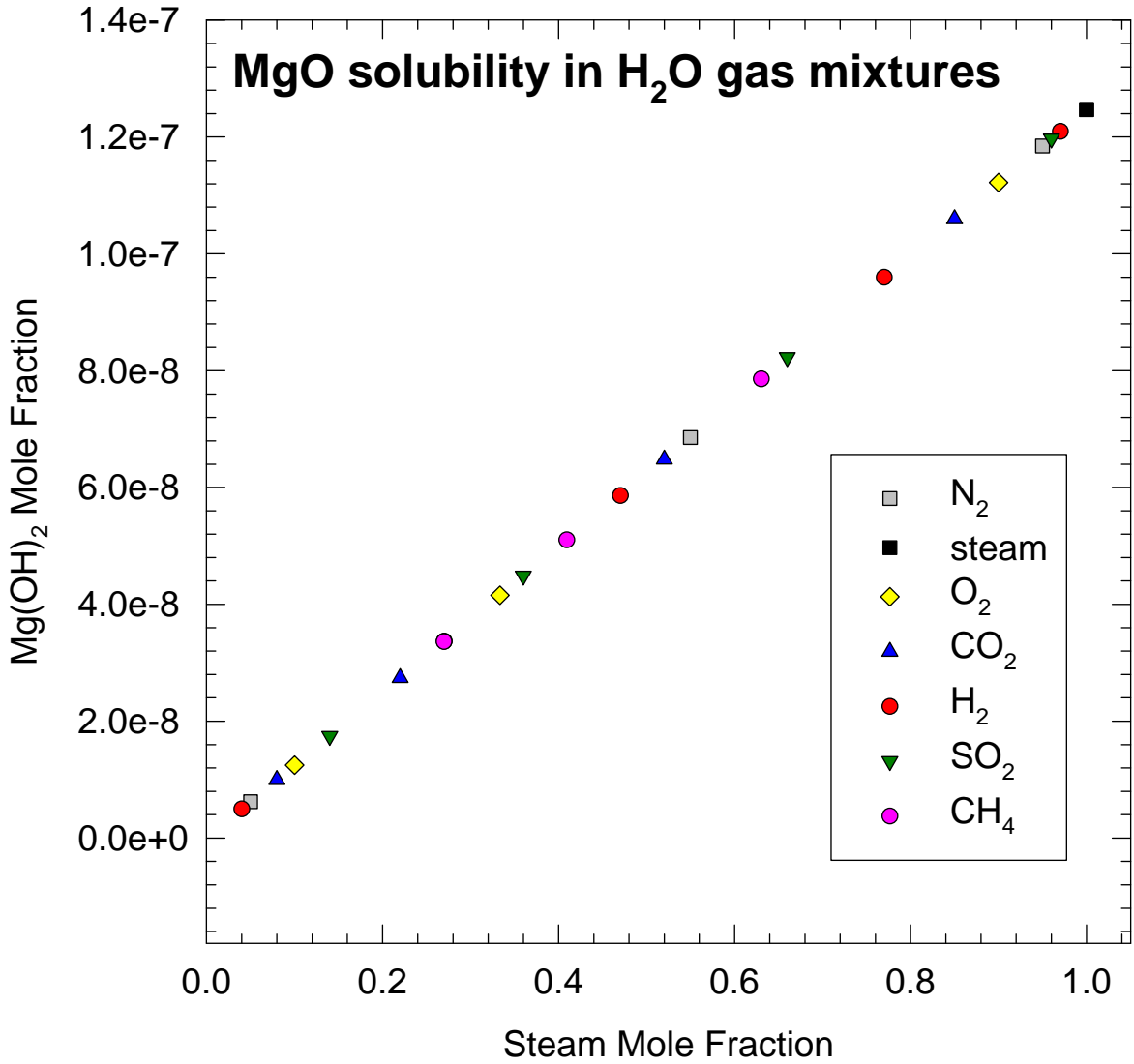

MgGasMix.spw

# FeO solubility in H$_2$O gas mixtures

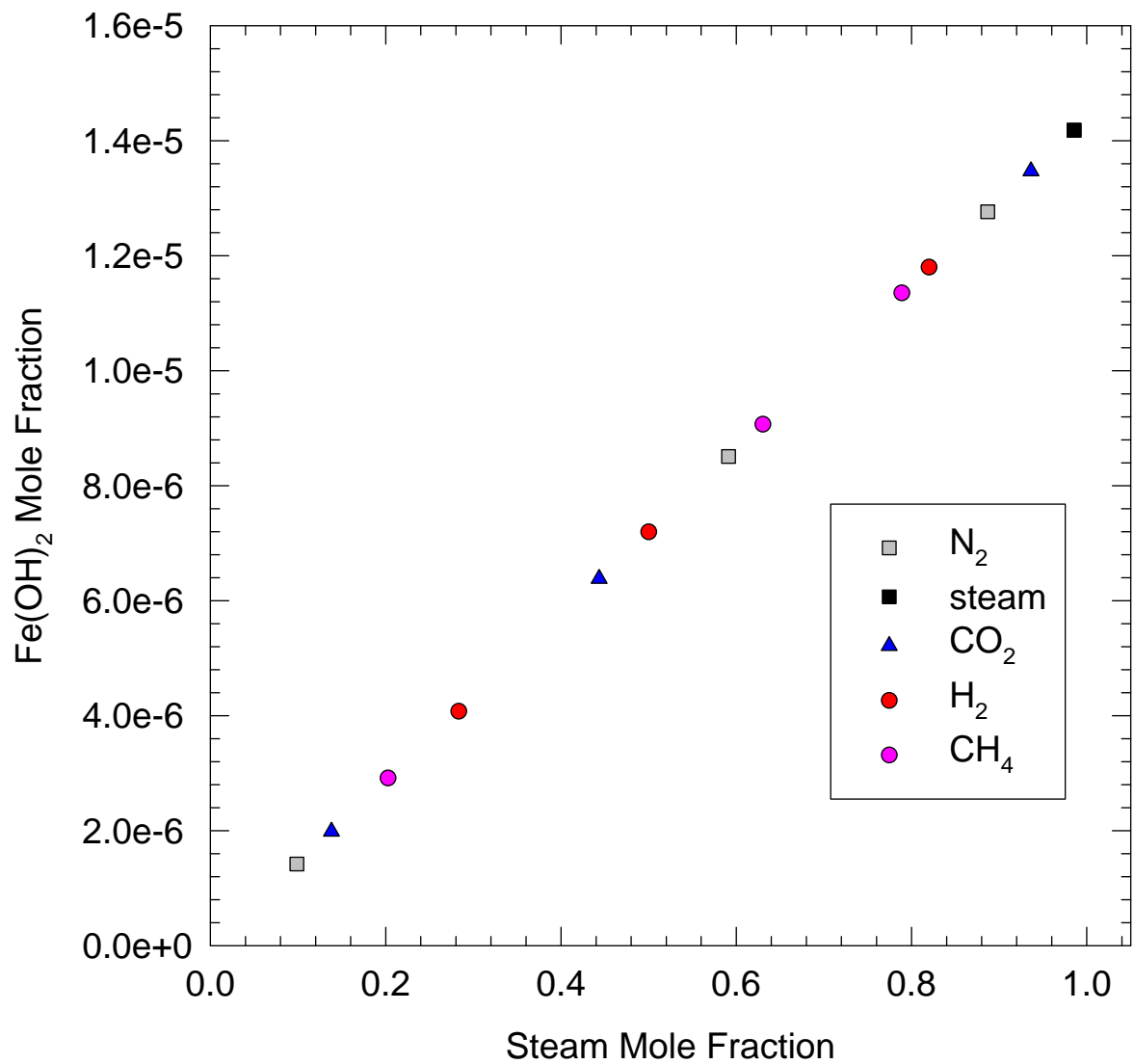

FeGasMix.spw

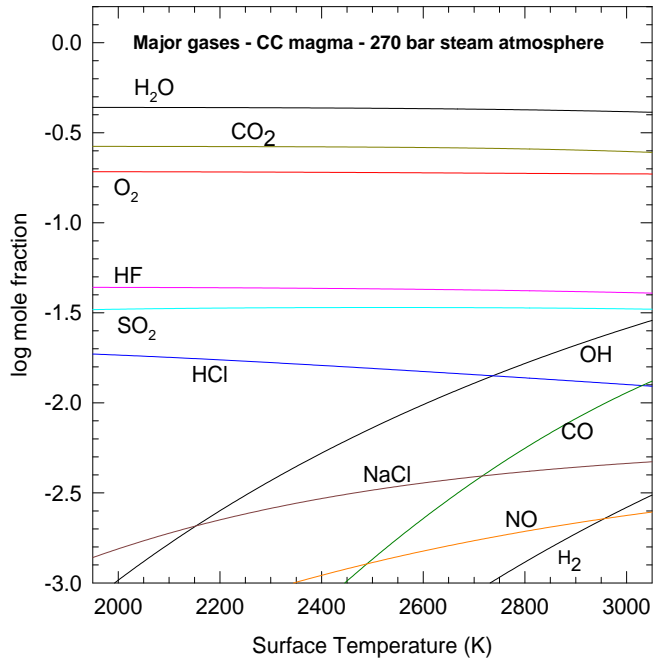

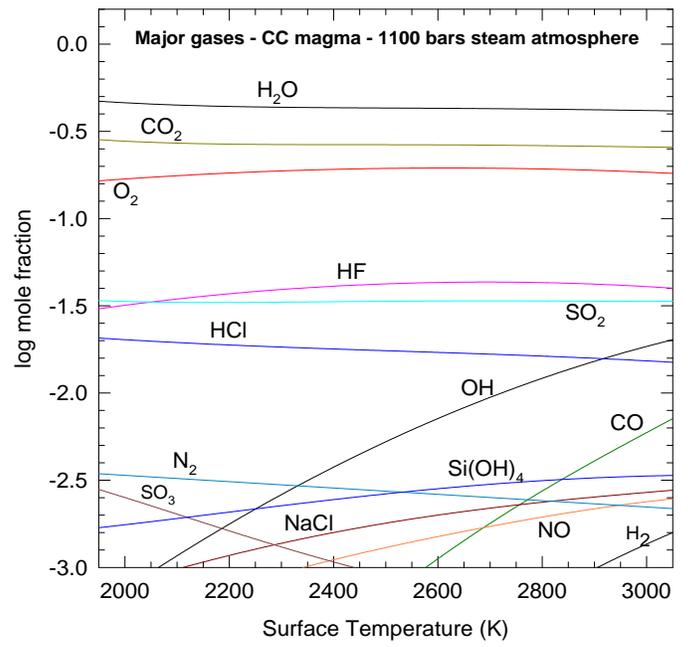

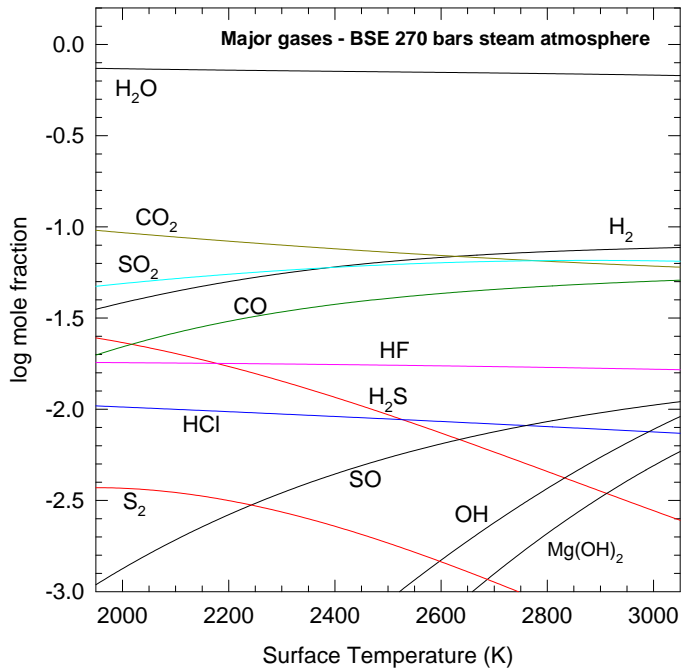

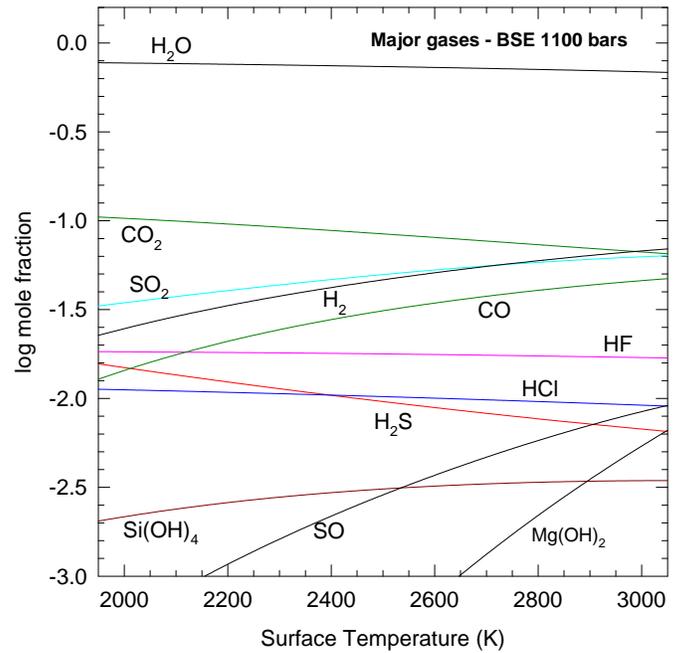

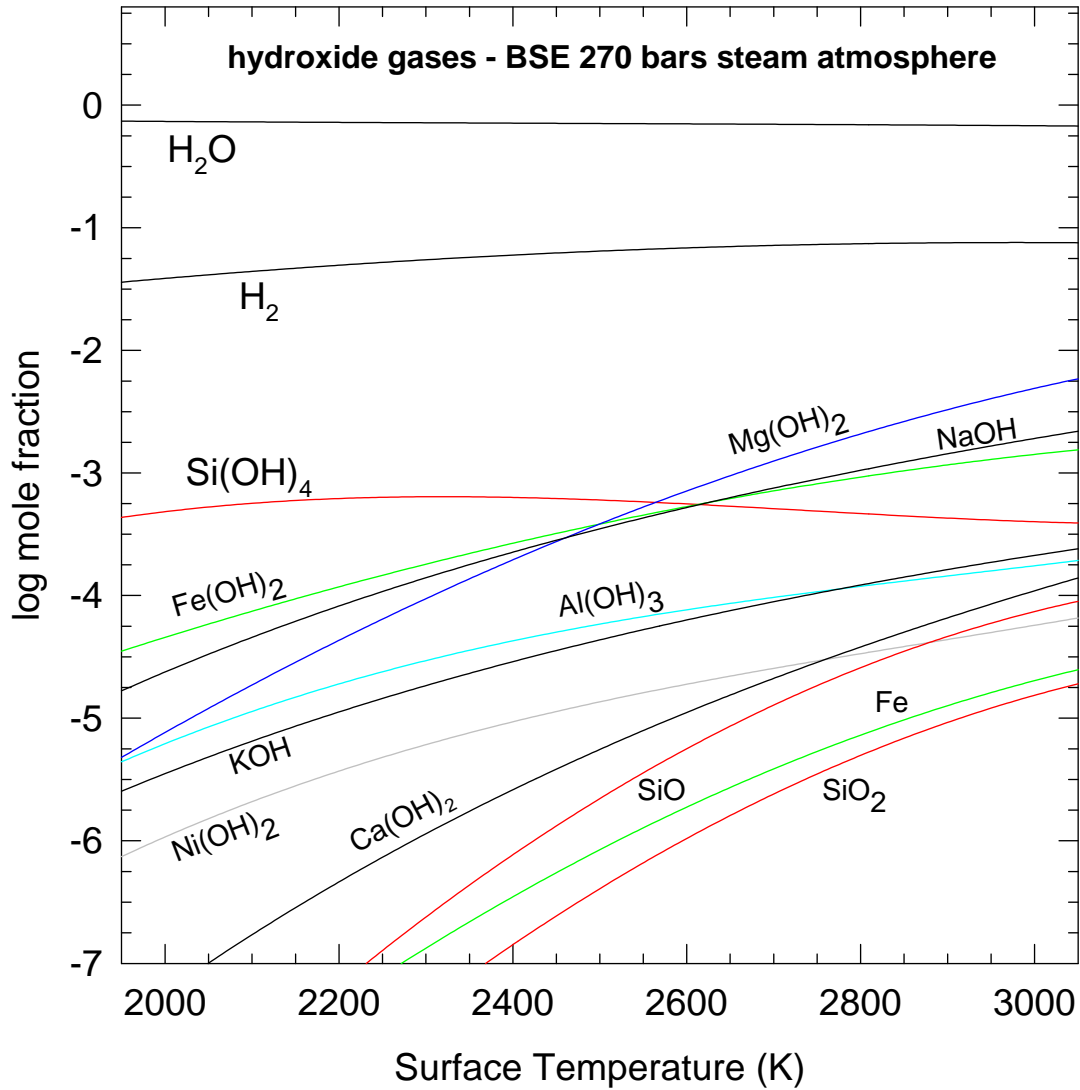

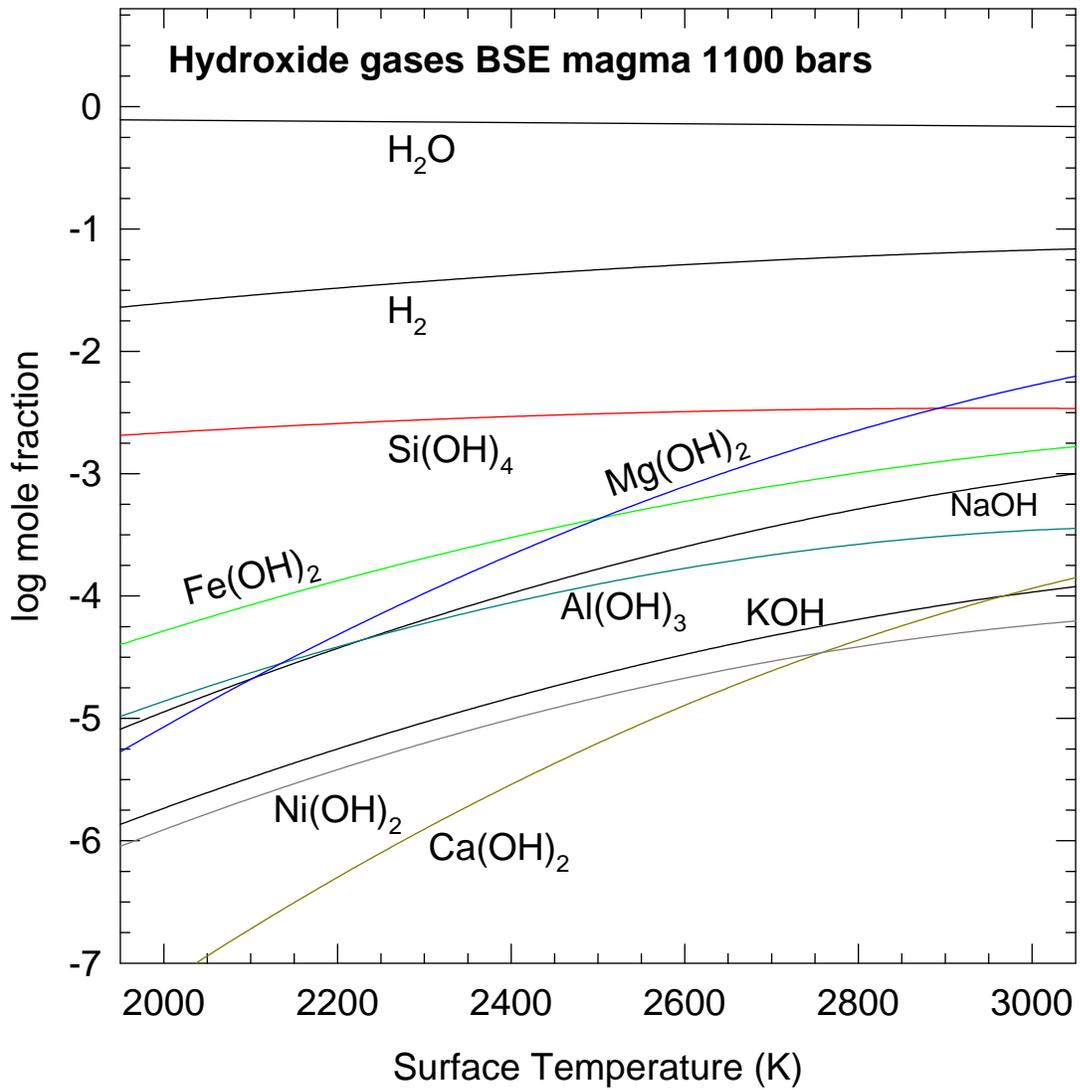

Hydroxide gases BSE magma 1100 bars

BSE1100H.spw

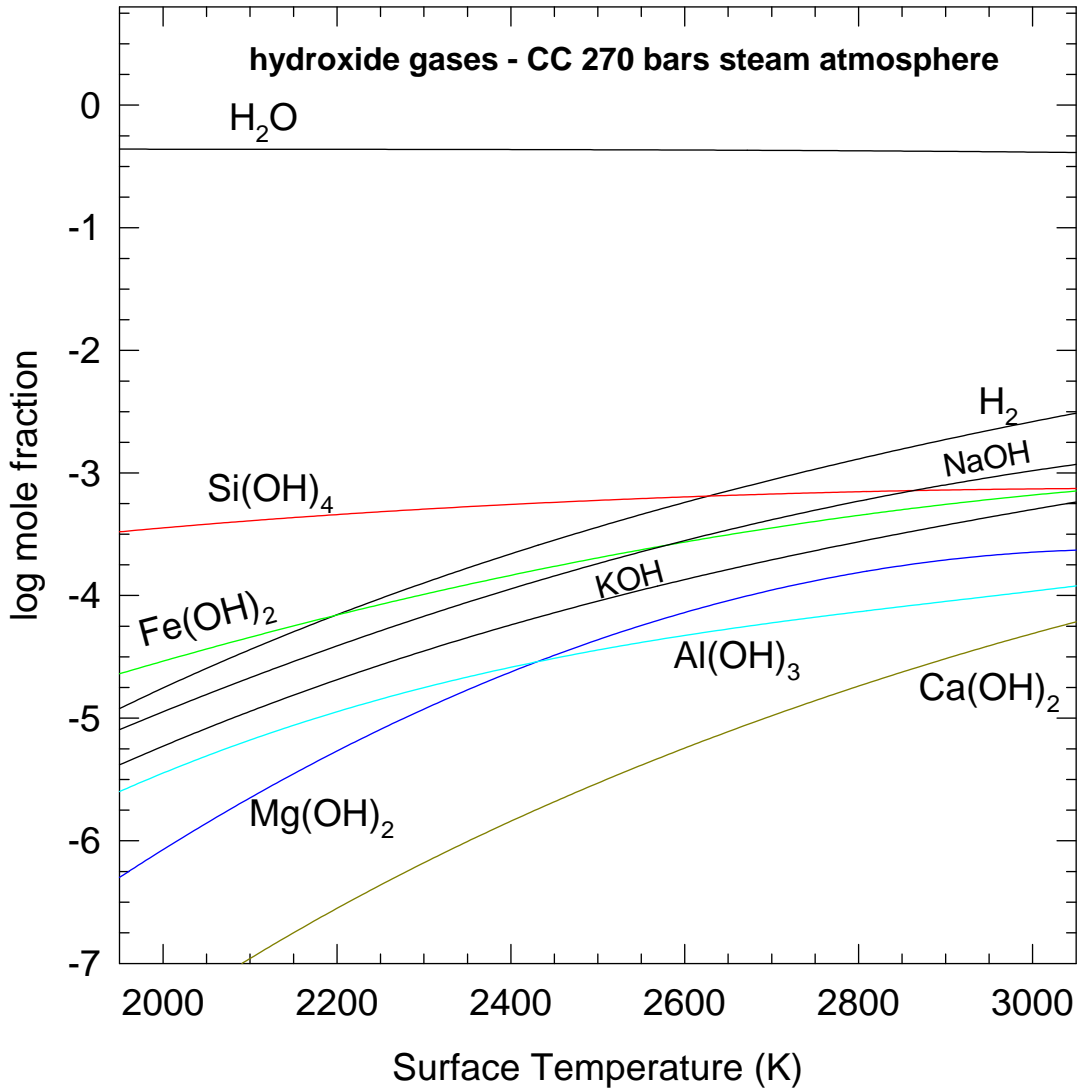

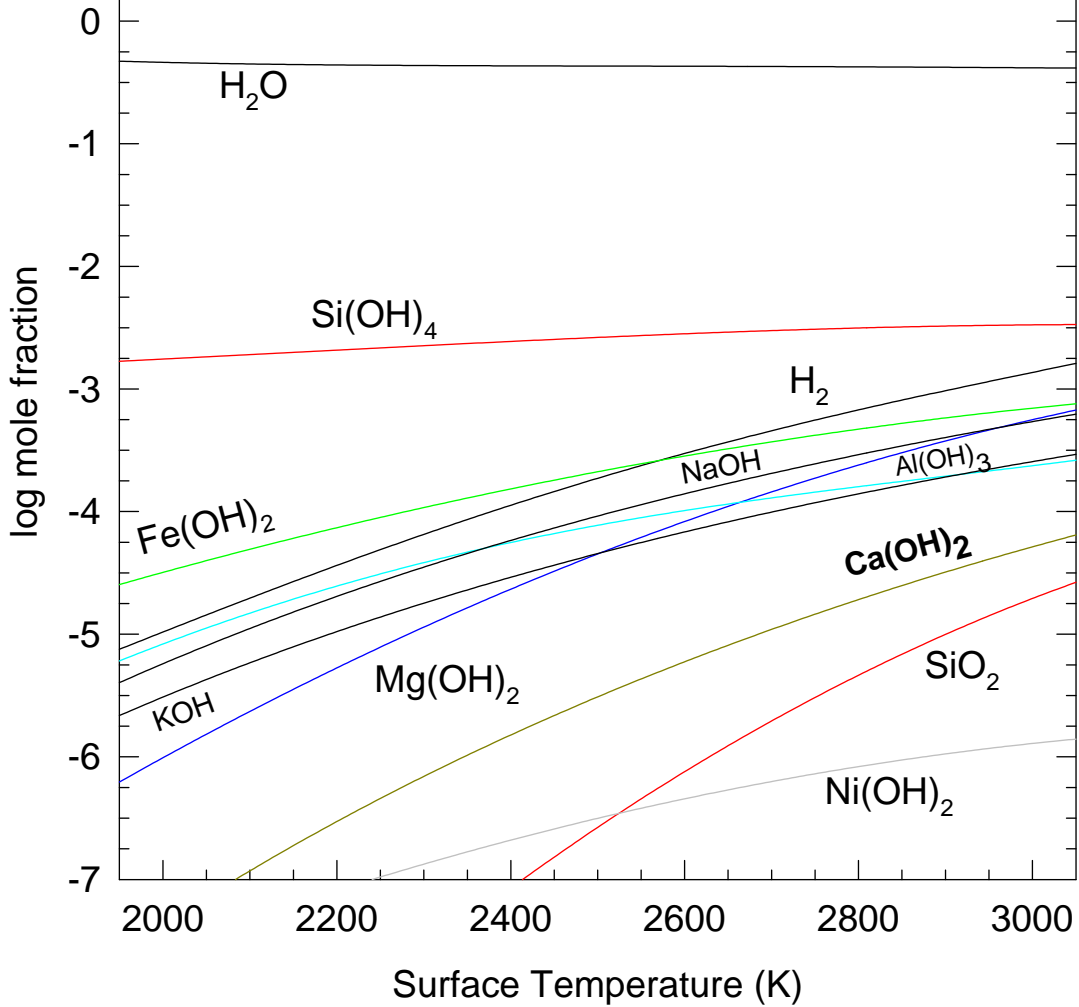

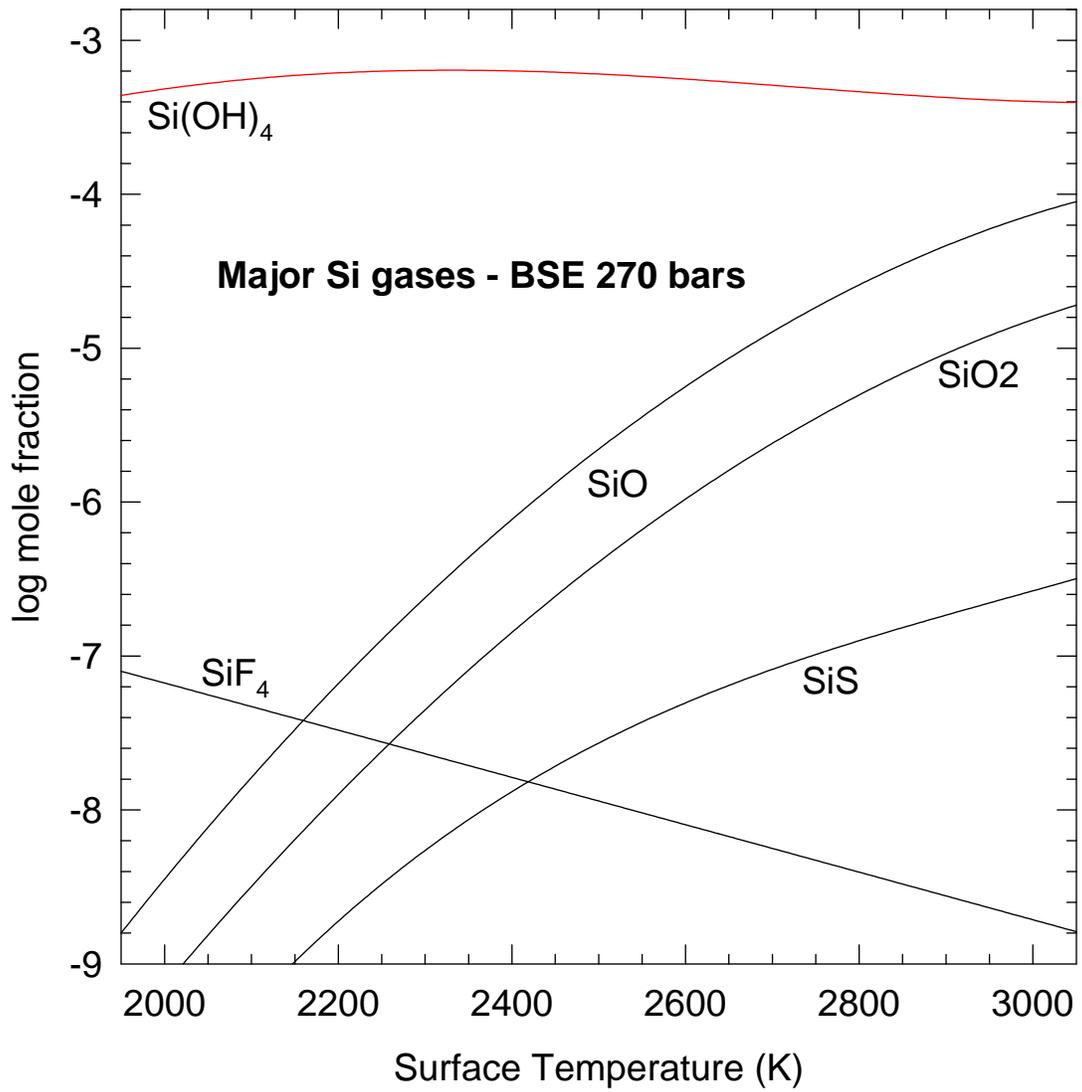

Major Si gases - BSE 270 bars

Si(OH)$_4$

SiO

SiO2

SiF$_4$

SiS

log mole fraction

Surface Temperature (K)

BSEMelts.spw

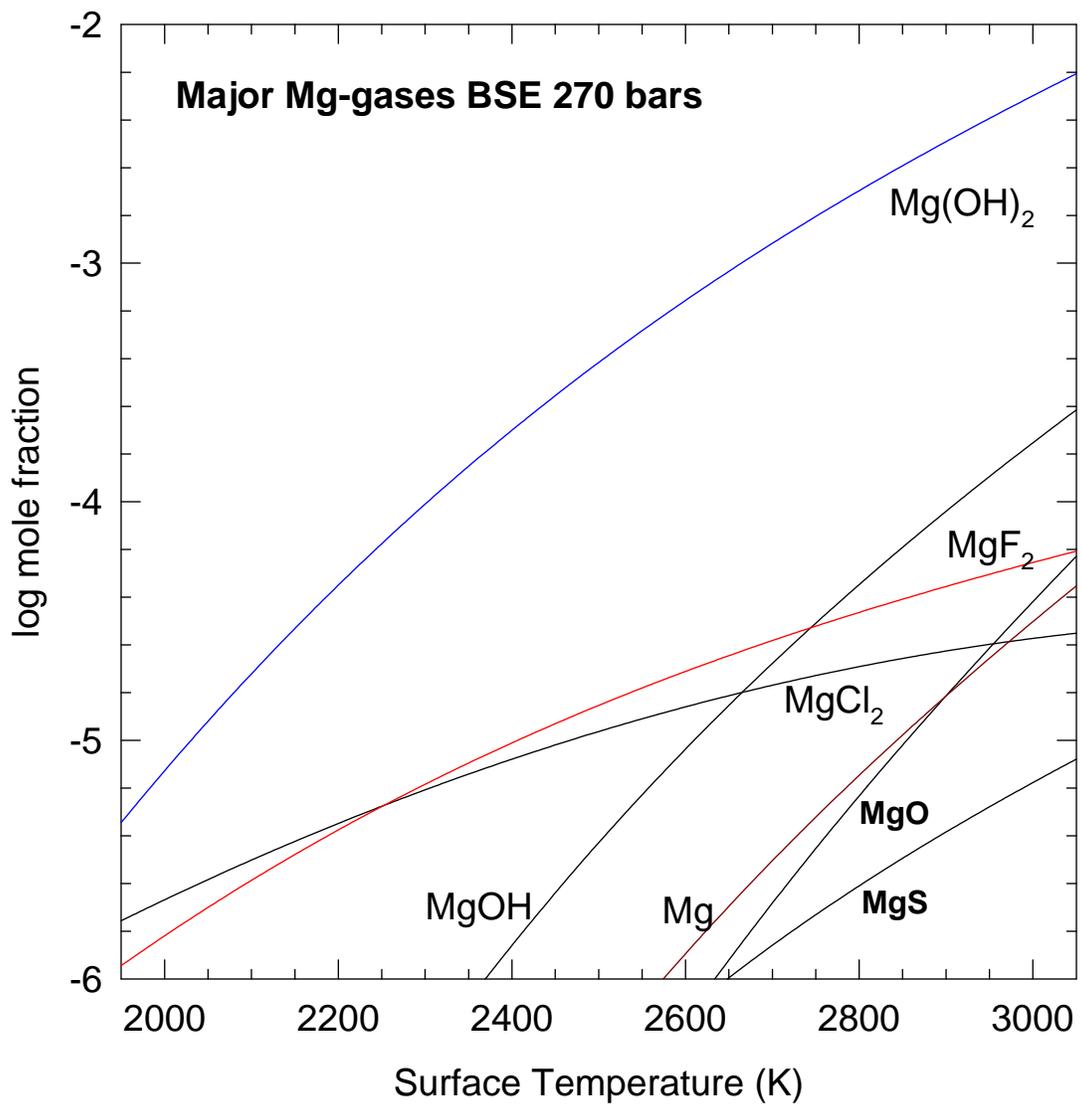

BSEMelts.spw

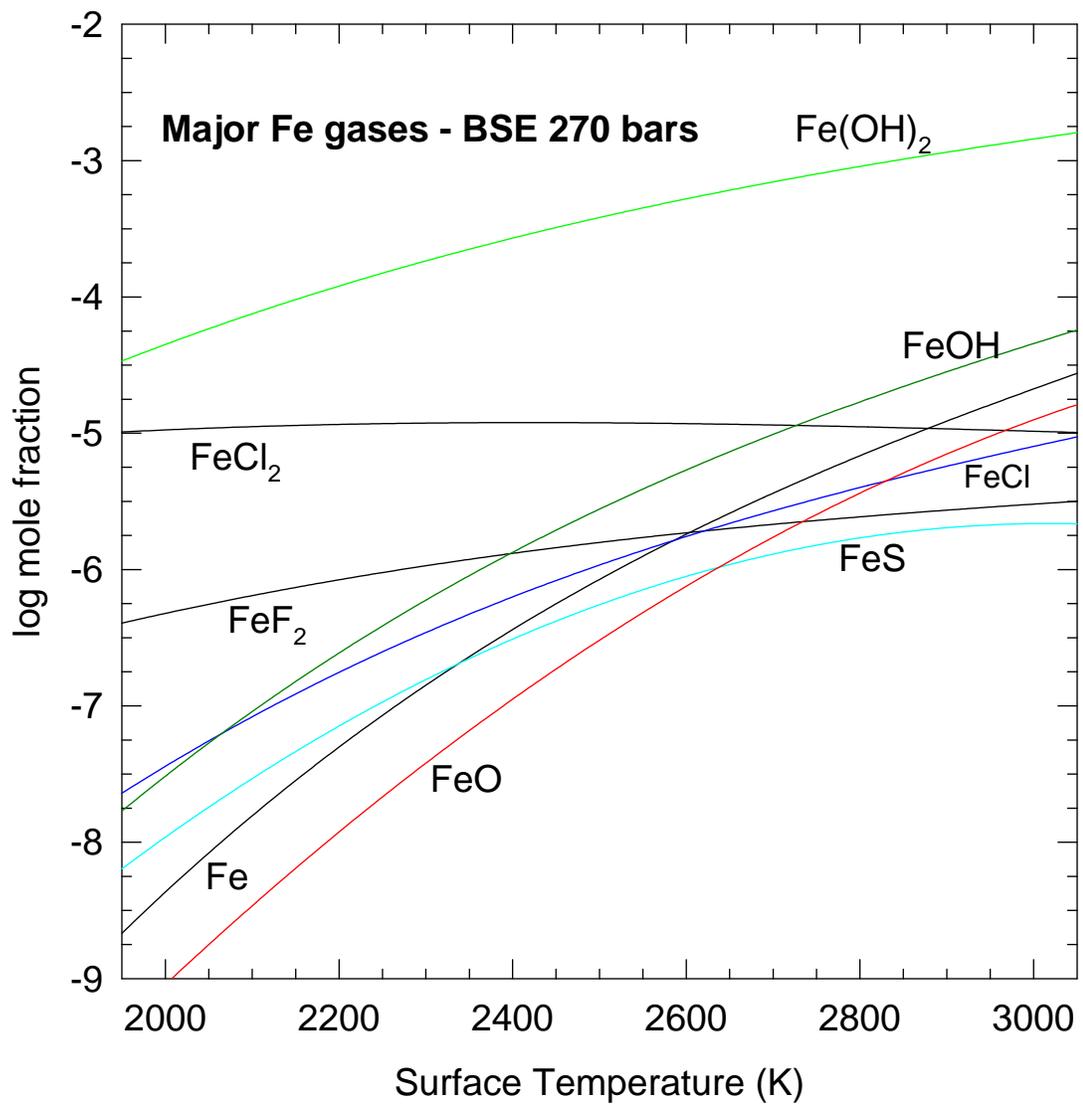

**Major Fe gases - BSE 270 bars**

Fe(OH)₂

FeOH

FeCl₂

FeF₂

FeCl

FeS

FeO

Fe

log mole fraction

Surface Temperature (K)

BSEMelts.spw

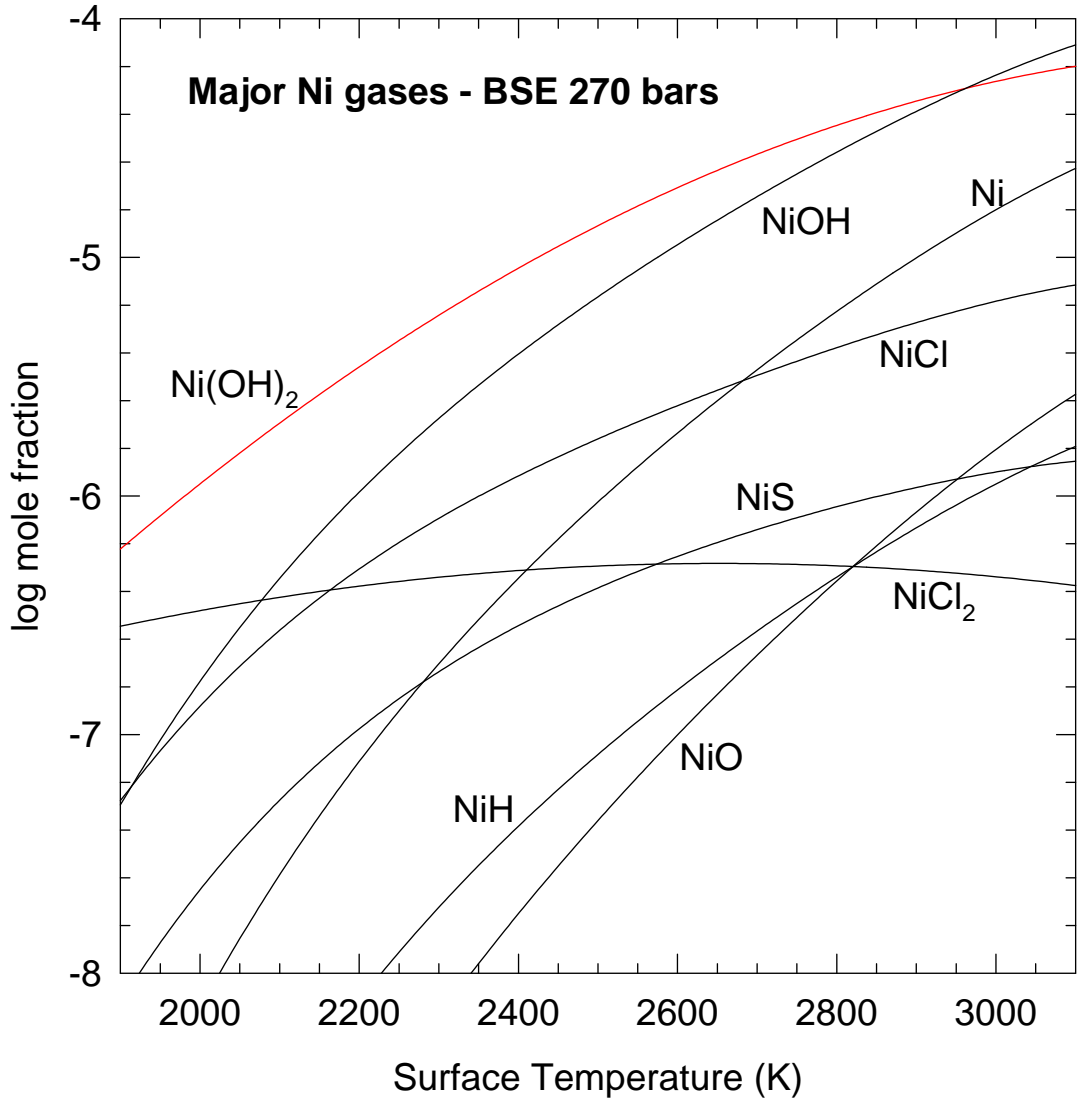

Major Ni gases - BSE 270 bars

Ni-BSE.spw

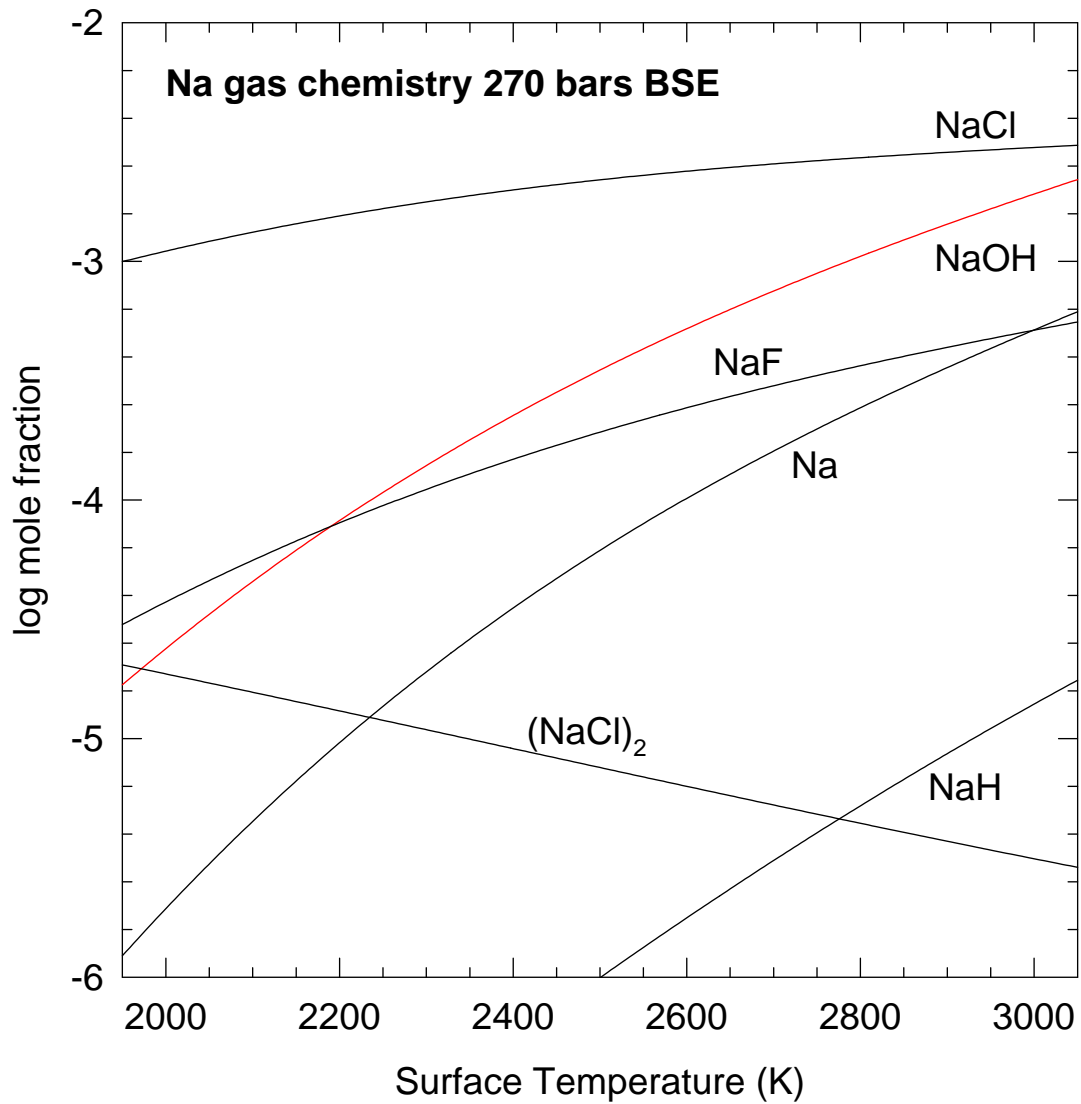

BSEMelts.spw

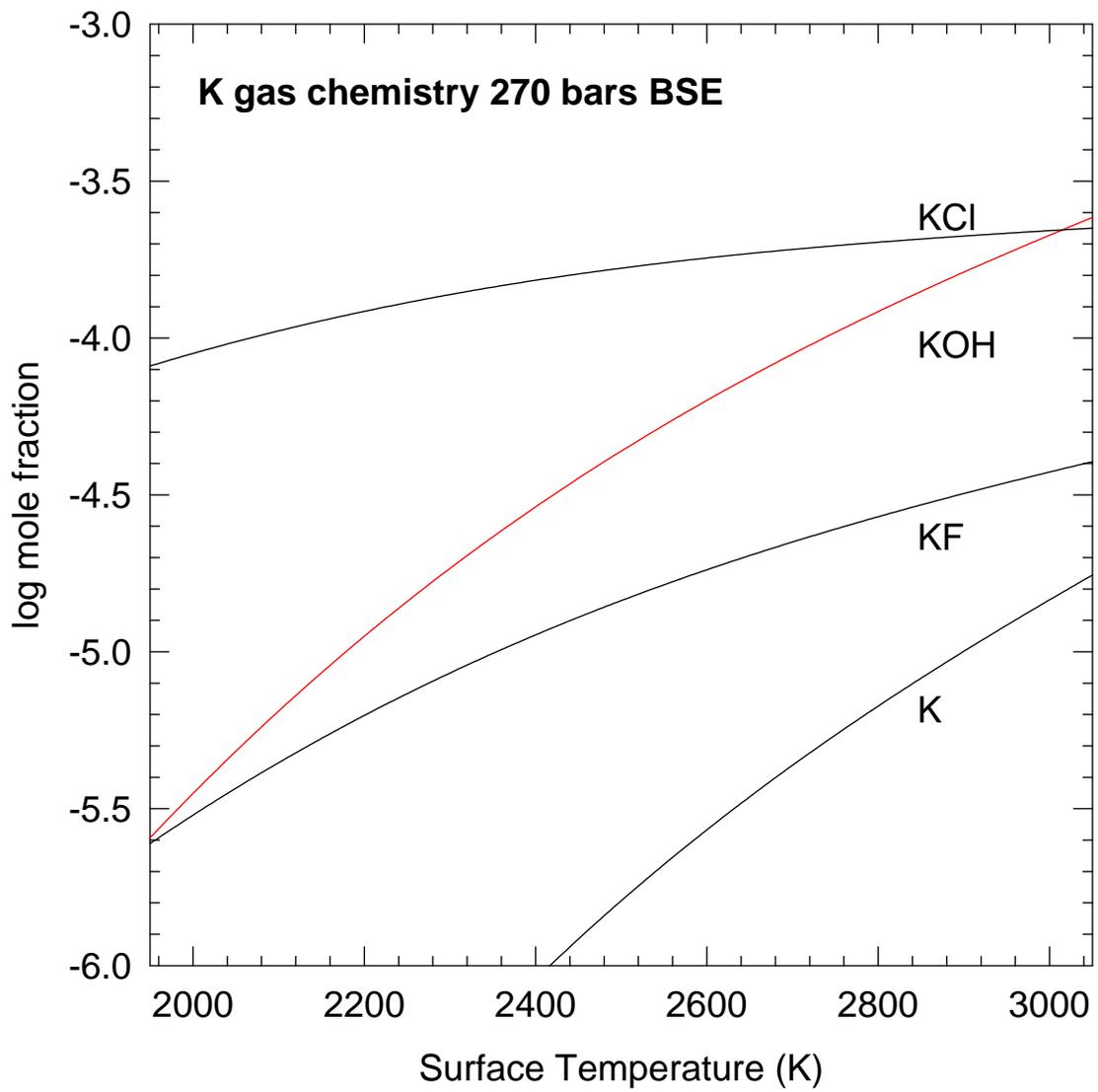

K gas chemistry 270 bars BSE

KCl

KOH

KF

K

log mole fraction

Surface Temperature (K)

BSEMelts.spw

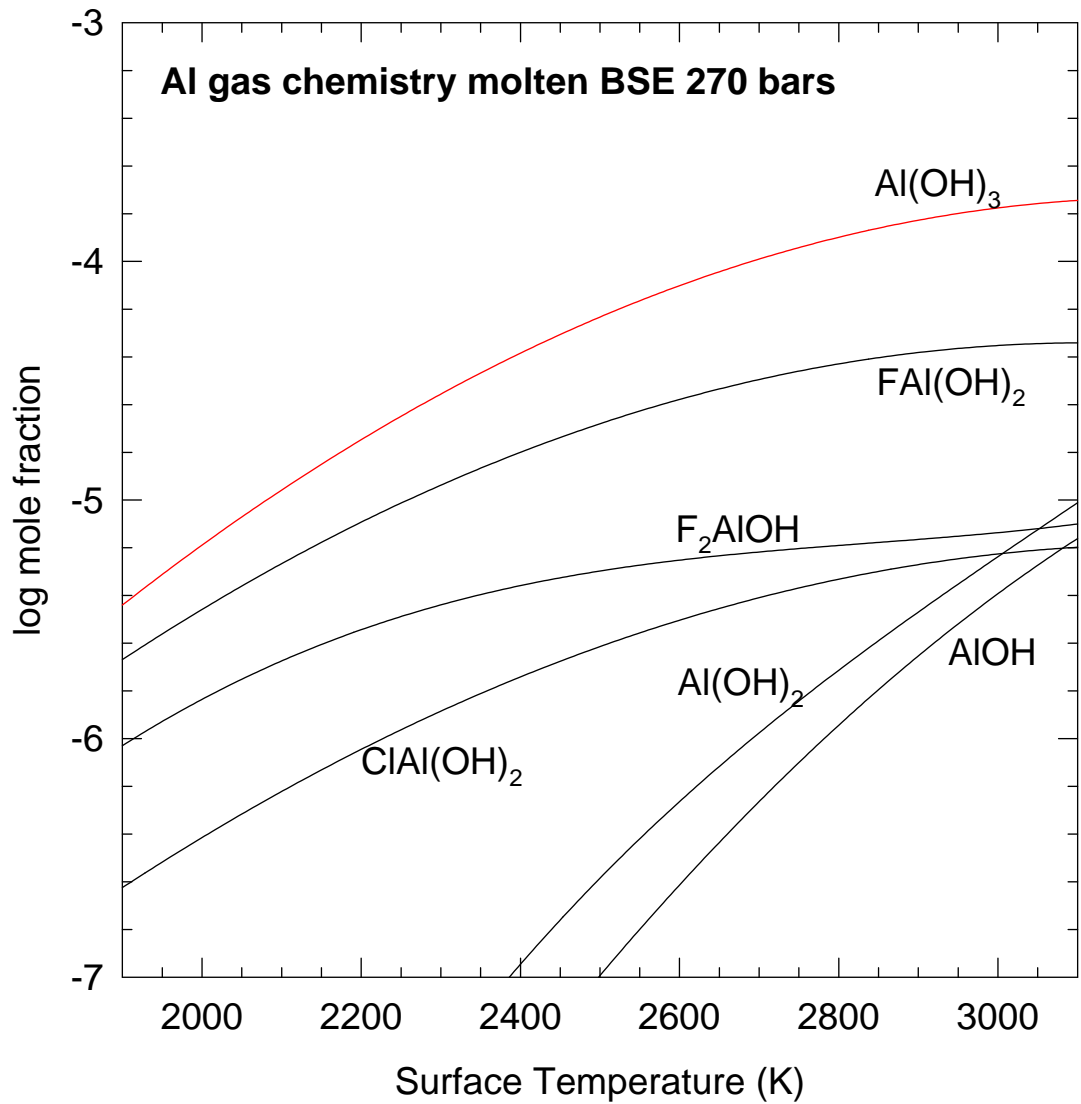

Al gas chemistry molten BSE 270 bars

Al-BSE.spw

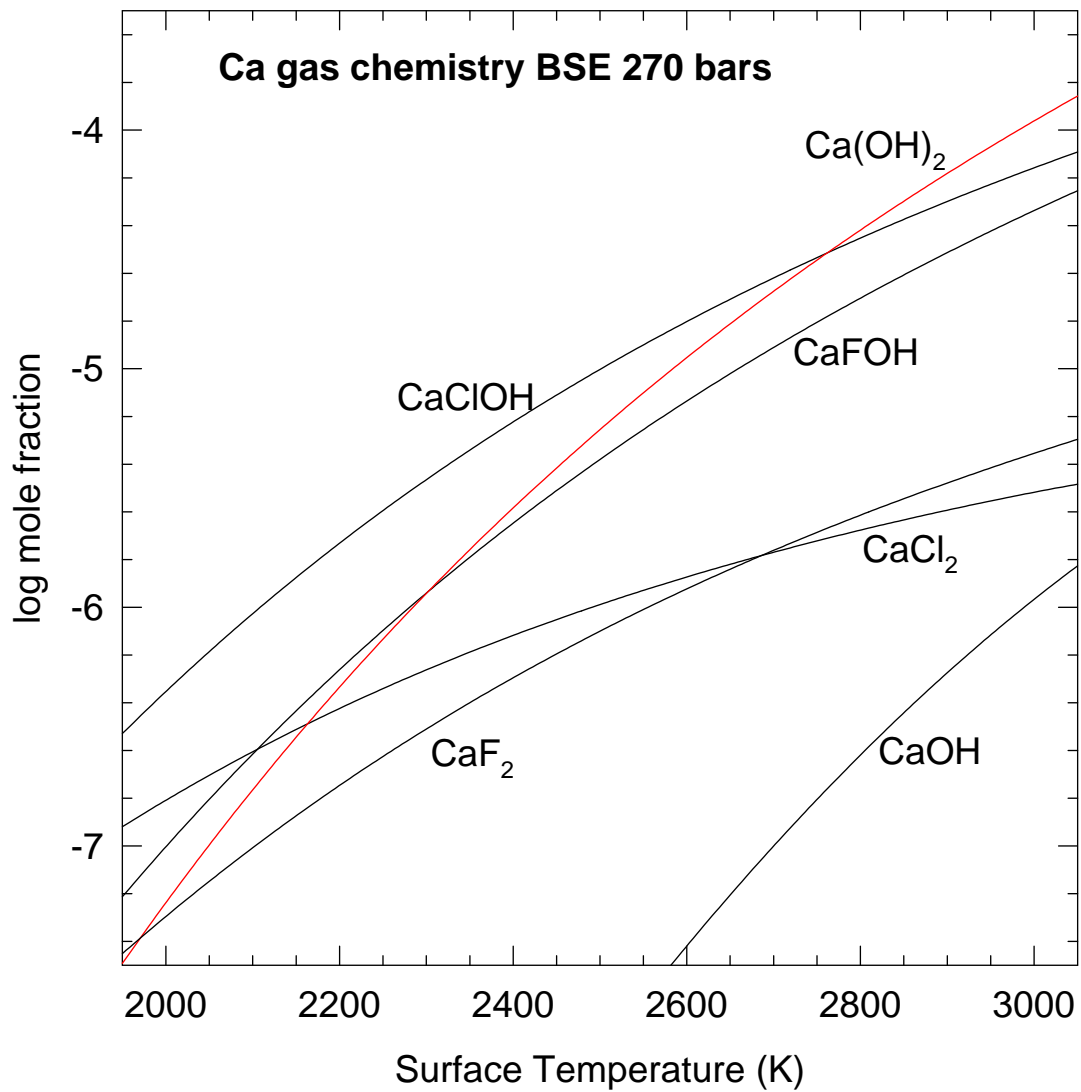

Ca gas chemistry BSE 270 bars

Ca-BSE.spw

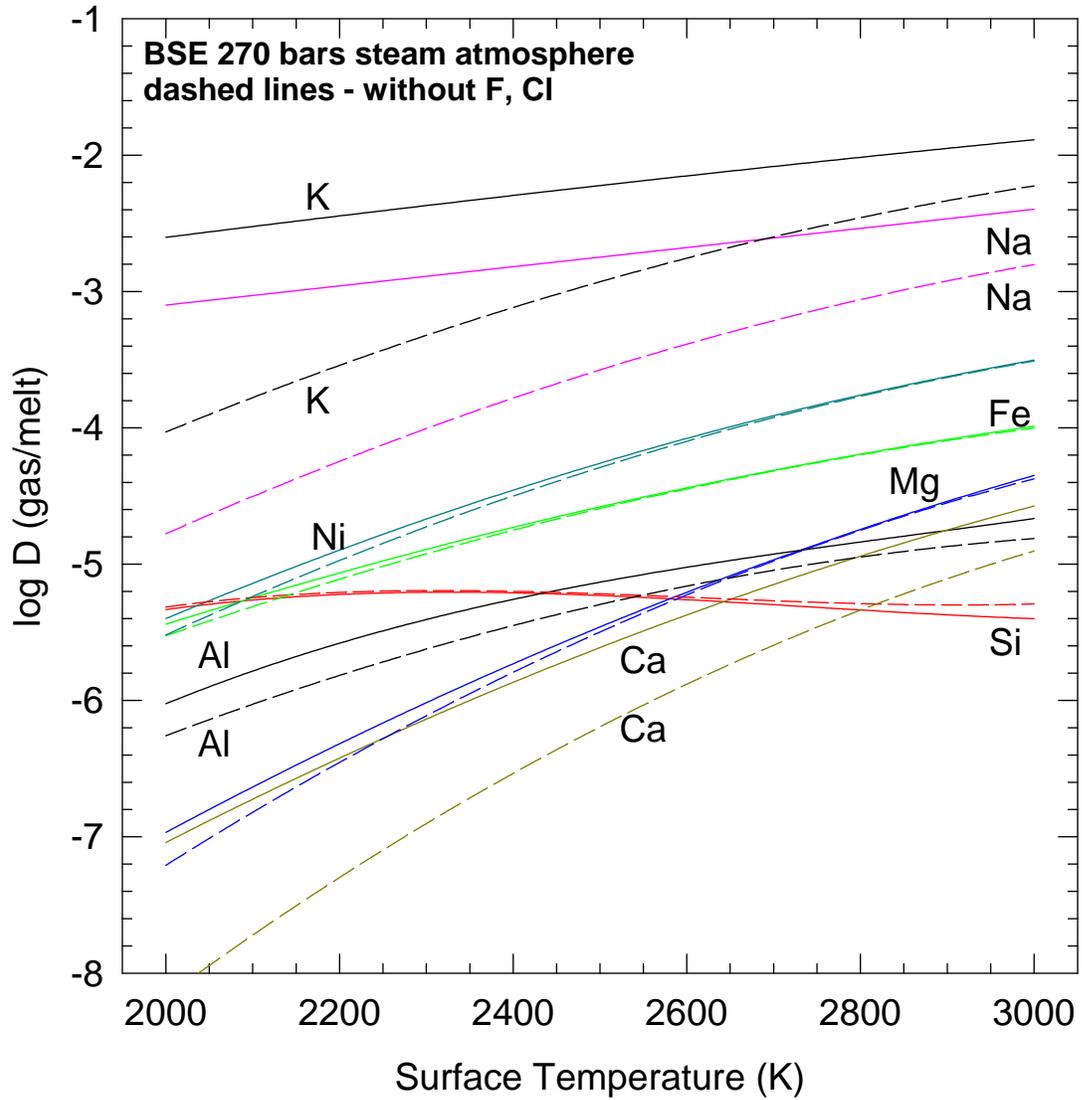

BSE-KD.spw

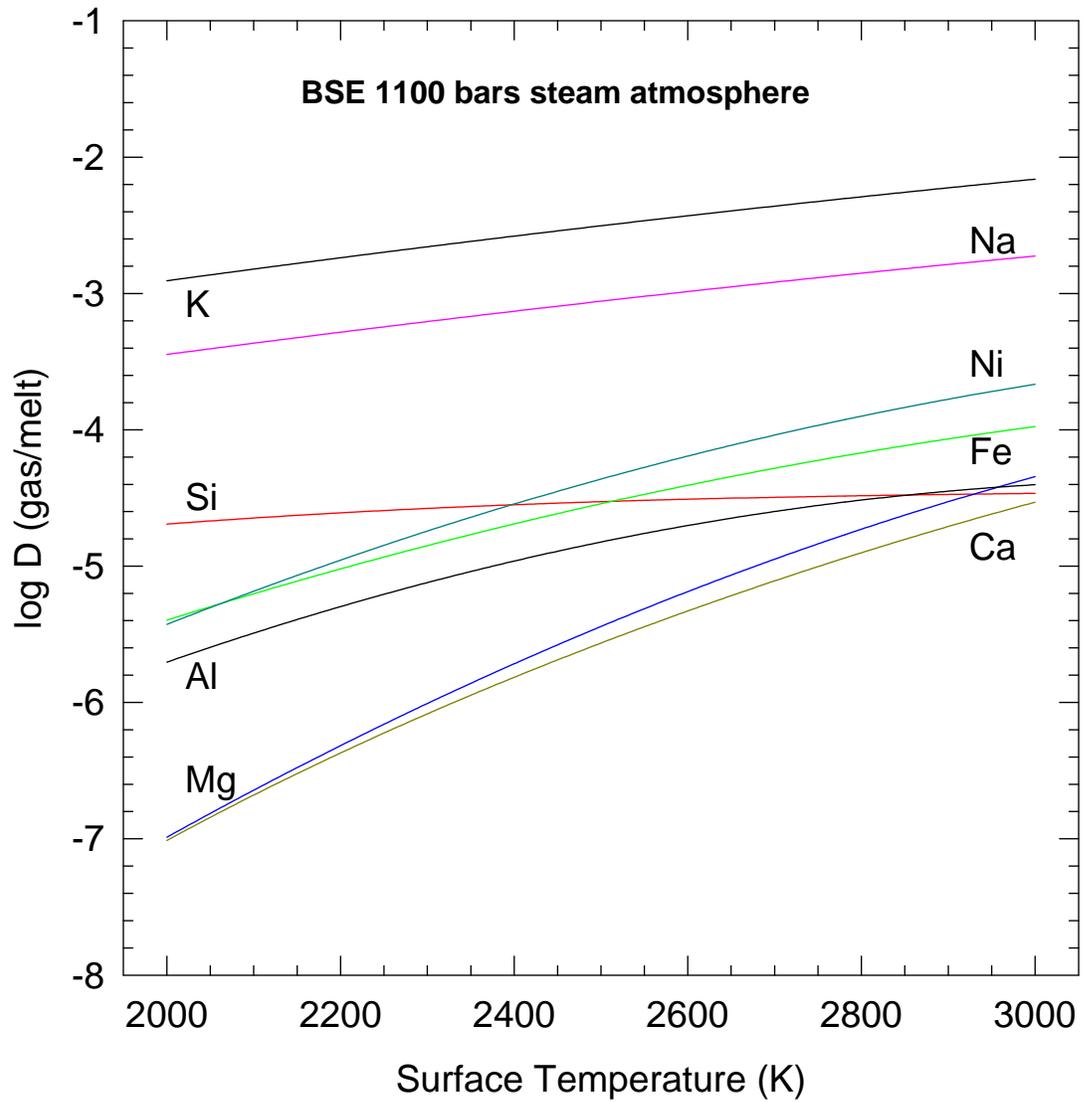

BSE1100D.spw

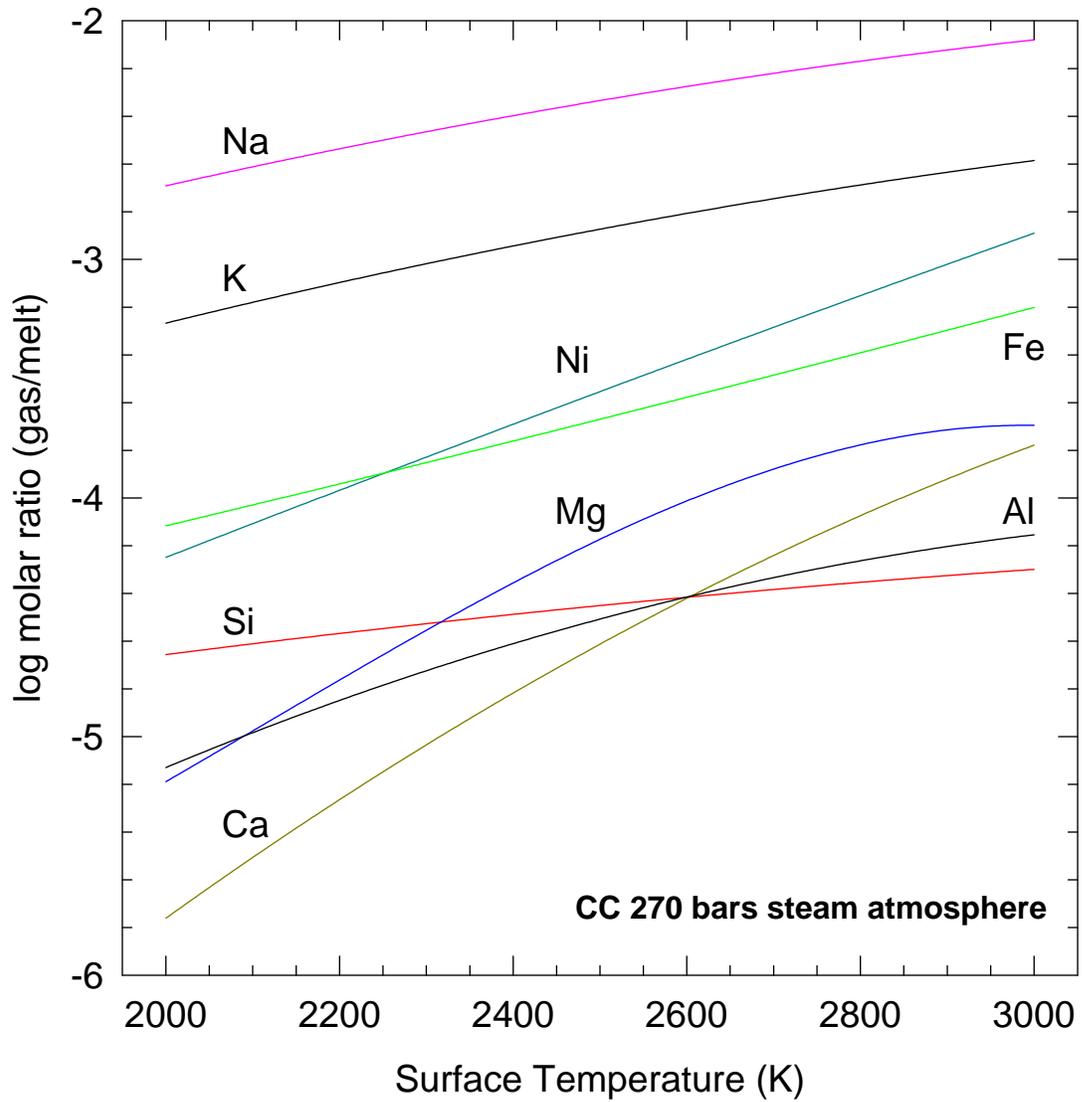

CC270KD.spw

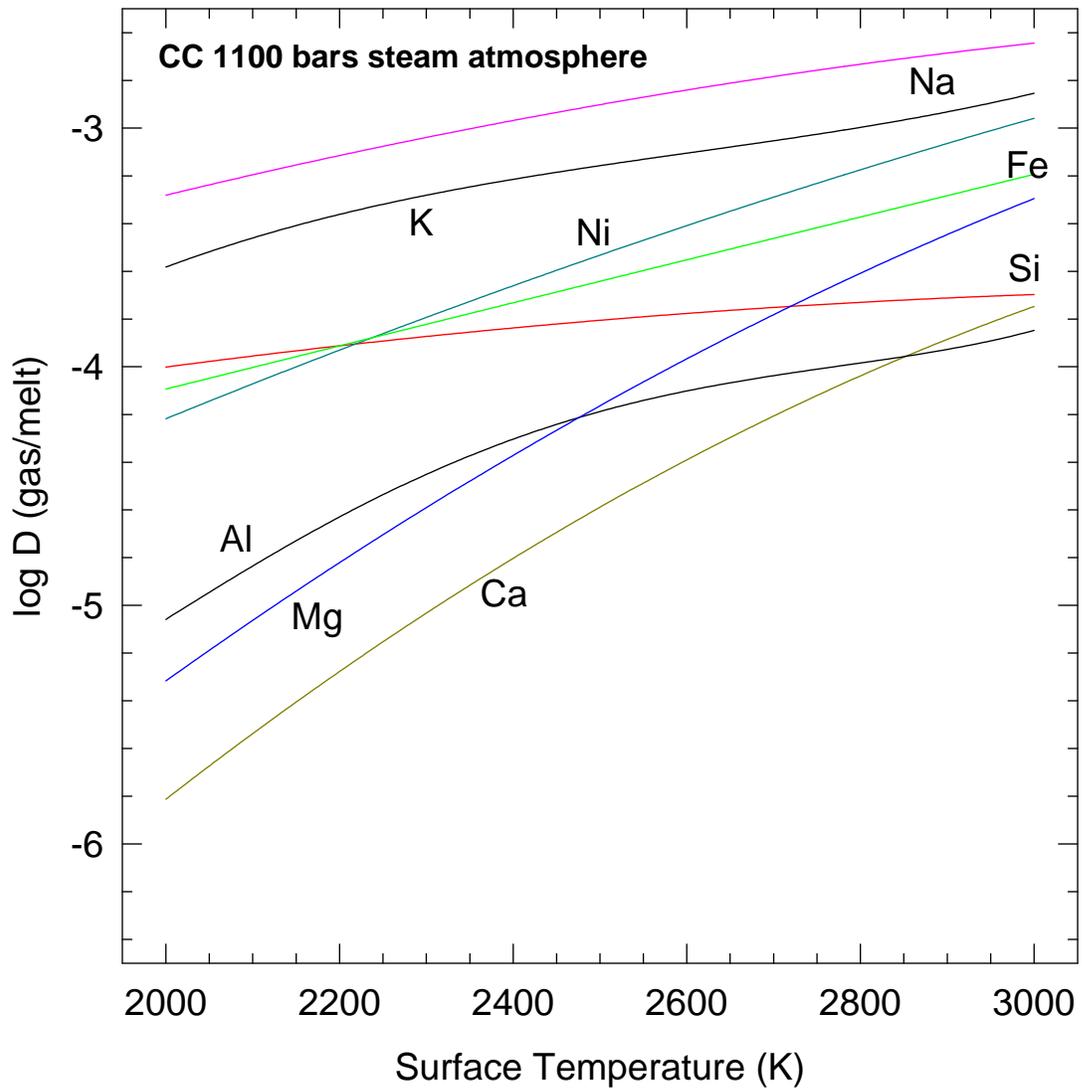

CC1100KD.spw

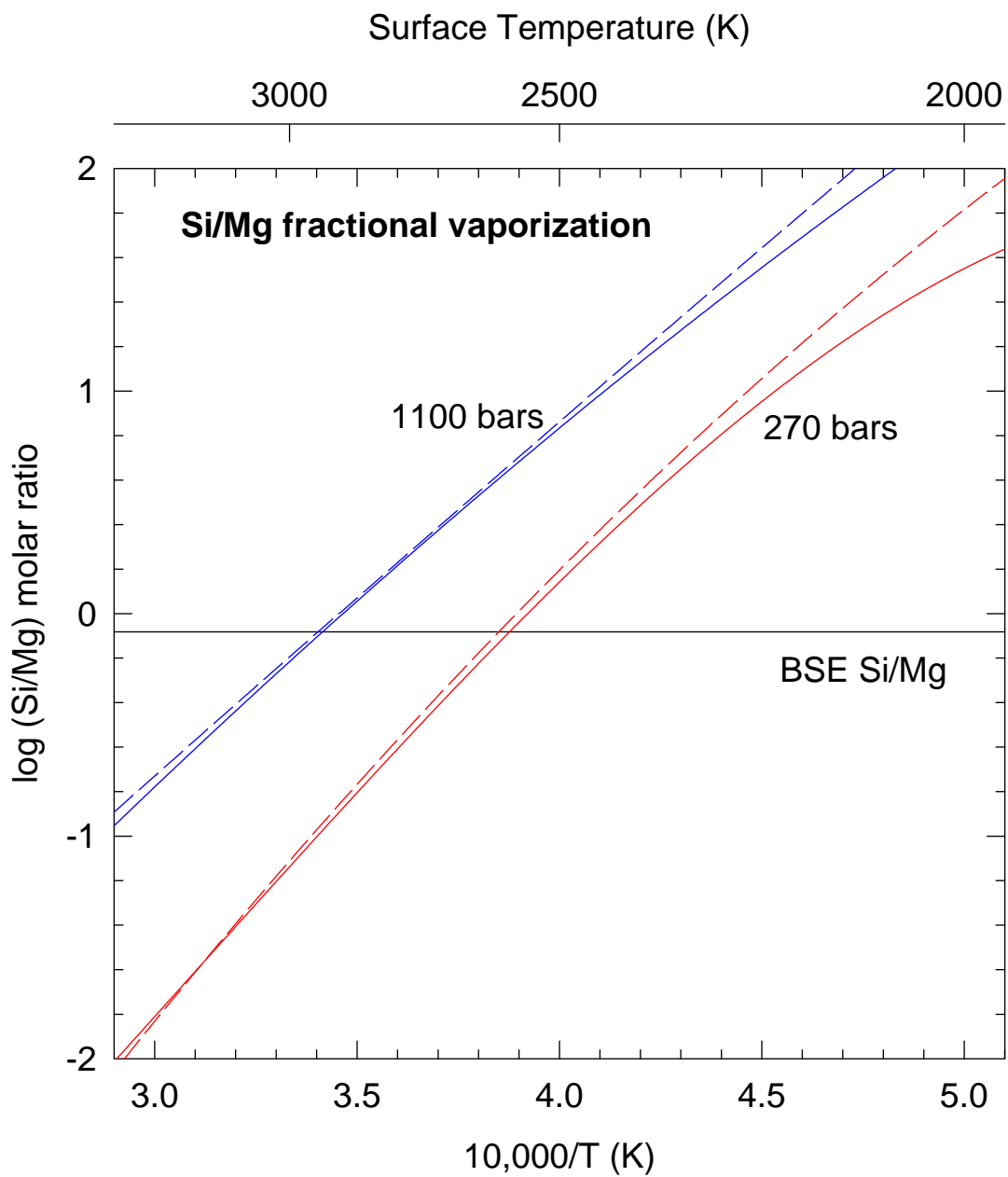

BSEMelts.spw

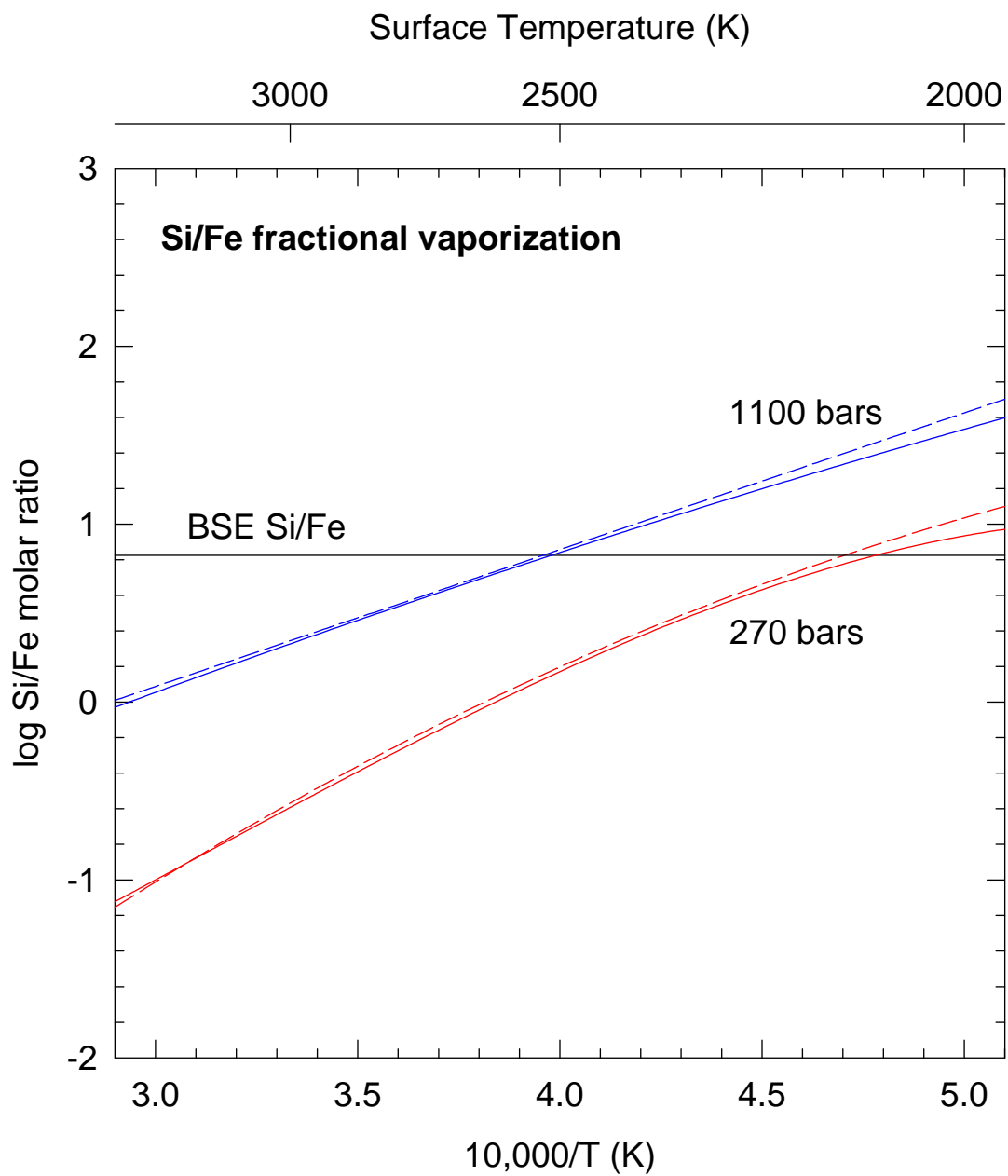

BSEMelts.spw

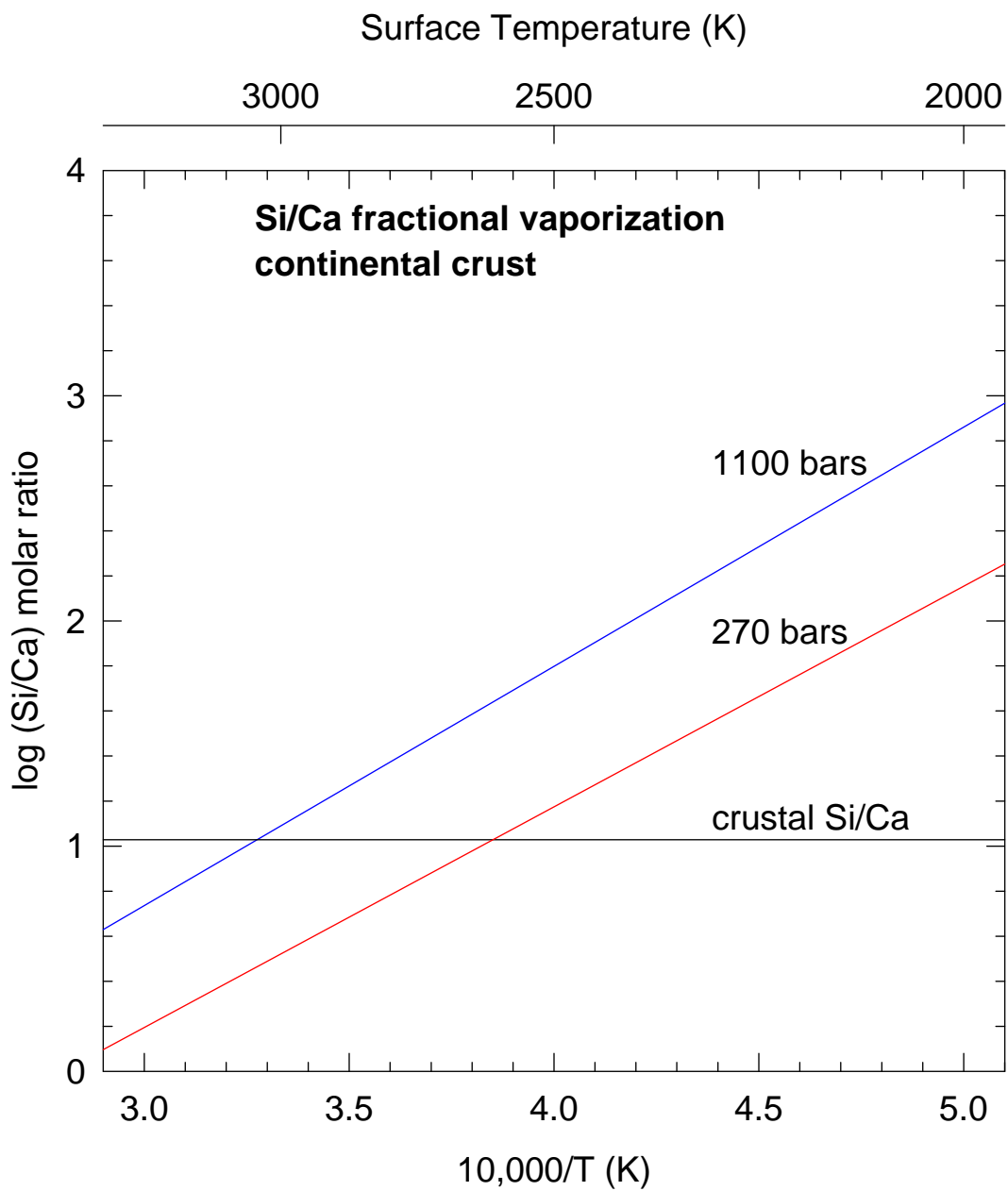

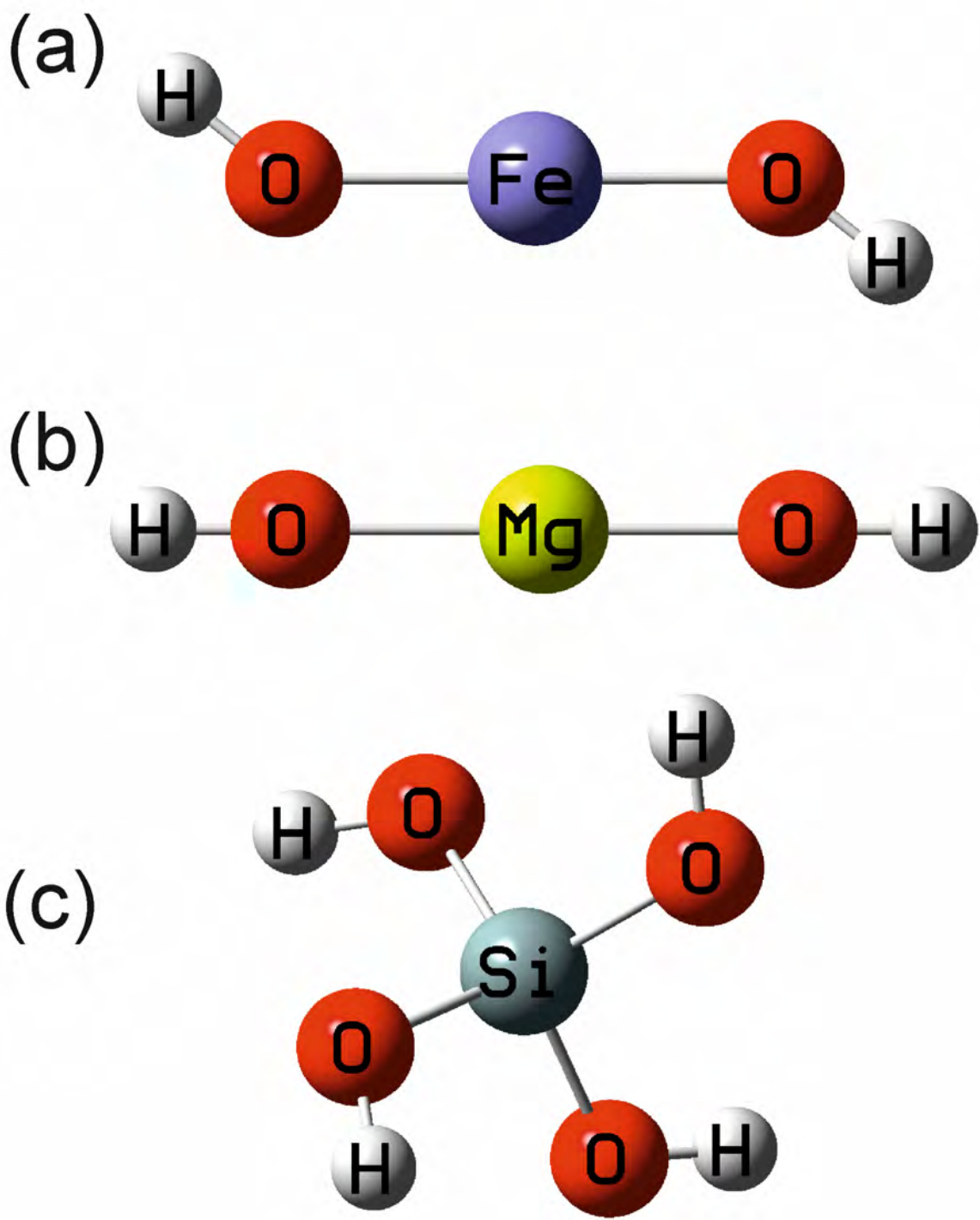

Figure 29. Optimised geometries at the CAM-B3LYP/6-311+g(2d,p) level of theory. Scale: the Fe-O bond length in $Fe(OH)_2$ is 1.77 Å.

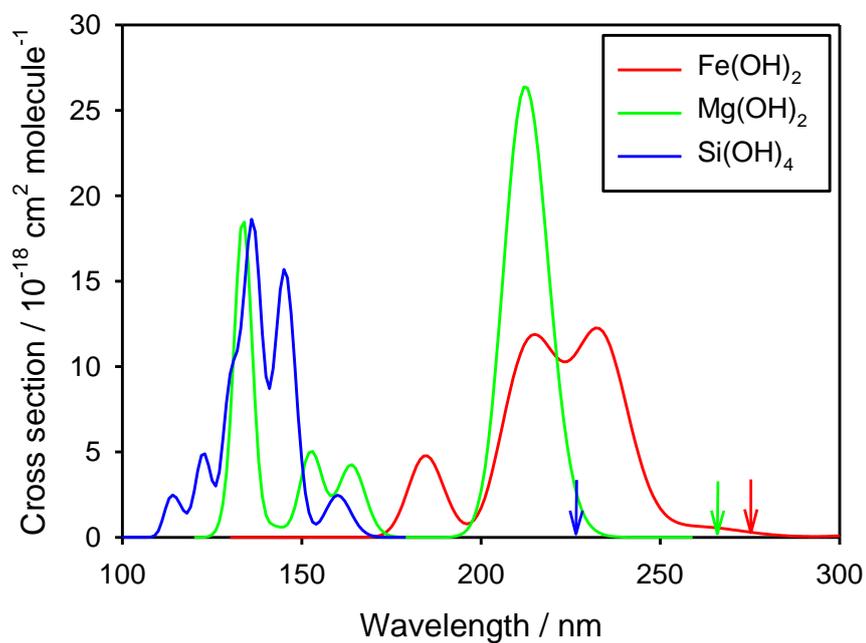

Figure 30. Calculated absorption cross sections as a function of wavelength for $Fe(OH)_2$ (red line), $Mg(OH)_2$ (green line) and $Si(OH)_4$ (blue line). The arrows in corresponding colours indicate the thermodynamic threshold for photolysis.

Table 1. Estimated partial pressures of Si-bearing gases

| $P_i$ (bar) | A | B | C | D |
|---|---|---|---|---|
| SiO | **$8.6 \times 10^{-6}$** | $1.6 \times 10^{-13}$ | $4.4 \times 10^{-14}$ | $2.9 \times 10^{-8}$ |
| $SiO_2$ | **$6.9 \times 10^{-7}$** | $6.4 \times 10^{-11}$ | $7.6 \times 10^{-12}$ | $6.9 \times 10^{-7}$ |
| SiO(OH) | $3.4 \times 10^{-11}$ | $6.7 \times 10^{-15}$ | $1.9 \times 10^{-14}$ | $1.3 \times 10^{-8}$ |
| $SiO(OH)_2$ | $3.3 \times 10^{-11}$ | $5.9 \times 10^{-9}$ | $4.9 \times 10^{-7}$ | $1.5 \times 10^{-3}$ |
| $Si(OH)_4$ | $3.9 \times 10^{-15}$ | **$4.1 \times 10^{-6}$** | **$3.0 \times 10^{-1}$** | **8.25** |
| $Si_2O(OH)_6$ | $7.5 \times 10^{-26}$ | $2.0 \times 10^{-11}$ | $5.4 \times 10^{-4}$ | $7.2 \times 10^{-3}$ |
| $\Sigma$all Si-gases | $9.3 \times 10^{-6}$ | $4.1 \times 10^{-6}$ | $3.0 \times 10^{-1}$ | 8.25 |

A: T = 2000 K, total P ($P_T$) = $4 \times 10^{-5}$ bar, liquid silica + steam
B: T = 1573 K, $P_T$ = 1 bar, cristobalite + 90% steam – 10% $O_2$ gas mixture
C: T = 1500 K, $P_T$ = 270 bar, cristobalite + steam
D: T = 2000 K, $P_T$ = 1,100 bar, liquid silica in steam (75%) – $CO_2$ (25%) atmosphere

Table 2. Composition of Earth's continental crust[a]

| Oxide | mole % |
| --- | --- |
| $SiO_2$ | 68.92 |
| $Al_2O_3$ | 9.92 |
| CaO | 6.46 |
| MgO | 6.08 |
| $Na_2O$ | 3.45 |
| $Fe_2O_3$[b] | 2.60 |
| $K_2O$ | 1.84 |
| $TiO_2$ | 0.56 |
| MnO | 0.09 |
| $P_2O_5$ | 0.08 |
| Total[c] | 100.01% |

[a]Major elements, Wedepohl (1995).
[b]25% $Fe^{3+}$ in the crust; most Fe in the BSE is $Fe^{2+}$.
[c]Also includes 0.008% $Cr_2O_3$ and 0.0062% NiO.
Volatiles that are not included in the sum are
1.91% $H_2O$, 0.64%C, 0.44% $CO_2$, 0.028% N,
0.18% F, 0.14% S, and 0.086% Cl.

| Oxide | mole % |
|---|---|
| MgO | 47.67 |
| $SiO_2$ | 39.48 |
| $FeO$[b] | 5.90 |
| CaO | 3.40 |
| $Al_2O_3$ | 2.30 |
| $Na_2O$ | 0.29 |
| NiO | 0.17 |
| $TiO_2$ | 0.14 |
| $Cr_2O_3$ | 0.13 |
| MnO | 0.10 |
| $K_2O$ | 0.0174 |
| $H_2O$ | 0.31 |
| $CO_2$ | 0.044 |
| N | $7.5 \times 10^{-4}$ |
| F | $6.9 \times 10^{-3}$ |
| Cl | $4.4 \times 10^{-3}$ |
| S | 0.0326 |
| $P_2O_5$ | $7.3 \times 10^{-3}$ |
| Total[c] | 100.00% |

Table 3. Composition of the Bulk Silicate Earth (BSE)[a]

[a]Computed from data in (Palme & O'Neill 2014).
[b]Most Fe in the BSE is $Fe^{2+}$; the crust is 25% $Fe^{3+}$

Table 4. Hydroxide gas partial pressures at 220.64 bars steam

| Gas | 1000 K | 1500 K | 2000 K |
|---|---|---|---|
| $Si(OH)_4$ | 0.029 | 0.20 | 0.59 |
| $Mg(OH)_2$ | $2.0 \times 10^{-10}$ | $2.7 \times 10^{-5}$ | 0.010 |
| $Fe(OH)_2$ | $1.5 \times 10^{-6}$ | $3.5 \times 10^{-3}$ | 0.11 |
| $Ca(OH)_2$ | $1.7 \times 10^{-10}$ | $1.2 \times 10^{-4}$ | 0.040 |
| $Al(OH)_3$ | $2.3 \times 10^{-7}$ | $5.0 \times 10^{-4}$ | 0.024 |
| $Ni(OH)_2$ | $1.3 \times 10^{-7}$ | $1.6 \times 10^{-3}$ | 0.19 |

Table 5. Gas/magma molar distribution coefficients for rocky elements (log D)[a]

| Element | 2000 | 2200 | 2400 | 2600 | 2800 | 3000 |
|---------|------|------|------|------|------|------|
| \multicolumn{7}{c}{BSE 270 bars steam atmosphere} | | | | | | |
| Si | –5.332 | –5.229 | –5.199 | –5.268 | –5.334 | –5.400 |
| Mg | –6.967 | –6.319 | –5.729 | –5.205 | –4.758 | –4.347 |
| Fe | –5.443 | –5.061 | –4.727 | –4.441 | –4.197 | –3.984 |
| Na | –3.116 | –2.950 | –2.806 | –2.673 | –2.541 | –2.405 |
| K | –2.607 | –2.440 | –2.295 | –2.158 | –2.023 | –1.889 |
| Al | –6.023 | –5.584 | –5.250 | –5.030 | –4.835 | –4.667 |
| Ca | –7.046 | –6.415 | –5.862 | –5.377 | –4.949 | –4.569 |
| Ni | –5.416 | –4.870 | –4.451 | –4.090 | –3.772 | –3.491 |
| \multicolumn{7}{c}{BSE 1100 bars steam atmosphere} | | | | | | |
| Si | –4.692 | –4.609 | –4.550 | –4.509 | –4.482 | –4.465 |
| Mg | –6.989 | –6.316 | –5.712 | –5.187 | –4.734 | –4.340 |
| Fe | –5.400 | –5.015 | –4.687 | –4.409 | –4.173 | –3.971 |
| Na | –3.450 | –3.278 | –3.127 | –2.988 | –2.854 | –2.721 |
| K | –2.909 | –2.732 | –2.576 | –2.432 | –2.294 | –2.158 |
| Al | –5.721 | –5.276 | –4.941 | –4.722 | –4.538 | –4.384 |
| Ca | –7.012 | –6.370 | –5.815 | –5.329 | –4.902 | –4.531 |
| Ni | –5.444 | –4.927 | –4.539 | –4.206 | –3.913 | –3.652 |
| \multicolumn{7}{c}{Continental crust 270 bars steam atmosphere} | | | | | | |
| Si | –4.661 | –4.562 | –4.481 | –4.419 | –4.364 | –4.292 |
| Mg | –5.188 | –4.761 | –4.367 | –3.992 | –3.791 | –3.691 |
| Fe | –4.108 | –3.953 | –3.765 | –3.570 | –3.381 | –3.208 |
| Na | –2.695 | –2.530 | –2.393 | –2.279 | –2.174 | –2.075 |
| K | –3.271 | –3.089 | –2.940 | –2.811 | –2.693 | –2.580 |
| Al | –5.136 | –4.844 | –4.592 | –4.427 | –4.278 | –4.144 |
| Ca | –5.766 | –5.256 | –4.814 | –4.425 | –4.079 | –3.774 |
| Ni | –4.258 | –3.950 | –3.692 | –3.426 | –3.154 | –2.885 |
| \multicolumn{7}{c}{Continental crust 1100 bars steam atmosphere} | | | | | | |
| Si | –4.000 | –3.913 | –3.837 | –3.775 | –3.729 | –3.697 |
| Mg | –5.320 | –4.822 | –4.354 | –3.975 | –3.618 | –3.287 |
| Fe | –4.084 | –3.926 | –3.734 | –3.545 | –3.363 | –3.200 |
| Na | –3.284 | –3.110 | –2.963 | –2.844 | –2.737 | –2.639 |
| K | –3.576 | –3.376 | –3.212 | –3.078 | –3.024 | –2.845 |
| Al | –5.057 | –4.639 | –4.284 | –4.120 | –3.974 | –3.849 |
| Ca | –5.816 | –5.276 | –4.791 | –4.396 | –4.047 | –3.740 |
| Ni | –4.227 | –3.912 | –3.661 | –3.416 | –3.178 | –2.953 |

[a] D = (moles in gas)/(moles in magma)

Table 6. Photodissociation coefficients for $Fe(OH)_2$, $Mg(OH)_2$ and $Si(OH)_4$

| Reaction | $\Delta H^{\ominus}_{0\,K}$ kJ mol$^{-1}$ | $\lambda_{threshold}$ nm | $J$ s$^{-1}$ |
|---|---|---|---|
| $Fe(OH)_2 + h\nu \rightarrow FeOH + OH$ | 435 | 275 | $2.3 \times 10^{-3}$ |
| $Mg(OH)_2 + h\nu \rightarrow MgOH + OH$ | 450 | 266 | $1.3 \times 10^{-3}$ |
| $Si(OH)_4 + h\nu \rightarrow Si(OH)_3 + OH$ | 529 | 226 | $4.4 \times 10^{-6}$ |